# BlockNet White Paper - Exploring the Blockchain Skills Concept and Best Practice Use Cases

Düdder, B., Fomin, V., Guerpinar, T., Henke, M., Ioannidis, P., Iqbal, M., Janavičienė, V., Matulevičius, R., Straub, N.

December, 2019

**Project acronym**: BlockNet
**Project name**: BlockChain Network Online Education for interdisciplinary European Competence Transfer

**Output type**: Intellectual Output

**Part1**
**Activity leader**: University of Tartu
**Contributors**: TU Dortmund University, Vilnius University, University of Copenhagen

**Acknowledgement:** The persons of University of Tartu involved in preparation of this document part are Raimundas Matulevičius, Mubashar Iqbal, and Abasi-Amefon Obot Affia. This document part is contributed and reviewed by Vladislav Fomin, Viktorija Janavičienė from Vilnius University, Michael Henke, Natalia Straub, Tan Guerpinar and Philipp Asterios Ioannidis from TU Dortmund Univeristy, and Boris Düdder from IT University of Copenhagen.

**Part2**
**Activity leader**: TU Dortmund University
**Contributors**: Technical University Dortmund, Vilnius University, University of Copenhagen

**Acknowledgement:** The persons of TU Dortmund University involved in preparation of this document part are Michael Henke, Natalia Straub, Tan Gürpinar, Philipp Asterios Ioannidis. This document part is contributed and reviewed by Boris Düdder from IT University of Copenhagen and Vladislav Fomin from Vilnius University, Raimundas Matulevičius, Mubashar Iqbal, and Abasi-Amefon Obot Affia from University of Tartu.

Disclaimer
The creation of these resources has been (partially) funded by the ERASMUS+ grant program of the European Union under grant no. 2018-1-LT01-KA203-047044.
Neither the European Commission nor the project's national funding agency DAAD are responsible for the content or liable for any losses or damage resulting of the use of these resources.

# Table of Contents

Disclaimer
The creation of these resources has been (partially) funded by the ERASMUS+ grant program of the European Union under grant no. 2018-1-LT01-KA203-047044.
Neither the European Commission nor the project's national funding agency DAAD are responsible for the content or liable for any losses or damage resulting of the use of these resources.

Disclaimer
The creation of these resources has been (partially) funded by the ERASMUS+ grant program of the European Union under grant no. 2018-1-LT01-KA203-047044.
Neither the European Commission nor the project's national funding agency DAAD are responsible for the content or liable for any losses or damage resulting of the use of these resources.

# Introduction

Blockchain technology (BCT) is arguably one of the most promising and most hyped innovations of recent years. Many visionaries focus solely on the big picture regarding a revolutionary paradigm shift from an internet of information to an internet of value. This highlights the need for a comprehensive understanding of how this new technology is affecting and improving existing structures and relationships in a business environment. Furthermore, increasing numbers of job adverts for Blockchain-related positions show that there is a high demand for experts and personnel with competence in the field of Blockchain. BCT builds a base for smart contracts which constitute the missing link for the realization of the vision of Industry 4.0/the Internet of Things (IoT).

Smart contract use cases express the imperative nature of interdisciplinary work. Due to the many different players involved in practice, education approaches in the field of BCT need to be designed interdisciplinary from the core in order to be successful in empowering and improving students' employability. Blockchain is a highly interdisciplinary area, bringing together new challenges and opportunities at the intersections of computer science, economics, engineering, finance, business and law. Despite this interdisciplinarity, there is a lack of educational programs, incorporating different perspective of blockchain from across multiple disciplines such as engineering, business, logistics, and finance. This issue is seen at national, European and global levels. This makes it necessary for universities to apply interdisciplinary courses giving students of different major comprehensive essential skills and knowledge of BCT and its application and impact on different business environments. Sufficient effort from academic institutions to educate and transfer accumulated knowledge about blockchain technology will foster the creation of professional workforce, help exploit the benefits of this innovation, and create other innovations. Graduates might face challenges of understanding advanced Blockchain technology in the later stage of their career. Thus, to educate about this new technology, academic institutions should provide sufficient interdisciplinary courses, especially in the fields of Supply Chain and Logistics, Business, Economics and Finance, as well as Computer Science and IT-security. This will help students acquire the knowledge and skills needed to exploit opportunities and to be prepared for changes in employment trends.

## Project goals and milestones

The BlockNet project includes the following milestones and intellectual outputs: firstly, literature review and the competencies requirements and job adverts will be analyzed for developing the interdisciplinary theoretical skill concept. Secondly, the insights gained will be empirically validated by case studies with European companies in order to identify the best practice use cases for BCT and required skills and competencies. Moreover, this provides a comprehensive and systematic analysis of skill and competence needed to be covered in the online course. Thirdly, based on this first-ever comprehensive competence assessment and use case analysis project BlockNet is going to design a didactical and organizational concept for interdisciplinary Blockchain SNOC, facilitating remote learning opportunities leveraging educational access. Fourthly, the multimedia learning material for interdisciplinary Blockchain SNOC will be produced and evaluated. Fifthly, the course will be technically realized, conducted and evaluated in blended mobility with the students of participating universities. In order to achieve the objectives and milestones, a consortium with four project partners and advisory board is formed, which brings together leading universities: Vilnius University (Lithuania), TU Dortmund University (Germany), University of Tartu (Estonia), University of

Disclaimer
The creation of these resources has been (partially) funded by the ERASMUS+ grant program of the European Union under grant no. 2018-1-LT01-KA203-047044.
Neither the European Commission nor the project's national funding agency DAAD are responsible for the content or liable for any losses or damage resulting of the use of these resources.

Copenhagen (Denmark), European project management competence and practical exposure (e.g. associations and industry partner workshops) with a very strong background and international network.

## Purpose of this Document

**Part 1:**

In order to explore the practical potential and needs of interdisciplinary knowledge and competence requirements of Blockchain technology, the project activity "Development of Interdisciplinary Blockchain Skills Concept" starts with the literature review identifying the state of the art of Blockchain in Supply Chain Management and Logistics, Business and Finance, as well as Computer Science and IT-Security. and the project activity further explores the academic and industry landscape of existing initiatives in education which offer Blockchain courses (learning programs, modules, courses, vocational trainings etc.). Moreover, job descriptions and adverts are analyzed in order to specify today's competence requirements from enterprises. To discuss and define the future required competence, expert workshops are organized to validate the findings by academic experts. Based on the research outcome and validation, an interdisciplinary approach for Blockchain competence is developed.

**Part2:**

This part focuses on the development of the Blockchain Best Practices activity while conducting qualitative empirical research based on case studies with industry representatives. Therefore, company interviews, based on the literature review and theoretical basis of Output 1, explore existing Blockchain use cases in different sectors. Required skills and knowledge will emerge. Due to the interdisciplinary importance of Blockchain technology, these skills will be defined by different perspectives of Blockchain from across multiple disciplines such as Supply Chain Management and Logistics, Business, Economics and Finance, and Computer Science. The use cases and companies for the interviews will be selected based on various sampling criteria (e.g., industry, IT maturity and Blockchain experience, future adoption and the transformative capabilities) to gain results valid for a broad scale. The analysis of the various use cases will be conducted and defined in a standardized format to identify the key drivers and competence requirements for Blockchain technology applications and their adoption. On the one hand, this approach ensures comparability, on the other hand, it facilitates the development of a structured and systematic framework.

Disclaimer
The creation of these resources has been (partially) funded by the ERASMUS+ grant program of the European Union under grant no. 2018-1-LT01-KA203-047044.
Neither the European Commission nor the project's national funding agency DAAD are responsible for the content or liable for any losses or damage resulting of the use of these resources.

# PART 1: Interdisciplinary Blockchain Skills Concept

Disclaimer
The creation of these resources has been (partially) funded by the ERASMUS+ grant program of the European Union under grant no. 2018-1-LT01-KA203-047044.
Neither the European Commission nor the project's national funding agency DAAD are responsible for the content or liable for any losses or damage resulting of the use of these resources.

## 1.1. Method

In this section, we will discuss the research method used to develop the Interdisciplinary Blockchain Skills Concepts. Firstly, we will overview the five major method steps. Secondly, we will discuss the competence modelling (see Section 1.1.2) mainly integrating two models - the Occupational Acting Competence Model (KMK) (see Section 1.1.3) and the Four Fields Competence Model (see Section 1.1.4), that are used to classify identified interdisciplinary skills. Finally, we will overview the Bloom's taxonomy used to define the skills concept (see Section 1.1.5).

### 1.1.1. Method to Develop Skills Concept

The research method to develop the Interdisciplinary Blockchain Skills Concept is illustrated in Fig. 1 and consists of the following five steps:

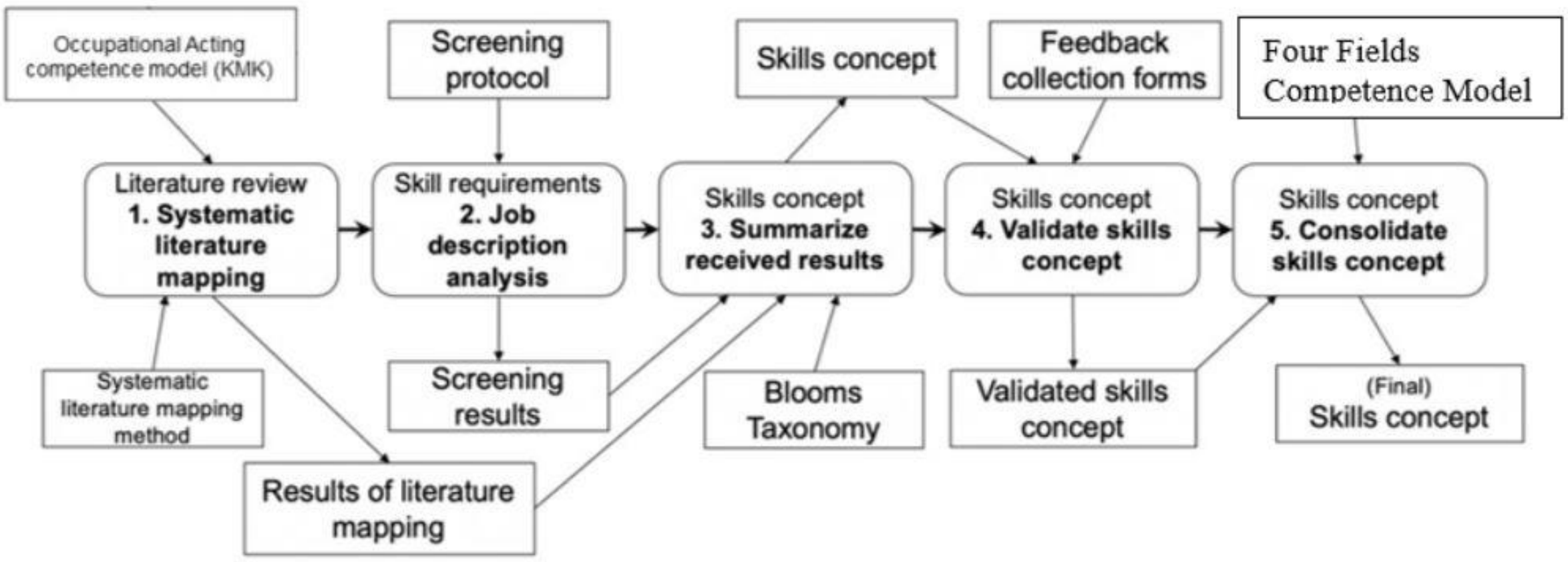

**Fig. 1**: Method to develop Interdisciplinary Blockchain Skills Concept

1. *Perform systematic literature mapping*. The goal of the first step is twofold: (*i*) to define the initial competence classes using the literature sources; (*ii*) to review the existing literature to explain what the interdisciplinary blockchain skills are reported in the fields of economics, finance and business, supply chain management, software engineering and security risk management and engineering. We have applied the systematic literature mapping method (Petersen *et al*. 2008) to guide the literature mapping process. In addition, we also used the KMK Competence Model (see Section 1.1.3), to structure the initial results to the competences. This step results in the literature mapping of the Skills concept. The step is discussed in Section 1.2.

2. *Analyze job descriptions*. The goal of the second step is to explore what the interdisciplinary Blockchain competences are requested from the enterprises. To run this step a screening protocol was prepared. The major input to the step was the job descriptions and adverts of the Blockchain related employment positions. The outcome of this step was the list of the interdisciplinary Blockchain competences found in these job descriptions and adverts. The step is discussed in Section 1.3.

3. *Summarize received results*. The goal of the third step is to summarize and aggregate the results received in step 1 and step 2. Both the literature mapping of the Skills concept and the interdisciplinary Blockchain competences found in the job descriptions

Disclaimer
The creation of these resources has been (partially) funded by the ERASMUS+ grant program of the European Union under grant no. 2018-1-LT01-KA203-047044.
Neither the European Commission nor the project's national funding agency DAAD are responsible for the content or liable for any losses or damage resulting of the use of these resources.

and adverts are aligned and revised using Bloom's taxonomy (see Section 1.1.5). The step results in the interdisciplinary Blockchain competence concept as discussed in Section 1.4.

4. *Validate skills concept*. The goal of the fourth step is to validate the Skills Concept. The Skills Concept defined in step 3 was adapted to the feedback collection form. Then this form was used to collect responses from the members of the project advisory board as well as to the leading universities of the partner countries. The step resulted in the validated Skills Concept. The step is discussed in Section 1.5.

5. *Consolidate skills concepts*. The goal of the fifth step is to consolidate the interdisciplinary Skills Concept. The output of the step 4 was carefully revised and mapped to the categories suggested in the Four Fields Competence Model (see Section 1.1.3). The outcome of this step is the final Interdisciplinary Blockchain Skills Concept as presented in Section 1.6.

Within BlockNet project, the competences for interdisciplinary groups of students will be determined in order to prepare them for future employments, where Blockchain technology becomes increasingly important. Although similar research work has already been carried out for related areas like industry 4.0, for example by Jerman et al. (2018) and Hecklau et al. (2017), no comparable study has yet been carried out for the Blockchain environment. Within BlockNet project, the competence requirements for interdisciplinary work in Blockchain environment are collected and structured in a domain-specific competence model and presented in 1.6.2. The competence model should serve to derive learning objectives for an interdisciplinary online course.

In the remaining of this section we will discuss the steps of the method to develop the Interdisciplinary Blockchain Skills Concept in detail (see Section 1.1.2). The used competence models (i.e., namely KMK Competence Model (see Section 1.1.3) and Four Fields Competence model (see Section 1.1.4) and Bloom's Taxonomy (see Section 1.1.5) are overviewed.

### 1.1.2. Competence Modelling

Over the course of decades, the term "competence" has evolved into a key concept of educational research and gained importance by fundamental changes of demands in the fields of life and work (Klieme and Leutner 2006). The use of the concept of competence in a variety of scientific disciplines, such as management or law (Mulder et al. 2009), has resulted in different conceptual definitions over time. Following Jerman et al. (2018) the presented definitions are only a brief overview of a large selection: Spencer and Spencer (1993) define the term competence, as "an underlying characteristic of an individual that is causally related to criterion-referenced effective and/or superior performance in a job or situation. Underlying characteristic means that competences are a fairly deep and enduring part of a person´s personality and can predict behaviour in a wide variety of situations and job tasks (Spencer & Spencer, p. 9). Boyatizs defines competences as "characteristics that are casually related to effective and/or superior performance in that job" (Boyatizs 1982, p. 23). Roberts defines competence as a range of personality traits, knowledge, experience, skills and values that are all essential to perform a job successfully (Roberts 1997). As the last definition presented, Klieme and Leutner see competences as "context-specific cognitive activity schedules that

Disclaimer
The creation of these resources has been (partially) funded by the ERASMUS+ grant program of the European Union under grant no. 2018-1-LT01-KA203-047044.
Neither the European Commission nor the project's national funding agency DAAD are responsible for the content or liable for any losses or damage resulting of the use of these resources.

relate functionally to situations and requirements in certain domains" (Klieme and Leutner 2006, p. 879).

Similar to the definition of competences, there is no standardized classification cluster of competences (Jerman et al. 2018, p, 3f; Hecklau et al. 2017). Many competence clusters are based on Roth, who defined the terms technical or professional competence, social competence and self competence in 1971 (Roth 1971). Following Reetz, these clusters can be transferred to the professional sector and supplemented by the dimension of methodical competences, since overarching action strategies and problem-solving skills are gaining increasing importance in the professional context (Reetz 1989a, 1989b, 1999; Baethge et al. 2006). Combining these four competences we create a superordinate competence - the "occupational acting competence". This competence is understood as "the willingness and ability of the individual to think through professional, social and private situations properly and to behave in an individual and socially responsible manner" (KMK 2011, p. 15).

In context of the BlockNet project, the occupational acting competence can be seen as the overarching competence goal and is titled as "occupational acting competence for Blockchain related interdisciplinary projects". The used KMK model is based on the presented four competences. KMK, however, collects competences in a more granular and differentiated way. This leads to the fact that learning competences, which are particularly important for the BlockNet learning course, can be considered separately.

Competence modelling refers to the description of the respective competences and their clustering. This is done at the level of competence fields and competence levels and can only be absolved in connection with an explicit domain (Nickolaus and Seeber 2012). This modelling task must be performed according to the respective professional contexts. This means that the four competence areas mentioned above are concretized and further differentiated according to the work tasks of the respective domain. Within BlockNet project, the application of Blockchain technology is focused in an interdisciplinary project context.

The development of competence models, competence levels and concepts for competence development are among the central tasks of research (Klieme and Leutner 2006). Schaper (2009) identifies three different types of competence models:

1. *Competence structure models* describe which competence facets and dimensions are needed to cope with different requirements in a domain. It illustrates the competences in a structured way and shows their connections. Two examples for a competence structure model are the KMK competence model or the four fields competence model based on Reetz.
2. *Competence level models* describe what different people in one domain can master and what specific requirements they can meet at different levels (e.g. PISA, Program for International Student Assessment).
3. *Competence development models* describe the development stages of the acquisition of competences in a domain.

The use of these competence models in educational research essentially enables the theoretical and empirical derivation and foundation of hypotheses about meaningful and necessary educational goals. It further enables the access to established foundations for the design of precise and valid measuring instruments and the orientation of educational goals towards the

Disclaimer
The creation of these resources has been (partially) funded by the ERASMUS+ grant program of the European Union under grant no. 2018-1-LT01-KA203-047044.
Neither the European Commission nor the project's national funding agency DAAD are responsible for the content or liable for any losses or damage resulting of the use of these resources.

requirements of professional contexts in everyday life (Schaper 2009). The first competence structure model used to collect Blockchain related competences is the KMK Competence model, which is presented in the following chapter.

### 1.1.3. KMK Competence Model

The KMK Competence Model (Fig. 2) was chosen as the competence structure model for differentiated competence modelling (KMK 2011). The model was used as a basis for the development of the BlockNet category system for the SLM and the analysis of job adverts. The KMK model was developed by the German *Kultusministerkonferenz* and is based on the model of Roth describing the vocational action ability as a combination of six competence fields. As shown in Fig. 2, the KMK competence model focuses on the three competences developed by Roth: professional competence, self-competence and social competence. These are elementary to enable a person to make professional decisions and can be described as follows:

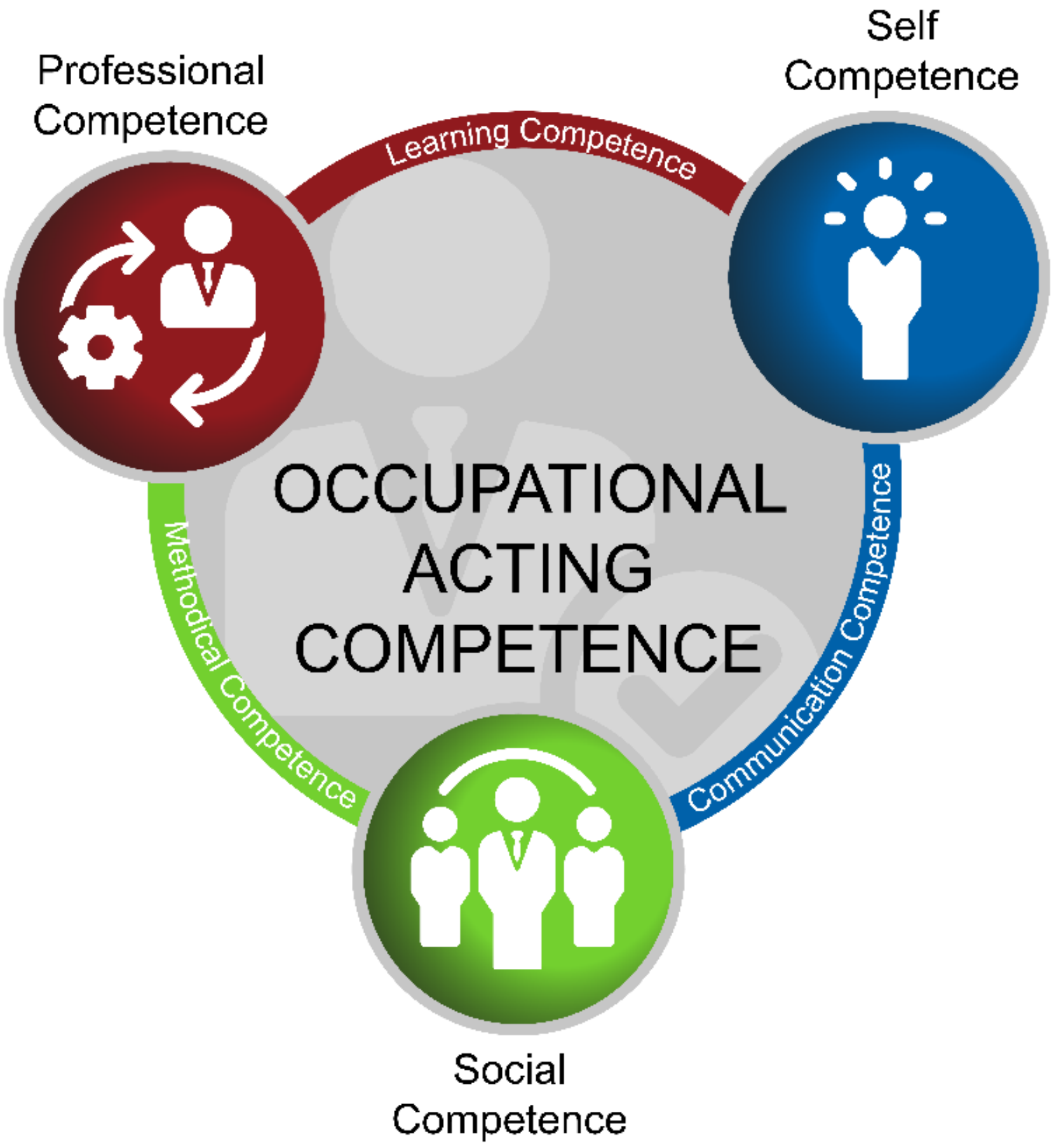

**Fig. 2:** KMK competence model (Source: Own figure based on KMK 2011, pp. 15-16)

- *Professional competence* - the willingness and ability to solve tasks and problems in a goal-oriented, appropriate, method-oriented and independent manner on the basis of technical knowledge and competences and to evaluate the result.
- *Self-competence* - the willingness and ability as an individual to recognize the opportunities for development, requirements and restrictions in family, work and public life. To think things through and to judge things, to develop one's own talents as well as to make and develop life plans. It includes attributes such as

Disclaimer
The creation of these resources has been (partially) funded by the ERASMUS+ grant program of the European Union under grant no. 2018-1-LT01-KA203-047044.
Neither the European Commission nor the project's national funding agency DAAD are responsible for the content or liable for any losses or damage resulting of the use of these resources.

independence, the ability to accept criticism, self-confidence, reliability, a sense of responsibility and duty. It includes in particular the development of sophisticated values and the self-determined attachment to values.

- *Social competence* - the willingness and ability to live and shape social relationships, to grasp and understand gifts and tensions and to deal with and communicate with others in a rational and responsible manner. This includes in particular the development of social responsibility and solidarity. (KMK 2011, p. 15)

These three types form the basis of the KMK competence model. However, they are enhanced by methodological, communicative and learning competences, which are an immanent part of professional, personal and social competences. This differentiation enables a focused survey of the individual competence requirements. The immanent competences can be described as follows:

- *Communication competence* - the willingness and ability to understand and shape communicative situations. This includes perceiving, understanding and presenting one's own intentions and needs as well as those of the partners.
- *Methodological competence* - the willingness and ability to take a goal-oriented, planned approach to the processing of tasks and problems (for example, when planning work steps).
- *Learning competence* - the willingness and ability to understand and evaluate information about facts and contexts independently and together with others and to classify them into mental structures. In addition, learning competence includes the ability and willingness to develop learning techniques and learning strategies at work and beyond the workplace and to use them for lifelong learning. (KMK 2011, p. 16).

Finally, yet importantly, the KMK model combines the competences necessary for professional decisions and the additional immanent competences as “occupational acting competence”, which represents the overarching competence of the model. It is understood as the willingness and ability of an individual to think through professional, social and personal situations properly and to behave in a responsible manner (KMK 2011).

The derived BlockNet competence model is developed on the basis of the presented KMK elements. In a next step, the Four Fields Competence Model was used to finalize the BlockNet competence model in its structure and is described in the following.

### 1.1.4. Four Fields Competence Model

As already described in the previous chapters, most competence clusters are based on the four competence fields division developed by Roth and Reetz. Roth defined the terms technical or professional competence, social competence and self competence already in 1971 (Roth 1971). It was Reetz, however, who transferred these competences to the professional sector. Since overarching action strategies and problem-solving competences gained an increasing importance in the professional sector, he extended the existing competences with the competence field of methodological competences (Reetz 1989a, 1989b, 1999; Baethge et al. 2006).

Disclaimer
The creation of these resources has been (partially) funded by the ERASMUS+ grant program of the European Union under grant no. 2018-1-LT01-KA203-047044.
Neither the European Commission nor the project’s national funding agency DAAD are responsible for the content or liable for any losses or damage resulting of the use of these resources.

There are several ways of classifying competences in the literature. The model developed by Heyse and Erpenbeck, called "Competence Atlas", is an enterprise-unspecific one-size-fits-all approach. Four superior dimensions are matched to 64 individual competences here (Heyse and Erpenbeck 2004). Although this model was developed as a universally valid tool for competence assessment, many scientific studies adapt the model to use domain- and subject-specific dimensions. Nevertheless, in many cases a similar structure is recognizable. In most cases, the described four fields are used to divide competences in scientific research and are presented in Fig. 3.

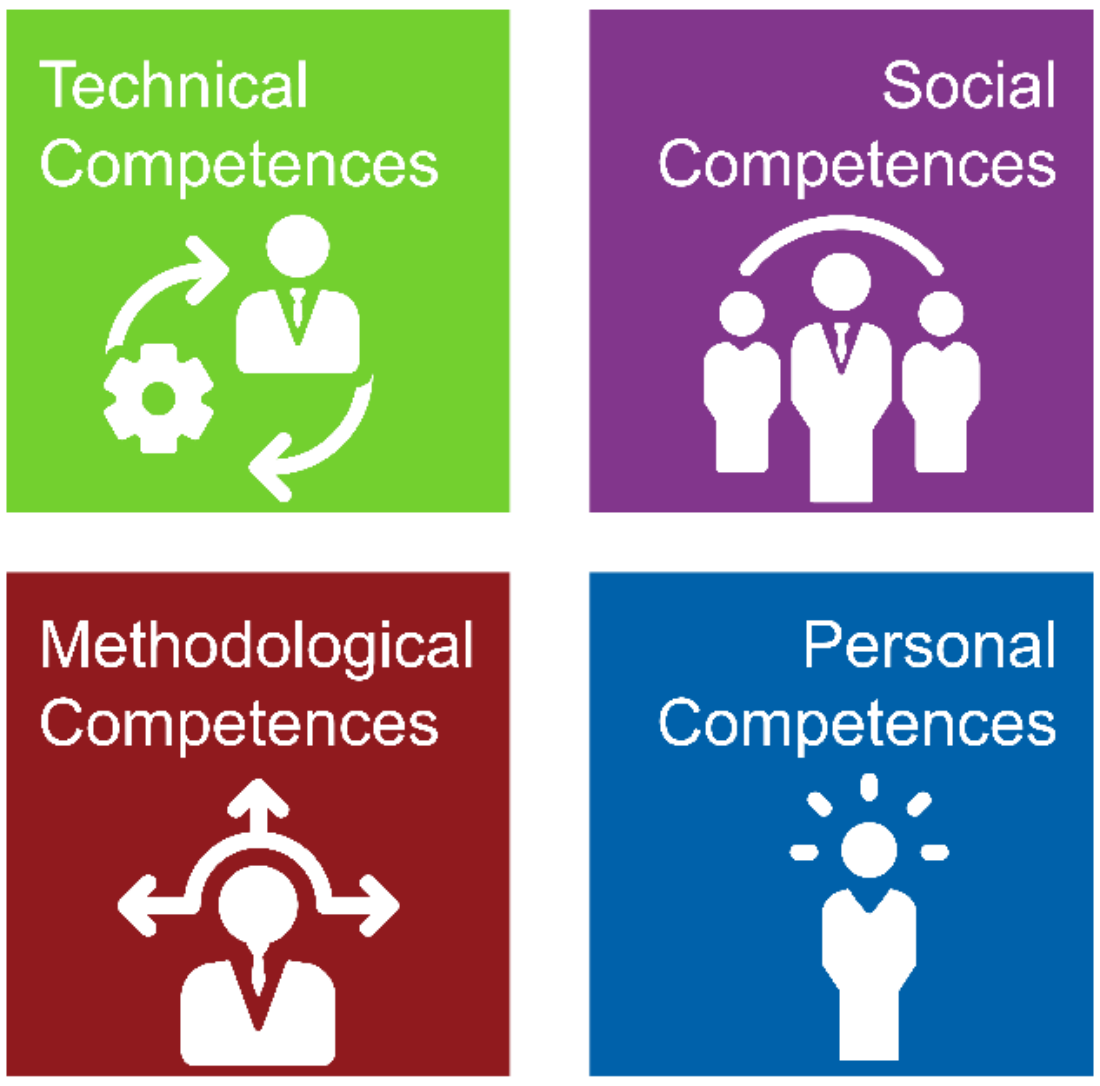

**Fig. 3:** Four Field Competence Model

Within the scope of the competence assessment for BlockNet project, competences for future experts in the area of Blockchain were identified and allocated to domain-specific competence dimensions following the KMK approach. For this purpose, the dimensions developed by Hecklau et al. (2017) and Jerman et al. (2018) for the adjacent area of industry 4.0 were used. Subsequently, the presented four dimensions were used to concisely present technical-, methodological-, social- and personal competences. The allocation to the four domain-specific dimensions leads to a leaner presentation, better comprehensibility and thus to an increase in acceptance.

Nyikes (2018) or Janjua et al. (2012) also provide examples for the before mentioned. To get an impression of how Janjua uses the presented four fields' model to classify competence items in the field of IT security, Fig. 4 can be observed.

Disclaimer
The creation of these resources has been (partially) funded by the ERASMUS+ grant program of the European Union under grant no. 2018-1-LT01-KA203-047044.
Neither the European Commission nor the project's national funding agency DAAD are responsible for the content or liable for any losses or damage resulting of the use of these resources.

| Technical Competences | Methodological Competences | Social Competences | Personal Competences |
|---|---|---|---|
| understanding IT security | creativity | seeing the big picture (overview competence, integration competence) | commitment to lifelong learning |
| coding capabilities | problem solving | the ability to lead | personal flexibility |
| understanding of processes | creative problem-solving competence | the ability to communicate effectively in complex situations | motivation for learning |
| technical capabilities | conflict resolution | network competence | adaptability |
| understanding the analogies of the operation of new technologies | the ability to act as mediators in decision-making processes | the ability to participate and work in a team | ability to work in stressful situations |
| the ability to solve complex challenges | analytical skills | language skills | social responsibility |
| | research skills | the ability to transfer knowledge to others | the successful determination of the dividing line between important and less important information |

**Fig. 4:** Janjua Four Field Competence Model

The BlockNet competence model developed on the basis of KMK- and four fields' model comprises both the status quo and future competence requirements. Thus, it enables the modelling of competence requirements, the creation of activity-related target and actual competence profiles, as well as competence diagnostics on the basis of a so-called *competence classification key* in the area of Blockchain. The classification key is based on Bloom's Taxonomy and explained in more detail in the following chapter.

### 1.1.5. Bloom's Taxonomy

Bloom's taxonomy (Bloom 1956; Anderson et al. 2001) presents classes of verbs that "demonstrate critical thinking" to characterize knowledge, skills, attitudes, behaviors and abilities. The extract of this taxonomy is illustrated in Fig. 5. Hence, the ***knowledge*** category characterizes how the learner could *remember* the previously learnt information. ***Comprehension*** is about *demonstrating* and understanding of the learnt fact, then the ***application*** concerns how the knowledge is actually *applied* in the situations. The next category – ***analysis*** – deals with the *breaking down* the ideas and objects into the smaller and more simpler components thus helping to find evidences for the generalization. Finally, the ***synthesis*** category is about *compiling* the different ideas into a new whole or alternative solution.

Disclaimer
The creation of these resources has been (partially) funded by the ERASMUS+ grant program of the European Union under grant no. 2018-1-LT01-KA203-047044.
Neither the European Commission nor the project's national funding agency DAAD are responsible for the content or liable for any losses or damage resulting of the use of these resources.

| KNOWLEDGE | COMPREHENSION | APPLICATION | ANALYSIS | SYNTHESIS | EVALUATION |
|---|---|---|---|---|---|
| List | Compare | Apply | Analyze | Arrange | Appraise |
| Name | Describe | Complete | Appraise | Assemble | Argue |
| Recall | Discuss | Construct | Categorize | Collect | Assess |
| Record | Explain | Demonstrate | Compare | Combine | Choose |
| Relate | Express | Dramatize | Contrast | Comply | Compare |
| | Identify | Employ | Debate | Compose | Conclude |
| | Recognize | Illustrate | Diagram | Construct | Estimate |
| | | Interpret | Differentiate | Create | Evaluate |
| | | Operate | Distinguish | Design | Interpret |
| | | | Examine | Devise | Judge |
| | | | Experiment | Formulate | Justify |
| | | | | Manage | Measure |
| | | | | Organize | Rate |
| | | | | | Revise |
| | | | | | Score |

Fig. 5. Extract of the Bloom’s Taxonomy
(adapted from (Bloom 1956; Anderson et al. 2001))

Disclaimer
The creation of these resources has been (partially) funded by the ERASMUS+ grant program of the European Union under grant no. 2018-1-LT01-KA203-047044.
Neither the European Commission nor the project’s national funding agency DAAD are responsible for the content or liable for any losses or damage resulting of the use of these resources.

# 1.2. Systematic Literature Mapping

The first step consists of performing of the systematic literature mapping. Its goal is to define the initial competence classes using the literature sources and to review the existing literature in order to explain the interdisciplinary Blockchain needs reported in the literature. The systematic literature was performed following the guidelines of Petersen *et al*. (2008). As illustrated in Fig. 6, it consists of five stages: definition of research questions (discussed in Section 1.2.1), conduct research (discussed in Section 1.2.2), screening of papers (discussed in Section 1.2.3), keywording using abstract and data extraction and mapping process (both discussed in Section 1.2.4)

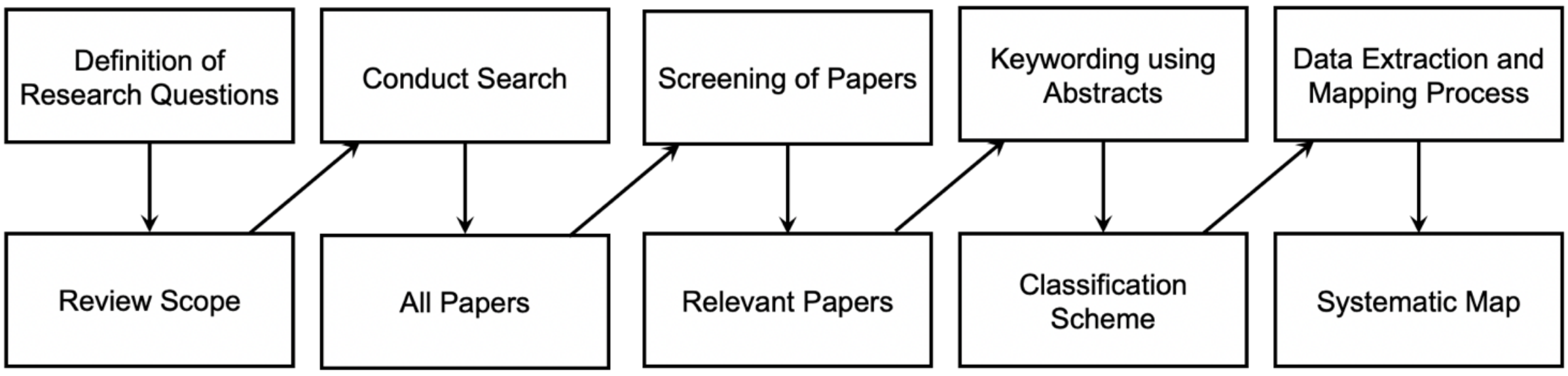

**Fig. 6**. Systematic Literature Mapping Method (adapted from (Petersen et al. 2008))

## 1.2.1. Definition of Research Question

In the first step of the systematic literature mapping, four different fields were examined blockchain applications could potentially developed, deployed, and used to support ongoing processes. The studied fields included (*i*) economics, finance (FinTech) and business, (*ii*) supply chain management, (*iii*) security risk and security risk management and engineering, and (*iv*) software engineering. Our goal is to find out what interdisciplinary blockchain skills are reported in the literature. We have formulated a research question which addresses all four targeted fields:

What are the most important research areas (skills) reported in the scientific articles and papers on

- the *economics*, *finance* (FinTech) and *business* in the blockchain applications?
- blockchain technology applied in *supply chain management*?
- the *security risk* and *security risk management and engineering* in the blockchain applications?
- the *software engineering* aspects in blockchain applications and platforms?

## 1.2.2. Conduct Search

The literature search was executed in five recognised digital libraries:

- IEEE: https://ieeexplore.ieee.org/
- ACM: https://dl.acm.org/
- Science Direct: https://www.sciencedirect.com/
- Springer: https://link.springer.com/
- Scopus: https://www.scopus.com/

Disclaimer
The creation of these resources has been (partially) funded by the ERASMUS+ grant program of the European Union under grant no. 2018-1-LT01-KA203-047044.
Neither the European Commission nor the project's national funding agency DAAD are responsible for the content or liable for any losses or damage resulting of the use of these resources.

The used keywords and search phrases are provided in Table 1.

**Table 1**: Keywords used to find the relevant literature

| Research field | Keywords and search phrases |
|---|---|
| Economics, finance (FinTech) and business | Blockchain applications for finance, blockchain applications for business, blockchain applications for economics, blockchain applications for fintech |
| Supply chain management | Blockchain AND "Supply Chain Management"; Blockchain AND "Supply Chain"; Blockchain AND Supply; Blockchain AND Procurement; "Smart Contract" AND "Supply Chain Management"; "Smart Contract" AND Procurement |
| Security risk and security risk management and engineering | Blockchain applications security, blockchain applications security (risks OR threats OR gaps OR issues OR challenges), permissioned blockchain applications security, private blockchain applications security, permissionless blockchain applications security, public blockchain applications security |
| Software engineering | 'Blockchain AND software AND (engineering)', Blockchain software engineering, |

### 1.2.3. Screening of Papers

Table 2 summarised the results found after conducting the search in the selected digital libraries using the defined keywords and search phrases. Initially 919 studies were collected. The paper inclusion and exclusion criteria are provided in Table 3. For instance, only the peer-reviewed papers are include only papers which consider the current trends of the blockchain-based application (in the fintech, supply chain management) or specific analysis of blockchain-based application components (in the field of security risk management and software engineering).

**Table 2**. Screening results

| | **Studies collected from defined sources** | **Study screening, Quality criteria** | **Included studies after full reading** |
|---|---|---|---|
| Economics, finance (FinTech) and business | 309 | 251 | **58** |
| Supply chain management | 160 | 110 | **50** |
| Security risk and security risk management and engineering | 141 | 73 | **67** |
| Software engineering | 309 | 256 | **53** |
| **Total** | **919** | **690** | **228** |

Next, the quality of the paper was assessed. The quality criteria are defined to assess the quality of the selected papers. The following quality assessment criteria are formulated based on the relevancy and research scope of our study, together with the guidelines of the Kitchenham and Charters (2007) quality criteria. These included the following questions:

1. Are the goals and purpose of a study is clearly stated?
2. Does the study describe security threats or risks in the blockchain-based applications?
3. Does the study provide the solution or architecture to mitigate security risks in the blockchain-based applications?
4. Did the study answer all the defined questions or problems?
5. How well are the research results presented?
6. How clear are the links between data, interpretation and conclusions?

Disclaimer
The creation of these resources has been (partially) funded by the ERASMUS+ grant program of the European Union under grant no. 2018-1-LT01-KA203-047044.
Neither the European Commission nor the project's national funding agency DAAD are responsible for the content or liable for any losses or damage resulting of the use of these resources.

The above questions are scored as follows: 1 = Fully satisfy the question, 0.5 = Partially satisfy the question, 0 = Not satisfy the question. All those studies who score 50% or more were included, and remaining will be excluded from the list. The quality assessment resulted in 690 papers.

**Table 3**: Inclusion/exclusion criteria

| **Inclusion criteria** | **Exclusion criteria** |
|---|---|
| Only the peer-reviewed literature | Literature that does not subject to peer review |
| Literature that defines current trends of the blockchain-based application or specific analysis of blockchain-based application components | Grey literature or informal studies with no concrete evidence |
| Studies published from 2015-2018 which describes security in the blockchain-based applications | Work in progress, proposals or draft versions |

Finally, the full paper reading was done. In the end 228 papers (see Table 2) were selected for the extraction of the interdisciplinary blockchain skills. The selected papers are listed in Appendix I.1.

### 1.2.4. Keywording, Data Extraction and Mapping Process

Keywording in each studied field depend on the specific characteristics of the respective field. However, the biggest attention was placed on the blockchain types used in the field, the application domains and the contribution facets. For instance, the keywording and mapping process in the economics, finance (FinTech) and business field are illustrated in Table 4, where the blockchain type facet is confronted with the application domain, contribution facet, and FinTech. In Table 5, the supply chain management keywording and mapping process is illustrated. Here the addressed challenges are the aspects to be confronted with the blockchain solutions.

Table 6 presents (first part) the quantity of applications domains & technology solutions based on the different blockchain platforms found when considering security risk management and engineering research field. The second part of Table 6 presents the security risks which are mitigated by introducing the blockchain-based applications, and the third part overview the security risks which appear within the blockchain-based applications. In Table 7 we encompass the security risks along with the blockchain-based applications research areas to show which security risks are more frequently occurring on different blockchain-based applications. The consequent Table 8 presents the domains and major keywords found while considering the software engineering research field.

The keywording, data extraction and mapping process has resulted in the list of the interdisciplinary blockchain skills regarding each studies research fields. In Appendix 1.2 the mapping between the literature sources and the interdisciplinary blockchain skills is provided. In the listings below we give the interdisciplinary blockchain skills (*and the number of literature sources where the competence was found*):

Disclaimer
The creation of these resources has been (partially) funded by the ERASMUS+ grant program of the European Union under grant no. 2018-1-LT01-KA203-047044.
Neither the European Commission nor the project's national funding agency DAAD are responsible for the content or liable for any losses or damage resulting of the use of these resources.

1. Economics, finance (FinTech) and business:
   - Compare blockchain platforms (literature reviews, investigations, overviews, mapping) (*12 articles*)
   - Compare blockchain platforms with illustrated application (literature reviews, investigations, overviews) (*5 articles*)
   - Explain blockchain technology (describe, explain, recognize, case studies) (*6 articles*)
   - General research frameworks for blockchain systems with theoretical models (*2 articles*)

**Table 4**: Blockchain type facet *vs* Application domain, Contribution facet, and FinTech

| | | **Blockchain type facet** | | | | | |
|---|---|---|---|---|---|---|---|
| | | **Blockchain (technology, app.)** | **Bitcoin** | **Crypto-currencies** | **Distributed ledger** | **Smart contract** | **General** |
| **Application domain** | **Business** | 10 | 2 | 1 | 1 | 5 | |
| | **Management** | 4 | | | | 3 | |
| | **Finance** | 11 | 3 | 4 | 1 | 3 | 3 |
| | **Information technology systems** | 16 | 1 | 2 | 1 | 9 | 1 |
| | **Economics** | 10 | 1 | 1 | | 5 | |
| | **General** | 3 | | | | 1 | |
| **Contribution facet** | **Research framework** | 8 | 1 | 2 | 2 | 4 | |
| | **Literature review, mapping** | 3 | | 1 | | | |
| | **Investigation/ solution /development of new app, model, system, schema, paradigm, approach, platform** | 23 | 2 | | 2 | 8 | 1 |
| | **Implemented/ designed** | 6 | | | | 3 | 1 |
| | **Discussion/ description/ explanation/ evaluation/ identification/ analysis** | 9 | 2 | 3 | | 6 | 1 |
| **FinTech** | **Fin regulation** | 1 | | | | | 1 |
| | **Fin innovation** | 3 | 1 | | 1 | | |
| | **Fin market** | 3 | | | | | 1 |
| | **Fin service industry (banking, insurance, accounting)** | 3 | | | 1 | | |
| | **ICO (initial coin offering)** | 2 | | | | | |
| | **Crowdfunding** | 1 | | | | | |
| | **Auditing** | 4 | | | | 1 | |

Disclaimer
The creation of these resources has been (partially) funded by the ERASMUS+ grant program of the European Union under grant no. 2018-1-LT01-KA203-047044.
Neither the European Commission nor the project's national funding agency DAAD are responsible for the content or liable for any losses or damage resulting of the use of these resources.

- Discuss and compare different blockchain models, scheme and solutions with constructed/illustrated application (suggestions, proposals, methods for blockchain use in economics, business and finance) (*26 articles*)
- Analysis (analyze, examine, categorize of blockchain technology effectiveness, influence) (*3 articles*)
- Analysis (test on distributed ledger, cryptography, consensus protocol, smart contract) (*4 articles*)

**Table 5**: Blockchain challenges *vs* Blockchain solutions in the supply management chain field

| | | Blockchain solution | | | | | | | |
|---|---|---|---|---|---|---|---|---|---|
| | | **Not mentioned** | **General** | **New concept** | **Ethereum** | **Hyperledger Fabric** | **Hyperledger Sawtooth** | **BigChainDB** | **Corda** |
| **Addressed challenges** | **Connectivity and interdisciplinarity** | 1 | 1 | 2 | 1 | 1 | | | |
| | **Counterfeit products and other fraud** | 2 | 1 | 3 | 3 | 2 | | | |
| | **Business related challenges** | 8 | 4 | 4 | 2 | 4 | | 1 | |
| | **Data/IT security** | 5 | 2 | 3 | 2 | 1 | 1 | | |
| | **Information asymmetry** | 11 | 5 | 3 | 6 | 6 | 1 | 1 | 1 |
| | **Unknown provenance/ transaction history** | 9 | 2 | 2 | 5 | 5 | | 1 | 1 |
| | **None** | 1 | 4 | | | 1 | | | |

2. Supply chain management:
   - Explain general capabilities for the use in SCM (*6 articles*)
   - Explain the interoperability of BCT and possible collaboration between unknown or untrusted parties (*6 articles*)
   - Explain how data can be secured by the use of BCT (*14 articles*)
   - Explain how information asymmetry can be addressed by BCT applications (*34 articles*)
   - Demonstrate BCT capabilities and apply them to business-related challenges (*23 articles*)
   - Demonstrate BCT capabilities and apply them to counterfeit and fraud prevention problem statements (*11 articles*)
   - Demonstrate BCT capabilities and apply them to provenance and track&trace problem statements (*25 articles*)

3. Security risk and security risk management and engineering:
   - Explain blockchain technology principles (*2 articles*)
   - Compare blockchain platforms to enable understanding of different system design choices (*4 articles*)
   - Discuss and compare different blockchain models and solutions to secure data and information (*28 articles*)
   - Describe privacy management principles using the blockchain solutions (*7 articles*)
   - Recognise security countermeasure implications (*6 articles*)
   - Explain access control (authentication, authorization and identity) models (*6 articles*)
   - Explain identity management principles using the blockchain solutions (*2 articles*)
   - Describe transaction protection and validation principles (*3 articles*)
   - Underline major encryption and signature schemes (*3 articles*)

Disclaimer
The creation of these resources has been (partially) funded by the ERASMUS+ grant program of the European Union under grant no. 2018-1-LT01-KA203-047044.
Neither the European Commission nor the project's national funding agency DAAD are responsible for the content or liable for any losses or damage resulting of the use of these resources.

– Explain trust management principles (*2 articles*)
– State major fair mining principles (*2 articles*)
– Identify security errors in smart contracts (*2 articles*)

4. Software engineering:
   – Explain blockchain technology (*1 article*)
   – Compare blockchain platforms to enable understanding of different system design choices (*14 articles*)

**Table 6**: Security risk management/engineering, keywording and mapping results (CPL – other customised permissionless; CP – other customised permissioned; HLF – Hyperledger Fabric)

| | | Blockchain platforms/ applications | | | | | |
|---|---|---|---|---|---|---|---|
| | | **Permissionless** | | | **Permissioned** | | **General** |
| | | **Bitcoin** | **Ethereum** | **CPL** | **HLP** | **CP** | |
| **Research areas** | *Application domains where blockchain is used* | | | | | | |
| | **Healthcare** | | 3 | 1 | 2 | 4 | 1 |
| | **Resource monitoring and digital rights management** | 1 | 3 | 2 | | 2 | 1 |
| | **Financial** | 2 | 1 | 1 | 1 | | |
| | **Smart vehicles** | 1 | | 1 | 1 | 2 | |
| | **Voting** | 1 | 1 | | 2 | | |
| | *Technology solutions where blockchain is used* | | | | | | |
| | **Security layer** | 6 | 7 | 1 | | 1 | |
| | **Internet of things** | 2 | 2 | 1 | 2 | 2 | |
| **Security risks mitigated by blockchain app.** | **Data tampering attacks** | 7 | 8 | 4 | 7 | 5 | 1 |
| | **DoS/ DDoS** | 7 | 7 | 5 | 3 | 2 | 1 |
| | **Man in the Middle attack** | 3 | 6 | 2 | 2 | | 1 |
| | **Identify theft/ Hijacking** | 1 | | 3 | | | 1 |
| | **Spoofing attack** | 2 | | 1 | | 1 | |
| | **Other risks/ threats** | 6 | 4 | 2 | 1 | 2 | 2 |
| **Security risks which appear within blockchain app.** | **Sybil attack** | 5 | 1 | 1 | 4 | 1 | 1 |
| | **Double spending attack** | 4 | 1 | 2 | 2 | | 1 |
| | **51% attack** | 3 | 3 | 1 | | | 1 |
| | **Deanonymisation attack** | 2 | 1 | 3 | | | 1 |
| | **Replay attack** | 2 | 4 | 1 | | | |
| | **Quantum computing threat** | | 1 | 1 | 2 | | 1 |
| | **Selfish mining attack** | 1 | | 2 | 1 | | |
| | **SC re-entrancy attack** | | 2 | | | | 1 |
| | **Other risks/ threats** | 6 | 1 | 6 | 3 | 1 | 3 |

Disclaimer
The creation of these resources has been (partially) funded by the ERASMUS+ grant program of the European Union under grant no. 2018-1-LT01-KA203-047044.
Neither the European Commission nor the project's national funding agency DAAD are responsible for the content or liable for any losses or damage resulting of the use of these resources.

- Discuss and compare different blockchain models and solutions to secure data and information (*6 articles*)
- Describe development processes for blockchain solutions (*6 articles*)
- Explain the foundations and algorithms of blockchain systems (*8 articles*)
- Discuss software quality goals and their impact on blockchain system development (*2 articles*)
- Describe the software requirements elicitation process of blockchain systems (*2 articles*)
- Discuss various application and business models for blockchain systems with respect to system choices (*11 articles*)
- Develop a testing plan for concrete blockchain solutions (*4 articles*)

Disclaimer
The creation of these resources has been (partially) funded by the ERASMUS+ grant program of the European Union under grant no. 2018-1-LT01-KA203-047044.
Neither the European Commission nor the project's national funding agency DAAD are responsible for the content or liable for any losses or damage resulting of the use of these resources.

**Table 7**: Security Risks based on the research areas

| | Applications | | | | | Technology | | Other |
|---|---|---|---|---|---|---|---|---|
| | **Healthcare** | **Resource monitoring** | **Financial** | **Smart vehicles** | **Voting** | **Security layer** | **Internet of things** | |
| ***Security risks which are mitigated by introducing blockchain applications*** | | | | | | | | |
| Data tampering attack | 6 | 5 | 1 | 4 | 3 | 2 | 5 | 6 |
| DoS/ DDoS attack | | 5 | 1 | 3 | 1 | 7 | 3 | 5 |
| Man in the middle | 1 | 4 | 1 | 1 | 1 | 2 | 2 | 2 |
| Identity theft/ Hijacking | 1 | 2 | | | | | 1 | 1 |
| Spoofing attack | | | | | 1 | | 1 | 2 |
| Other risks/ threats | 2 | | 1 | | 1 | 5 | 5 | 3 |
| ***Security risks which appear within the blockchain applications*** | | | | | | | | |
| Sybil attack | 1 | 1 | 1 | 1 | 2 | 1 | 1 | 5 |
| Double spending attack | | 4 | 2 | | | 2 | | 2 |
| 51% attack | | 4 | | | 1 | 1 | | 2 |
| Deanonymisation attack | | 2 | 1 | 1 | 1 | 1 | 1 | |
| Replay attack | | 2 | 1 | | | 4 | | |
| Quantum computing attack | 1 | | | | | 2 | | 2 |
| Selfish mining attack | | 1 | 1 | | | 2 | | |
| SC re-entrance attack | | | | | | 3 | | |
| Other risks/ threats | | 11 | 5 | | | 2 | 1 | 1 |

**Table 8**: Domains and keywords in the software engineering research field

| Software Engineering: Application Domains | | |
|---|---|---|
| **Business** | 1 | 2% |
| **Construction** | 1 | 2% |
| **Energy** | 2 | 4% |
| **General** | 33 | 61% |
| **ITC** | 11 | 20% |
| **Payment** | 1 | 2% |
| **Smart City** | 1 | 2% |
| **Supply Chain** | 1 | 2% |
| **Transportation** | 2 | 4% |
| **Voting** | 1 | 2% |

| Software Engineering: Keywords | | |
|---|---|---|
| **Application** | 11 | 20% |
| **Design** | 14 | 26% |
| **Development process** | 6 | 11% |
| **Foundations** | 8 | 15% |
| **Quality** | 2 | 4% |
| **Requirements** | 2 | 4% |
| **Security** | 6 | 11% |
| **Technology** | 1 | 2% |
| **Testing** | 4 | 7% |

Disclaimer
The creation of these resources has been (partially) funded by the ERASMUS+ grant program of the European Union under grant no. 2018-1-LT01-KA203-047044.
Neither the European Commission nor the project's national funding agency DAAD are responsible for the content or liable for any losses or damage resulting of the use of these resources.

## 1.3. Job Description Analysis

The second step of the method (Fig. 1) to explore and define the interdisciplinary blockchain skills concept is the job description collection and analysis. Firstly, we have defined the template to collect the job descriptions (see Section 1.3). Then the data was collected (as discussed in Section 1.3.2) and analysed (as discussed in Section 1.3.3). This step resulted in the list of the interdisciplinary blockchain skills derived from the job descriptions.

### 1.3.1. Data Collection Template

Goal of the activity is to collect information from the job providers with the emphasis to understand needed market skills and requirements. Table 9 presents the template for collection of job descriptions. It includes: source (and if available short description), title of position, sector, field of the position/ application, skill demand (requirements for position), responsibilities, prerequisite, and comments (if any other).

**Table 9**: Template for collecting job descriptions

| | |
|---|---|
| Source (and if available short description) | Name of the company, potentially short description of the company and why it is relevant for the blockchain project. URL from where the information is extracted |
| Title of position | Title in of the position as given in the advertisement |
| Sector | Section (broader) where the company is operating |
| Field of the position/ Application | Field of the position (narrower), application domain, etc. |
| Skill demand (requirements for position) | Skills, requirements, profile of the position, description of what is required |
| Responsibilities | Responsibilities of the position, description of what should be done |
| Prerequisite | Previous experience, knowledge required, etc. |
| Comments (if any other) | Any comment if relevant |

### 1.3.2. Data Collection

Job advertisements (i.e., data for analysis) were collected in the countries of the project partners. More specifically they three largest recruiters (i.e., their advertising online systems) in Lithuania, Germany, Estonia and Denmark were inquired to collect the blockchain-related job advertisements. In addition, we have inquired the pan-European job advertising database and have requested all the BlockNet advisory board members to provide us their recent job advertisements on the blockchain-related job positions. The search process resulted in 85 job advertisements from Lithuania (12 advertisements), Germany (34 advertisements), Estonia (15 advertisements), Denmark (1 advertisement), and pan-European (23 advertisements) job online databases.

Disclaimer
The creation of these resources has been (partially) funded by the ERASMUS+ grant program of the European Union under grant no. 2018-1-LT01-KA203-047044.
Neither the European Commission nor the project's national funding agency DAAD are responsible for the content or liable for any losses or damage resulting of the use of these resources.

### 1.3.3. Result Analysis

While analysing the received results we have thoroughly reviewed each found advertisement, extracted the job skills and classified these skills to the categories of the KMK competence model. In total 47 interdisciplinary blockchain skills are elicited and classified to professional (15 skills), communicational (7 skills), self-competence (7 skills), learning (4 skills), methodological (2 skills), social (7 skills) and professional decision making (5 skills).

Table 10 presents 15 *professiona*l skills found in 73 advertisements. These skills are observed in different work related areas including information technology sector (found in 28 advertisements), financial (in 10 advertisements), crypto-currency trading sector (in 9 advertisements), consulting (in 7 advertisements) and other (e.g., automotive, banking, client innovation, education, energy, engineering, law, sales, supply chain management and tourism).

Table 11 presents 7 *communication* skills found in 36 advertisements. These skills are observed in information technology sector (found in 11 advertisements), crypto-currency trading (in 8 advertisements), consulting (in 5 advertisements), finance (in 5 advertisements) and other (education, sales, supply chain management and tourism).

Seven *self-competence* skills are observed in 31 advertisements as summarized in Table 12. The job sectors include information technology (in 12 advertisements), consulting (in 6 advertisements), finance (in 6 advertisements) and other (automotive, banking, cryptocurrency trading, engineering, tourism).

Table 13 presents the learning skills found in 12 advertisements. These are observed in information technology sector (found in 7 advertisements), client innovation, consulting, crypto currency trading, and finance.

We have observed 2 methodical skills: *Apply innovative development methods*, and *Research and design new methods of working*. These are found in 3 advertisements.

Seven *social* aspects are found in 35 advertisements (see Table 14). These are observed in information technology (found in 13 advertisements), consulting (in 5 advertisements), finance (in 6 advertisements) and other (automotive, banking, client innovation, cryptocurrency trading, engineering, supply chain management and tourism).

Table 15 summarizes the skills found for professional decision making. These are observed in 19 job sectors which include banking, client innovation, consulting, cryptocurrency trading, engineering, finance, information technology and tourism.

Disclaimer
The creation of these resources has been (partially) funded by the ERASMUS+ grant program of the European Union under grant no. 2018-1-LT01-KA203-047044.
Neither the European Commission nor the project's national funding agency DAAD are responsible for the content or liable for any losses or damage resulting of the use of these resources.

**Table 10**: Demand for *professional skills* in different job domains

| | Automotive | Banking | Client innovation | Consulting | Crypto Curr. Trade | Education | Energy | Engineering | Finance | IT | Law | Sales | Supply Chain Mng | Tourism |
|---|---|---|---|---|---|---|---|---|---|---|---|---|---|---|
| **Number (percentage) of advertisements** | **3 (100%)** | **2 (100%)** | **1 (100%)** | **7 (100%)** | **9 (100%)** | **3 (100%)** | **2 (100%)** | **1 ( 100%)** | **10 (100%)** | **28 (100%)** | **1 (100%)** | **3 (100%)** | **2 (100%)** | **1 (100%)** |
| Lead and manage the team | | | | 1 (14%) | | | | | 3 (30%) | | | | | |
| Developing blockchain applications from beginning to the end; | | | | 1 (14%) | 3 (33%) | 1 (33%) | | | 4 (40%) | 10 (36%) | | | 1 (50%) | |
| Demonstrate knowledge of blockchain technology principles (cryptocurrencies, wallets, smart contracts, separate platforms) | 3 (100%) | | 1 (100%) | 3 (43%) | 6 (67%) | 2 (66%) | 1 (50%) | | 2 (20%) | 4 (14%) | | | | 1 (100%) |
| Apply (different) programming languages | 3 (100%) | 1 (50%) | | 3 (43%) | | 1 (33%) | 2 (100%) | | 4 (40%) | 11 (39%) | | 1 (33%) | 2 (100%) | 1 (100%) |
| Apply requirements engineering for the blockchain applications | | | | | | | | | 1 (10%) | 1 (4%) | | 1 (33%) | | |
| Demonstrate knowledge of regulatory standards, rules, laws, regulations, management standards | | 1 (50%) | | | | 1 (33%) | | | 2 (20%) | | | 1 (33%) | | |
| Develop and manage databases using data management systems (SQL, etc.) | | | | | 1 (11%) | | | 1 (100%) | 2 (20%) | 2 (7%) | | | | |
| Perform auditing, accounting and taxation processes | | 1 (50%) | | | | | | | 2 (20%) | 1 (4%) | 1 (100%) | 1 (33%) | | |
| Demonstrate knowledge of system architectures, frameworks, different layers | | 1 (50%) | | 3 (43%) | 4 (44%) | | | 1 (100%) | | 9 (32%) | | | 1 (50%) | |
| Manage system and software security, security risk management | | | | | 2 (22%) | 1 (33%) | | | 2 (20%) | | | | | |
| Apply testing, monitoring, controlling and developing automated tests | | | | | | | | | | 1 (4%) | | | | |
| Educate/train internal (e.g., colleagues, developers, testers, etc.) and external (e.g., customers, support teams, and etc.) stakeholders. | | | | 1 (14%) | | | | | | 2 (7%) | | | | 1 (100%) |
| Maintain systems | | | | 2 (27%) | | | | | | 2 (7%) | | | | 1 (100%) |
| Demonstrate knowledge of financial operations, sales, payments, transactions | | | | 1 (14%) | | | | | | 1 (4%) | | 2 (67%) | | |
| Demonstrate knowledge of network protocols | | | | | | | | 1 (100%) | | 2 (7%) | | | | |

Disclaimer
The creation of these resources has been (partially) funded by the ERASMUS+ grant program of the European Union under grant no. 2018-1-LT01-KA203-047044.
Neither the European Commission nor the project's national funding agency DAAD are responsible for the content or liable for any losses or damage resulting of the use of these resources.

**Table 11**: Demand for *communication skills* in different job domains

| | Consulting | CryptoCurr. trading | Education | Finance | IT | Sales | Supply Chain Management | Tourism |
|---|---|---|---|---|---|---|---|---|
| **Number (and percentage) of advertisements** | **5 (100%)** | **8 (100%)** | **3 (100%)** | **5 (100%)** | **11 (100%)** | **1 (100%)** | **1 (100%)** | **1 (100%)** |
| Demonstrate communication skills to external stakeholders (external users, customers, advisors, and etc.) | 5 (100%) | 4 (50%) | 1 (33%) | 3 (60%) | 9 (82%) | - | 1 (100%) | 1 (100%) |
| Demonstrate communication skills to the internal stakeholders (internal users, customers, advisors, and etc.) | 3 (60%) | 3 (38%) | | 2 (40%) | 6 (55%) | - | 1 (100%) | 1 (100%) |
| Practice good written communication (documentation) skills | 2 (40%) | 6 (75%) | 2 (67%) | 1 (20%) | 9 (82%) | - | 1 (100%) | 1 (100%) |
| Demonstrate good verbal communication skills | 1 (20%) | 6 (75%) | 2 (67%) | 1 (20%) | 9 (82%) | 1 (100%) | 1 (100%) | 1 (100%) |
| Practice good skills to educate (e.g., internal and external) stakeholders | 2 (60%) | 5 (63%) | | 1 (20%) | 6 (55%) | | 1 (100%) | 1 (100%) |
| Operate good ability to communicate complex problems | 1 (20%) | 2 (25%) | 3 (100%) | 1 (20%) | 6 (55%) | | 1 (100%) | 1 (100%) |
| Demonstrate good presentation skills | 1 (20%) | 2 (25%) | 1 (33%) | | 7 (64%) | | 1 (100%) | 1 (100%) |

Disclaimer
The creation of these resources has been (partially) funded by the ERASMUS+ grant program of the European Union under grant no. 2018-1-LT01-KA203-047044.
Neither the European Commission nor the project's national funding agency DAAD are responsible for the content or liable for any losses or damage resulting of the use of these resources.

**Table 12**: Demand for *self-competence* skills in different job domains

| | Automotive | Banking | Consulting | CryptoCurr. Trading | Engineering | Finance | IT | Tourism |
|---|---|---|---|---|---|---|---|---|
| **Number (and percentage) of advertisements** | **2 (100%)** | **1 (100%)** | **6 (100%)** | **1 (100%)** | **2 (100%)** | **6 (100%)** | **12 (100%)** | **1 (100%)** |
| Demonstrate work independence and self-organisation | | 1 (100%) | 4 (67%) | 1 (100%) | 2 (100%) | 2 (33%) | 4 (33%) | 1 (100%) |
| Practice new ideas (open-minded) | | | | | | 2 (33%) | 4 (33%) | |
| Sketch/imply creative solutions | | | 3 (50%) | | | 1 (17%) | 2 (17%) | |
| Apply critical thinking | | | | | | 2 (33%) | 1 (8%) | |
| Be proactive and take initiative | 1 (50%) | 1 (100%) | 1 (17%) | | | 2 (33%) | 1 (8%) | |
| Be responsible, trusted, and committed | 1 (50%) | | 1 (17%) | | | 1 (17%) | 1 (8%) | |
| Construct high quality results | | | | | | 1 (17%) | 2 (17%) | |

**Table 13**: Demand for *learning skills* in different job domains

| | Client innovation | Consulting | Crypto Curr. Trading | Finance | IT |
|---|---|---|---|---|---|
| **Number (and percentage) of advertisements** | **1 (100%)** | **1 (100%)** | **2 (100%)** | **1 (100%)** | **7 (100%)** |
| Ability to learn quickly | 1 (100%) | | | 1 (100%) | |
| Interest in new technology | | 1 (100%) | 1 (50%) | 1 (100%) | 4 (57%) |
| Interest in continuing learning | 1 (100%) | | 1 (50%) | | 2 (29%) |
| Be open-minded (take into account feedback) | | | | | 1 (14%) |

Disclaimer
The creation of these resources has been (partially) funded by the ERASMUS+ grant program of the European Union under grant no. 2018-1-LT01-KA203-047044.
Neither the European Commission nor the project's national funding agency DAAD are responsible for the content or liable for any losses or damage resulting of the use of these resources.

Erasmus+

**Table 14**: Demand for *social skills* in different job domains

| | Automotive | Banking | Client innovation | Consulting | Crypto Curr. trade | Engineering | Finance | IT | Supply Chain Management | Tourism |
|---|---|---|---|---|---|---|---|---|---|---|
| **Number (and percentage) of advertisements** | **3 (100%)** | **1 (100%)** | **1 (100%)** | **5 (100%)** | **2 (100%)** | **2 (100%)** | **6 (100%)** | **13 (100%)** | **1 (100%)** | **2 (100%)** |
| Demonstrate strong (inter-) organisational skills | 3 (100%) | | 1 (100%) | 4 (80%) | 1 (50%) | 2 (100%) | 4 (67%) | 9 (69%) | 1 (100%) | 2 (100%) |
| Demonstrate ability to work in international teams | 1 (33%) | | | 1 (20%) | | | 1 (17%) | 2 (15%) | | |
| Practice to support colleagues with expert knowledge | | | | 1 (20%) | | | 2 (33%) | 3 (23%) | | 1 (50%) |
| Use social media means | | | | | 1 (50%) | | | | | |
| Establish good social relationships with the customers | | | | 1 (20%) | | | 1 (17%) | 2 (15%) | | |
| Manage team and organise work | | 1 (100%) | | | | | 1 (17%) | 1 (8%) | | |

**Table 15:** Demand for *professional decision making* in different job domains

| | Banking | Client innovation | Consulting | Crypto Curr. Trade | Engineering | Finance | IT | Tourism |
|---|---|---|---|---|---|---|---|---|
| **Number (and percentage) of advertisements** | **1 (100%)** | **1 (100%)** | **4 (100%)** | **2 (100%)** | **2 (100%)** | **4 (100%)** | **4 (100%)** | **1 (100%)** |
| Apply analytical methods to solve problem | | 1 (100%) | 4 (100%) | 1 (50%) | 2 (100%) | 3 (75%) | 3 (75%) | 1 (100%) |
| Apply evidence-based approaches for problem solving | | | 1 (100%) | | | 1 (25%) | | |
| Demonstrate ability to prioritise | | | | 1 (50%) | | | 1 (25%) | |
| Practice creative solution | 1 (100%) | | | | | | | |
| Maintain and support solution | | | | | | | 1 (25%) | |

Disclaimer
The creation of these resources has been (partially) funded by the ERASMUS+ grant program of the European Union under grant no. 2018-1-LT01-KA203-047044.
Neither the European Commission nor the project's national funding agency DAAD are responsible for the content or liable for any losses or damage resulting of the use of these resources.

Erasmus+

## 1.4. Summary of the Skills Concept

The goal of the third step (see Fig. 1) is to summarise and aggregate interdisciplinary blockchain skills developed in step 1 (systematic literature mapping, see Section 1.2) and step 2 (job description analysis, see Section 1.3).

**Table 16:** Aggregated Interdisciplinary Blockchain Skills Concept

| PROFESSIONAL COMPETENCE | |
|---|---|
| Leadership | **Lead and manage** the team |
| | **Maintain blockchain-based systems** |
| | Perform **auditing, accounting** and **taxation** processes |
| | **Educate/train** internal (e.g., colleagues, developers, testers, etc.) and external (e.g., customers, support teams, and etc.) stakeholders. |
| | Demonstrate knowledge of **financial operations, sales, payments,** and **transactions** |
| General | Explain the **foundations, algorithms**, **components** and **principles** (e.g., cryptocurrencies, wallets, smart contracts, separate platforms) of blockchain systems |
| | Demonstrate blockchain technology capabilities and apply them to **business-related** challenges |
| | **Compare blockchain platforms** to enable understanding of different system design choices |
| *Supply chain management, Finance, Business and Economics* | Explain **general capabilities of** blockchains in supply chain management |
| | Discuss and compare different **blockchain models, scheme and solutions** with constructed/illustrated application (suggestions, proposals, methods for blockchain use in economics, business and finance) |
| | Explain the **interoperability** of blockchain technology and possible collaboration between unknown or untrusted parties |
| | Explain how **information asymmetry** can be addressed by the blockchain-based applications |
| | Demonstrate blockchain capabilities and apply them to **counterfeit and fraud prevention** problem statements |
| | Demonstrate blockchain capabilities and apply them to **provenance and track&trace** problem statements |
| *Blockchain-based Application Development* | Describe **development processes** for blockchain solutions |
| | Demonstrate knowledge of **system architectures, frameworks, different layers** |
| | Discuss software quality goals and their impact on blockchain system development |
| | Knowledge of **regulatory standards, rules, laws, regulations, management standards** |
| | Describe the software **requirements** elicitation and engineering process of blockchain systems |
| | Develop and manage **databases** using data management systems (also use of SQL, etc.) |
| | Knowledge of **network protocols** |
| | Apply (different) **programming languages** |
| | Develop a **testing** plan for concrete blockchain solutions |
| *Privacy Manage-ment* | Describe **privacy management** principles using the blockchain solutions |
| | Explain **identity management** principles using the blockchain solutions |
| *Security Engineering and Security Management* | Explain how data, information and processes can be **secured** by the use of the blockchain technology |
| | Recognise **security countermeasure** implications |
| | Explain **access control** (authentication, authorization and identity) models |
| | Describe **transaction protection and validation** principles |
| | Underline major **encryption and signature** schemes |
| | Explain **trust management** principles |
| | State major **fair mining** principles |
| | Identify **security errors** in smart contracts |

Disclaimer
The creation of these resources has been (partially) funded by the ERASMUS+ grant program of the European Union under grant no. 2018-1-LT01-KA203-047044.
Neither the European Commission nor the project's national funding agency DAAD are responsible for the content or liable for any losses or damage resulting of the use of these resources.

The majority of the aggregation effort was done regarding the Professional skills as these were elicited in both step 1 and 2. Table 16 presents the revised list of the professional skills. Other competence categories remain rather the same as defined in the second step (see Section 1.3.3) The aggregated results are then used to define the feedback evaluation form.

## 1.5. Validation of Skills Concept

In this section the validation activities and results of the validation of the interdisciplinary blockchain skills concept are presented. Firstly, the validation goals are introduced (see Section 1.5.1). Then we will discuss the feedback collection form (see Section 1.5.2) and respondents' profile (see Section 1.5.3). Then we will present the validation results, which includes both the aggregated results collected using the feedback collection form (see Section 1.5.4) and the subjective comments given by the respondents (see Section 1.5.5).

### 1.5.1. Validation goals

The major objective of the validating the interdisciplinary blockchain skills concept is to evaluate whether the skills concept is important for the potential employees and educators. Therefore, we have requested to assess the level of each competence along the criteria of <u>knowledge</u> (ability to remember previously learned information and demonstrate an understanding of the facts), <u>application</u> (ability to apply knowledge to actual situations), and <u>evaluation</u> (ability to evaluate (i.e., make and defend judgments) based on internal evidence or external criteria).

### 1.5.2. Feedback Collection Form

The feedback evaluation form is provided in Appendix 1.3. Firstly, it includes several questions about the respondents' profile and gathers information about the respondents work place (e.g., university or industry company), job position, country (Denmark, Estonia, Germany, Lithuania or other) and experience in working with / teaching the Blockchain technology.

Next the feedback collection form includes seven sections, which consider professional (33 questions), self-competence (8 questions), social (7 questions), communication (8 questions), methodical (3 questions), learning skills (5 questions) and professional decision making issues (6 skills).

### 1.5.3. Respondents

We have invited members of the BlockNet project advisory board and people from at least two universities of the project partner's home countries.

We have received 16 answers (see Fig. 7): 13 answers came from the respondents who are working at the university and 3 from the respondents who are working in the company. *Job position* wise we have received answers from seven professors, two research associates, and one answer (each) from associate professor, student assistant, lecturer, scientist, CEO, developer, and head of blockchain solutions.

Our respondents also play different *roles* regarding the blockchain technology teaching, usage or development. This included teachers (Fintech, programming, technology, and etc.), curriculum developer, blockchain and business application, researchers, developer of research papers for the infrastructure projects and dApps (decentralized applications).

Disclaimer
The creation of these resources has been (partially) funded by the ERASMUS+ grant program of the European Union under grant no. 2018-1-LT01-KA203-047044.
Neither the European Commission nor the project's national funding agency DAAD are responsible for the content or liable for any losses or damage resulting of the use of these resources.

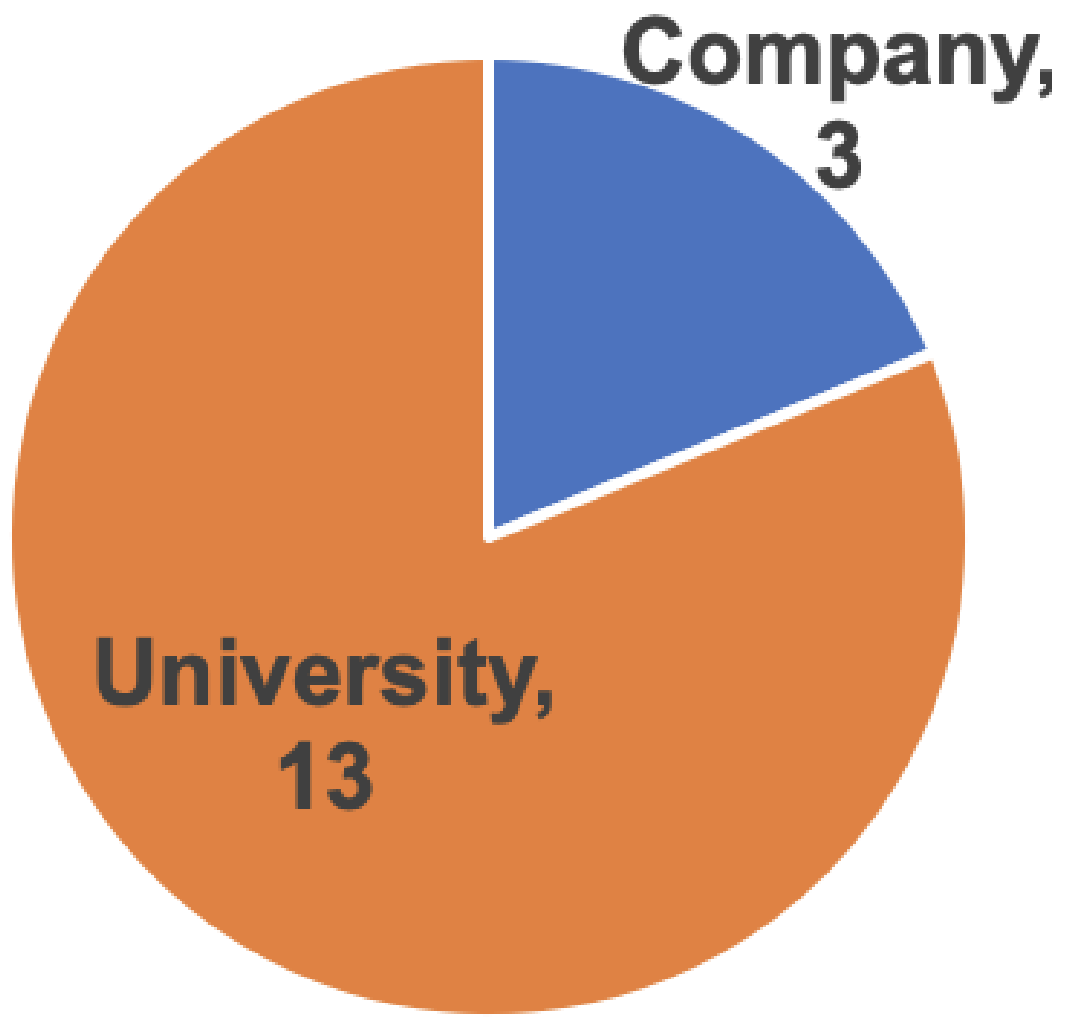

**Fig. 7:** Respondents' job profile

Finally, the respondents had different *experience* with the blockchain technology. As illustrated in Fig. 8, responses were collected from six novice (0-6 months of experience with blockchain technology), three beginners (7-12 months of experience with blockchain technology), four specialists (1-4 years of experience with blockchain technology) and three experts (more than 4 years of experience with blockchain technology).

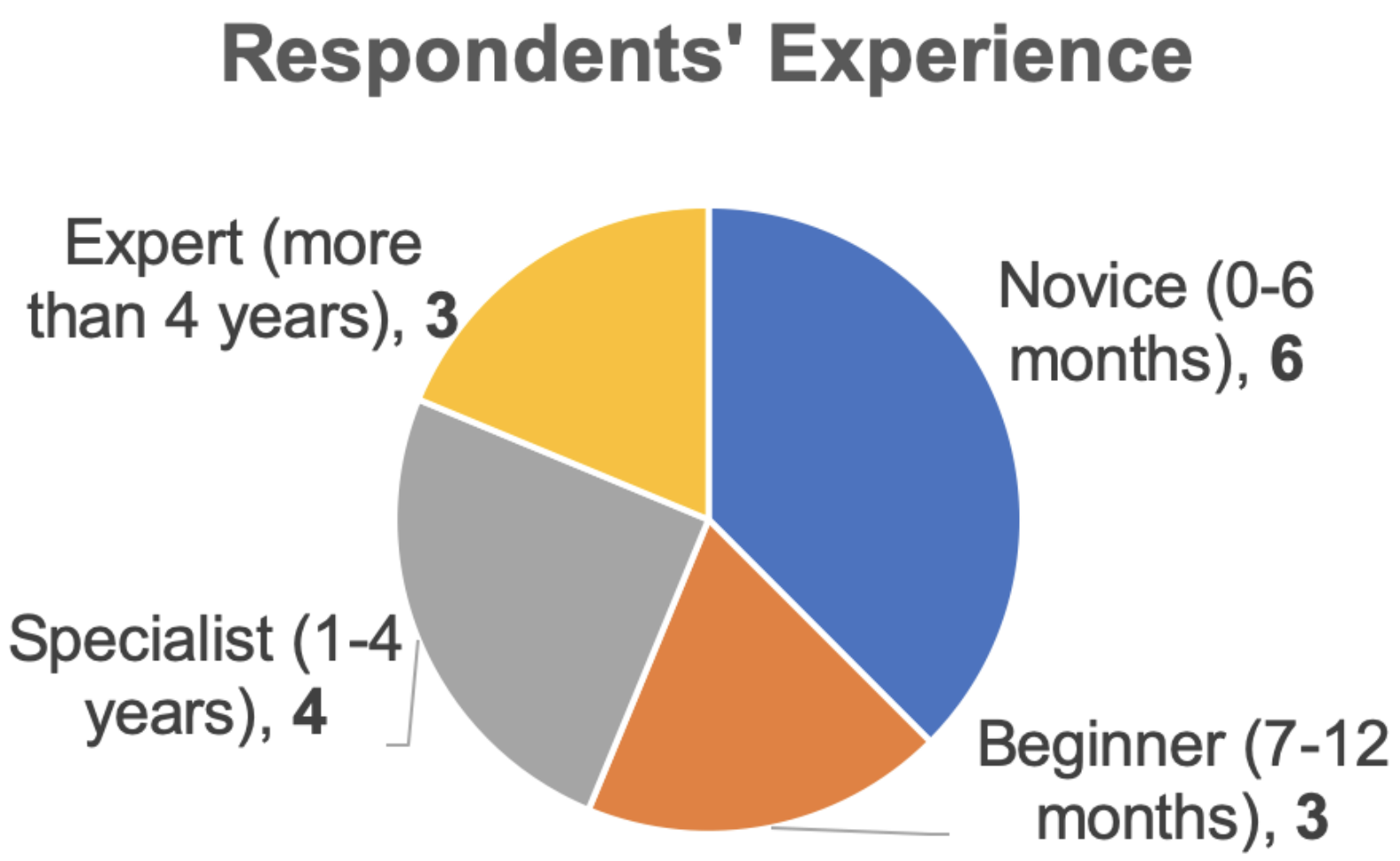

**Fig. 8**. Respondents' experience

As illustrated in Fig. 9 we have received six responses from Estonia, three responses both from Germany and Lithuania, two responses from Denmark and 1 response both from Lithuania and Latvia.

Disclaimer
The creation of these resources has been (partially) funded by the ERASMUS+ grant program of the European Union under grant no. 2018-1-LT01-KA203-047044.
Neither the European Commission nor the project's national funding agency DAAD are responsible for the content or liable for any losses or damage resulting of the use of these resources.

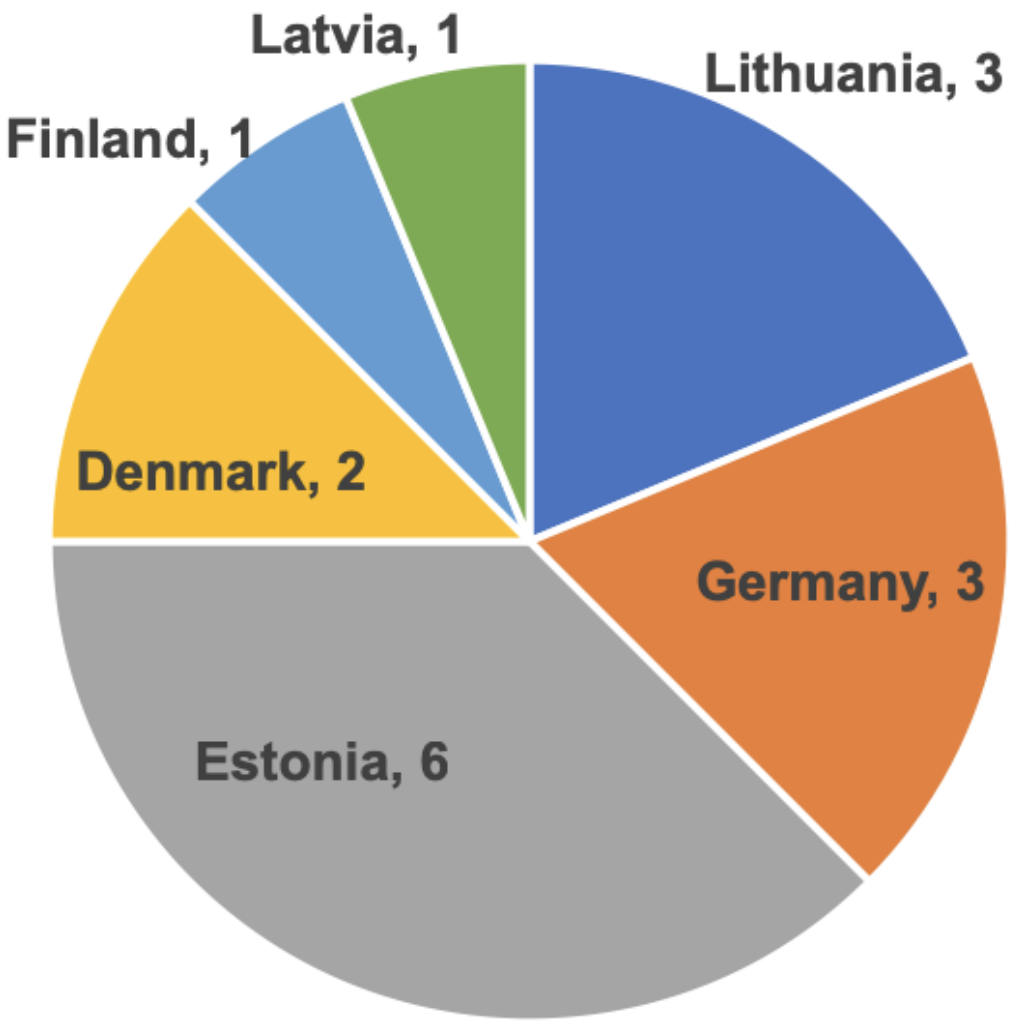

**Fig. 9**: Countries from where responses were received

## 1.5.4. Collected Feedback

### 1.5.4.1. Professional skills

Fig. 10 shows the feedback collected regarding *Professional skills – leadership group*. Respondents acknowledge the importance of <u>application</u> of the (2) *maintaining of blockchain applications*, (4) *education/training of the internal and external stakeholders* and (5) *knowledge of financial operations, sales, payments, and transactions*. Importance of <u>knowledge</u> is highlighted for (1) *leading and managing the team* and (3) *performing auditing accounting and taxation processes*.

Fig. 11 shows the feedback collected regarding *Professional skills – general group*. Criterion of <u>application</u> is highlighted for all items in this competence group.

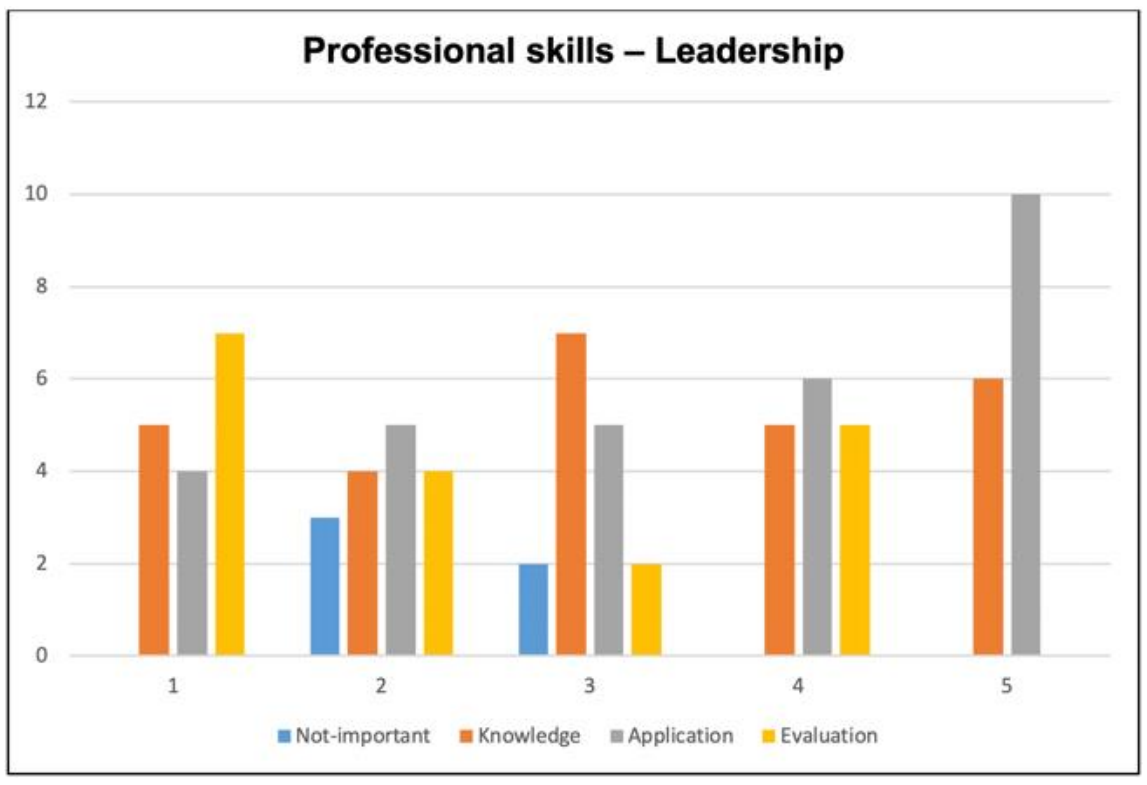

1. Lead and manage the team
2. Maintain blockchain-based systems
3. Perform auditing, accounting and taxation processes
4. Educate/train internal (e.g., colleagues, developers, testers, etc.) and external (e.g., customers, support teams, and etc.) stakeholders.
5. Knowledge of financial operations, sales, payments, and transactions

**Fig. 10**: Feedback on Professional skills - leadership

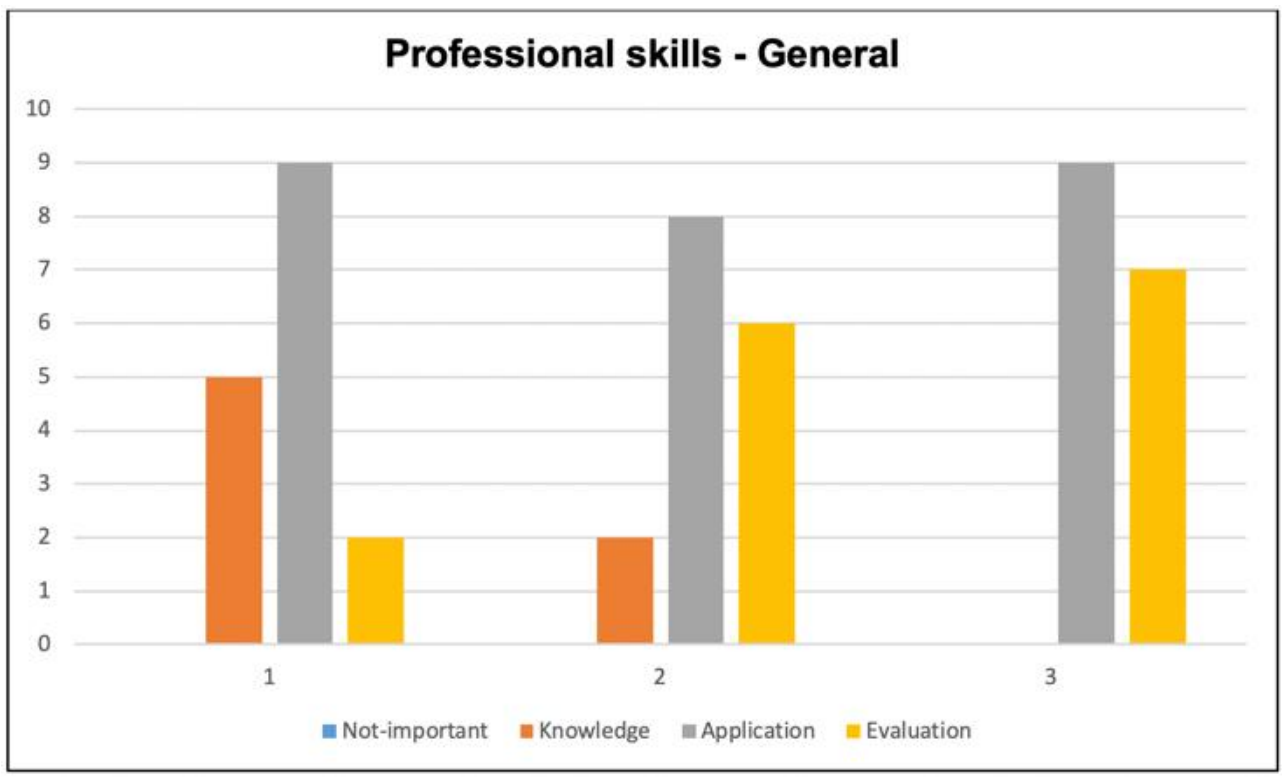

1. Explain the foundations, algorithms, components and principles (e.g., cryptocurrencies, wallets, smart contracts, separate platforms) of blockchain systems
2. Demonstrate blockchain technology capabilities and apply them to business-related challenges
3. Compare blockchain platforms to enable understanding of different system design choices

**Fig. 11**: Feedback on Professional skills - general

Disclaimer
The creation of these resources has been (partially) funded by the ERASMUS+ grant program of the European Union under grant no. 2018-1-LT01-KA203-047044.
Neither the European Commission nor the project's national funding agency DAAD are responsible for the content or liable for any losses or damage resulting of the use of these resources.

Fig. 12 shows the feedback collected regarding *Professional Skills – group of supply chain management, finance, business and economics*. Four items are estimated higher along the application criterion in this group. However, knowledge criterion is found higher for (1) *explanation of general capabilities of blockchain in supply chain management,* and competence (3) is found equal regarding application and evaluation criteria.

Fig. 13 shows the feedback collected regarding *Professional Skills – blockchain-based application development group*. Six skills are estimated higher regarding the application criterion in this group. Two items (i.e., (4) *knowledge of regulatory standards, rules, laws, regulations, management standards* and (5) *description of the software requirements elicitation and engineering process of blockchain systems*) are assessed equally along the knowledge and application criteria. One competence (i.e., (9) *development of a testing plan for concrete blockchain solution*) is assessed equally along the application and evaluation criteria.

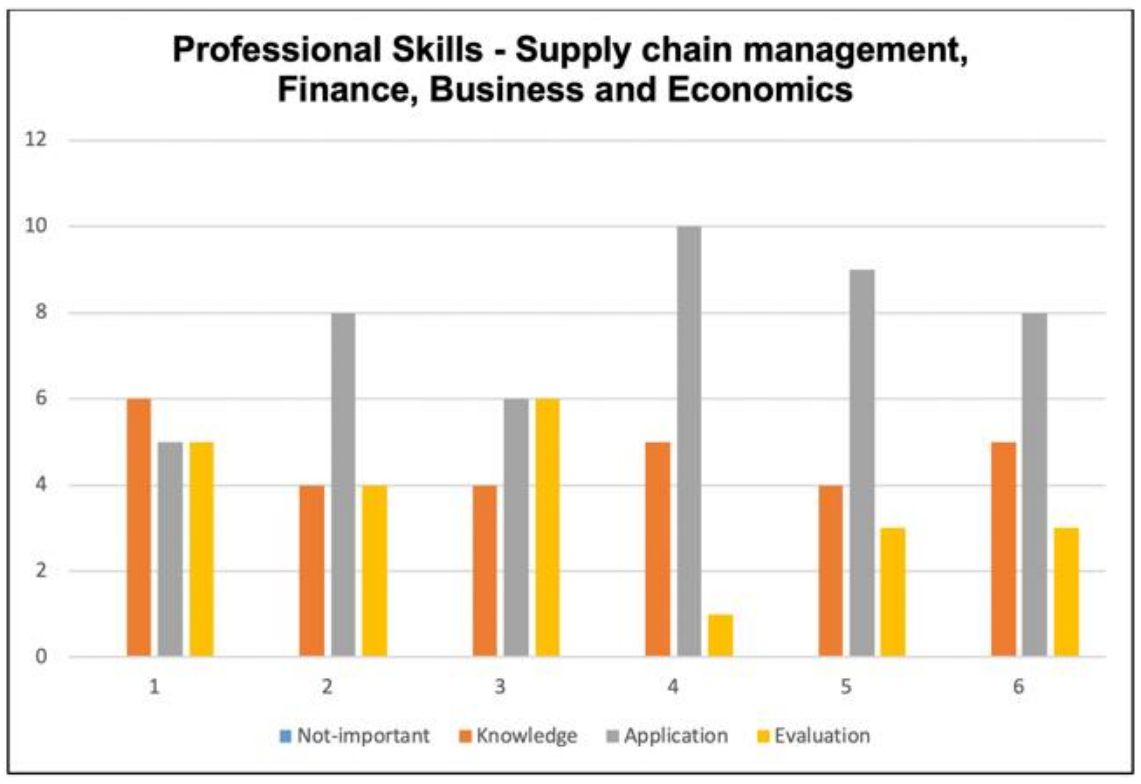

1. Explain general capabilities of blockchains in supply chain management
2. Discuss and compare different blockchain models, scheme and solutions with constructed/illustrated application (suggestions, proposals, methods for blockchain use in economics, business and finance)
3. Explain the interoperability of blockchain technology and possible collaboration between unknown or untrusted parties
4. Explain how information asymmetry can be addressed by the blockchain-based applications
5. Demonstrate blockchain capabilities and apply them to counterfeit and fraud prevention problem statements
6. Demonstrate blockchain capabilities and apply them to provenance and track&trace problem statements

**Fig. 12**: Feedback on Professional skills – supply chain management, financial and business and economics

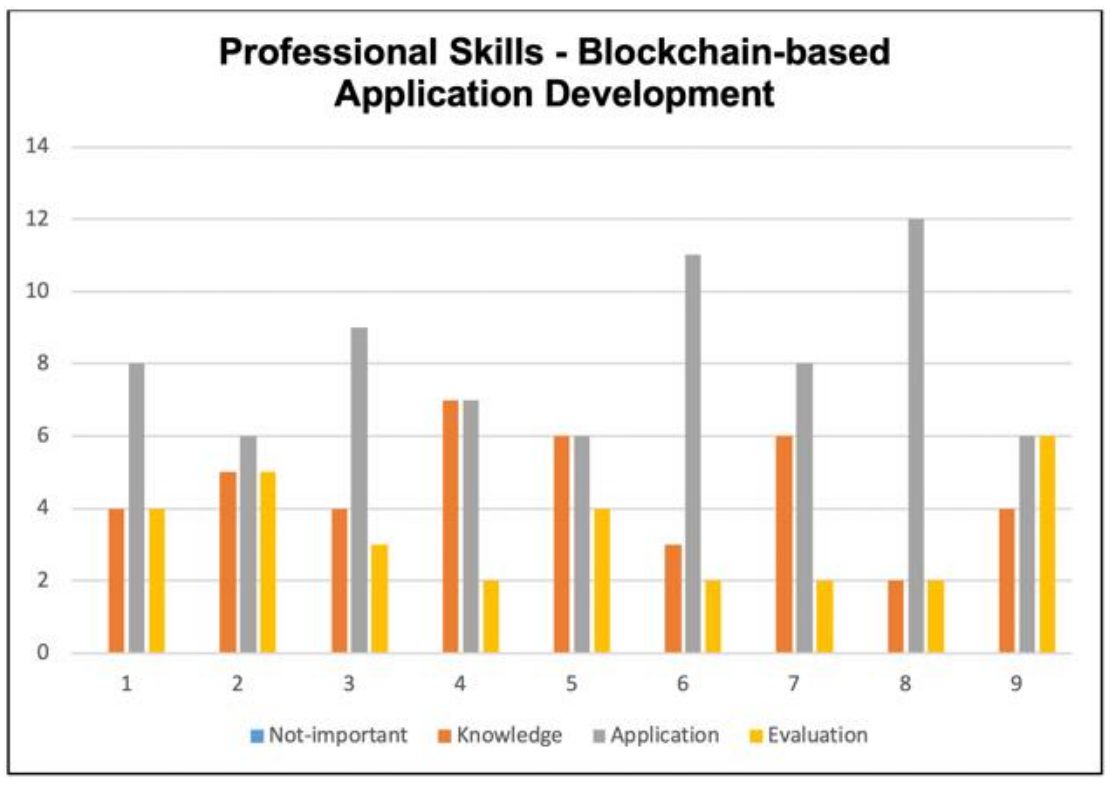

1. Describe development processes for blockchain solutions
2. Knowledge of system architectures, frameworks, different layers
3. Discuss software quality goals and their impact on blockchain system development
4. Knowledge of regulatory standards, rules, laws, regulations, management standards
5. Describe the software requirements elicitation and engineering process of blockchain systems
6. Develop and manage databases using data management systems (also use of SQL, etc.)
7. Knowledge of network protocols
8. Apply (different) programming languages
9. Develop a testing plan for concrete blockchain solutions

**Fig. 13**: Feedback on Professional skills – blockchain-based application development

Fig. 14 shows the feedback collected regarding *Professional Skills – privacy management group*. Competence on (2) *explaining identity management principles using blockchain solutions* is assessed higher along the application criterion. Another competence on (1) *describing privacy management principles using the blockchain solutions* is found equal along the knowledge and application criteria.

Fig. 15 shows the feedback collected regarding *Professional Skills – security engineering and security management group*. Respondents evaluated four skills higher regarding the application criterion and two items higher – knowledge criterion. Competence of (4) *describing*

Disclaimer
The creation of these resources has been (partially) funded by the ERASMUS+ grant program of the European Union under grant no. 2018-1-LT01-KA203-047044.
Neither the European Commission nor the project's national funding agency DAAD are responsible for the content or liable for any losses or damage resulting of the use of these resources.

*transaction protection and validation principles* is found equal regarding knowledge and application criteria.

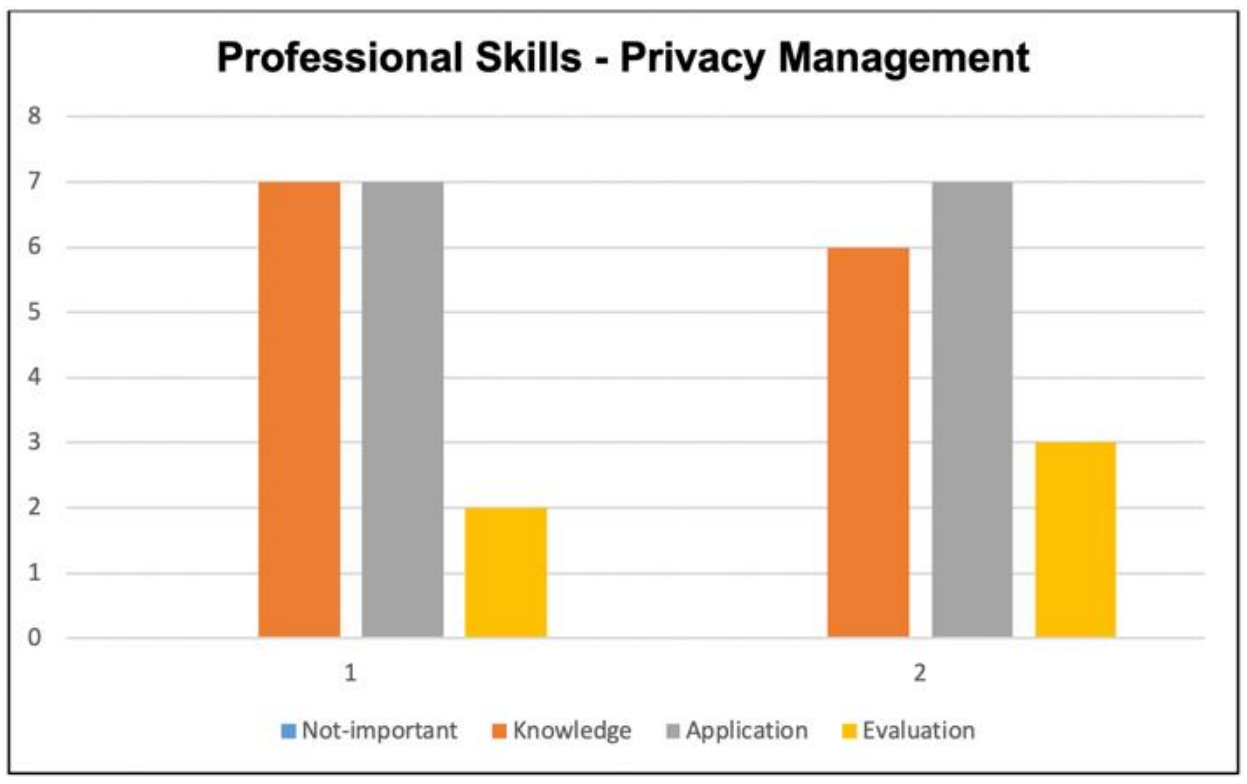

1. Describe privacy management principles using the blockchain solutions
2. Explain identity management principles using the blockchain solutions

**Fig. 14**: Feedback on Professional skills – privacy management

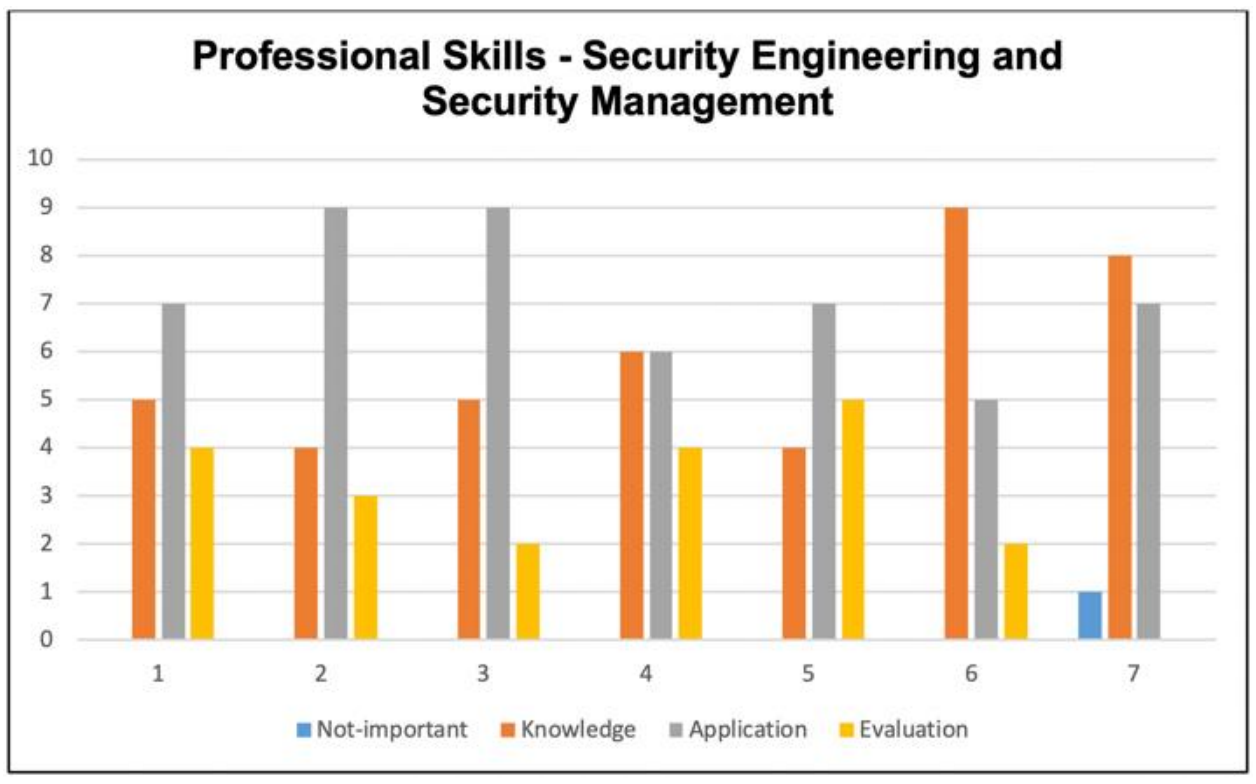

1. Explain how data, information and processes can be secured by the use of the blockchain technology
2. Recognise security countermeasure implications
3. Explain access control (authentication, authorization and identity) models
4. Describe transaction protection and validation principles
5. Underline major encryption and signature schemes
6. Explain trust management principles
7. State major fair mining principles

**Fig. 15**: Feedback on Professional skills – security engineering and security management

In the additional comments regarding the Professional competence, respondents indicated the need for "*more focus on the technology application to business models*", and focus "*on business applications of existing distributed protocols, rather than developing new ones*". It was also a proposal to "*remove the General group and distribute the affected questions to other categories*", because this groups seems to be too specific.

Respondents also commented that "*its ok to have wide knowledge and to be specialist only in some fields*". In addition, skills should be "*differentiated by learner categories, depending on their future roles*". This also means that "*blockchain/technology from a business, design or technical side, can be very different (e.g., in order to explain the technology)*".

Other comments highlight that "*the profile of the employee*" should set the sets the requirements for teaching. For example, "*admin, developer and manager all have different requirements*", thus "*professionals for various roles and levels will be required*" different skill teaching.

### 1.5.4.2. Other skills

**Self-competence skills**. Fig. 16 shows the feedback collected regarding *Self-competence skills*. Majority is assessed higher regarding the application criterion. One competence – (7) construction of high-quality results – is evaluated higher regarding the evaluation criterion.

In the additional comments, the respondents indicated that it is important to "*understand real problems and be critical that others solution, not blockchain might be a solution*", too. It is also important to develop "*the ability to see a failure as an opportunity to learn and improve*". Self-competence skills should also include "*tech-savvy and willingness for changes*" and

Disclaimer
The creation of these resources has been (partially) funded by the ERASMUS+ grant program of the European Union under grant no. 2018-1-LT01-KA203-047044.
Neither the European Commission nor the project's national funding agency DAAD are responsible for the content or liable for any losses or damage resulting of the use of these resources.

“*team-working skills*”, “*leadership, presentation, reporting and etc. skill*”. According to one respondent, self-competence skills “*are normal skills for bachelor students*”.

**Social skills**. Fig. 17 shows the feedback collected regarding *Social skills*. Respondents evaluated three social skills (i.e., (1) *demonstration of strong organizational skills*, (4) *usage of social media means*, and (6) *management team and organization of work*) higher regarding knowledge criterion. Other three skills – (2) *demonstration of ability to work in international teams*, (3) *practice to support colleagues with expert knowledge*, and (5) *establishment of good social relationships with customers* – are assessed higher regarding the application criterion.

In the additional comments, the respondents indicated that other skills should include skill on how to establish “*personal relations with the colleagues*” and “good social relationships with the team/team members”. Also, these depend “heavily on the work profile”. Finally, it was a comment that “many good specialists are introverts”.

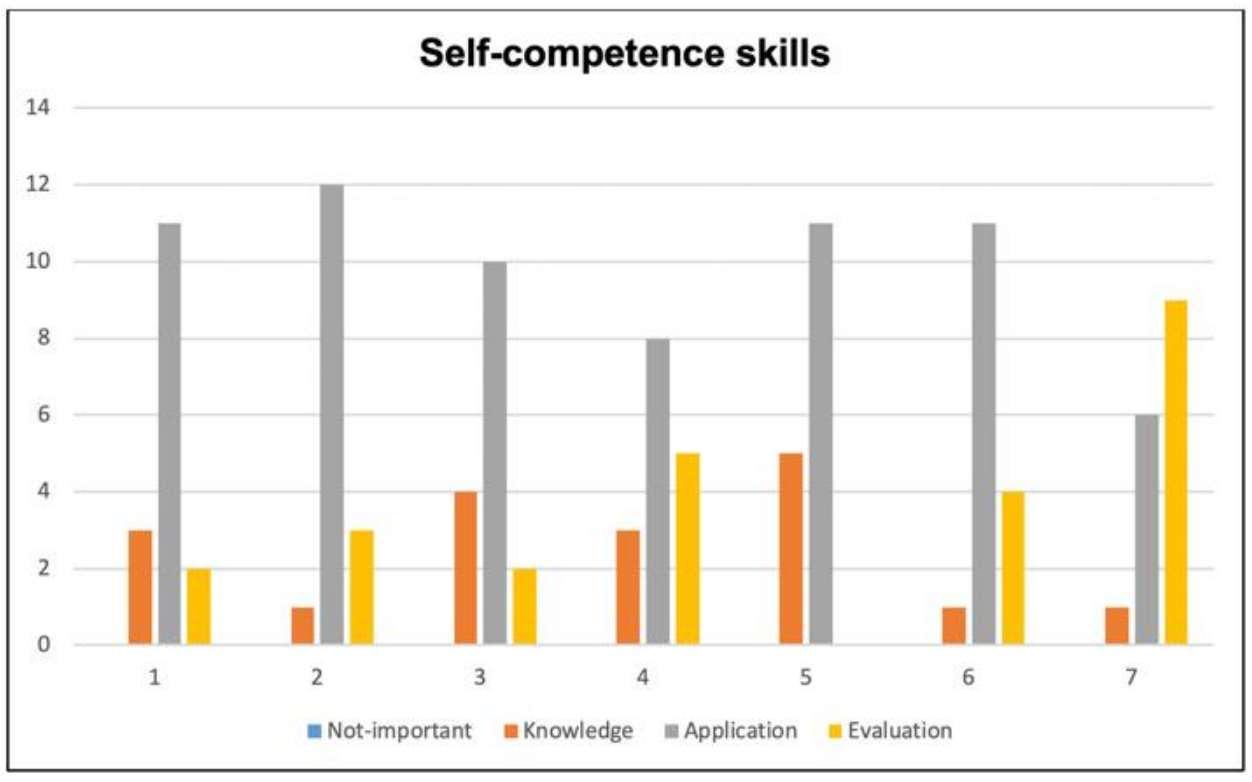

1. Demonstrate work independence and self-organisation
2. Practice new ideas (open-minded)
3. Sketch/imply creative solutions
4. Apply critical thinking
5. Be proactive and take initiative
6. Be responsible, trusted, and committed
7. Construct high quality results

**Fig. 16**: Feedback on Self-competence skills

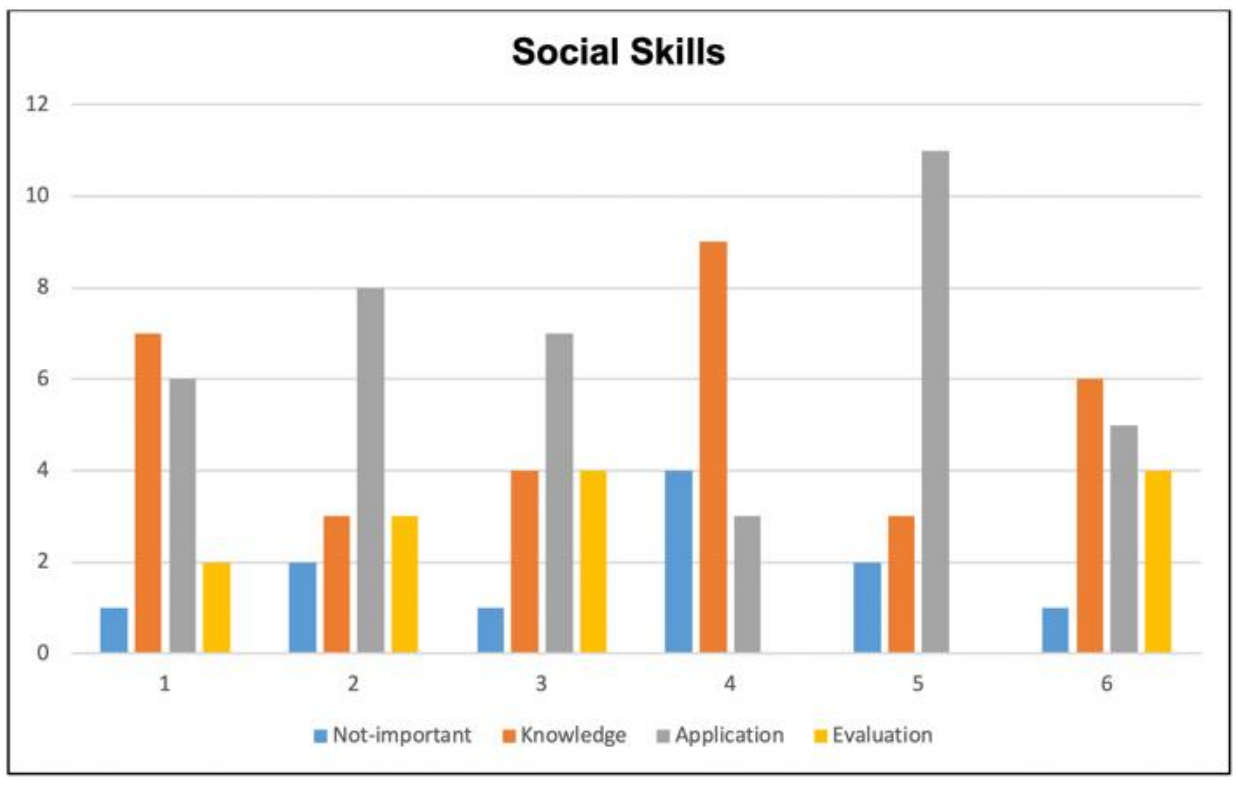

1. Demonstrate strong (inter-) organisational skills
2. Demonstrate ability to work in international teams
3. Practice to support colleagues with expert knowledge
4. Use social media means
5. Establish good social relationships with the customers
6. Manage team and organise work

**Fig. 17**: Feedback on Social skills

**Communication skills**. Fig. 18 shows the feedback collected regarding *Communication skills*. All communication competence (except for (3) items – *practice good written communication skills*, which is equally evaluated at knowledge and application criteria) is evaluated higher regarding the application criteria.

In the additional comments, respondents indicated importance of “ability to work in the interdisciplinary team” and ability to “communicate a failure or a critical argument”. It was also highlighted that this is “work profile dependent” and that “the higher the position (for example, group leader) the better should be the communications' skills”.

**Methodical skills**. Fig. 19 shows the feedback collected regarding *Methodical skills*. The competence of (1) *applying innovative development methods* is found higher regarding the application criterion. (2) *Research and design of new methods of working* is found rather equal regarding all three criteria.

Disclaimer
The creation of these resources has been (partially) funded by the ERASMUS+ grant program of the European Union under grant no. 2018-1-LT01-KA203-047044.
Neither the European Commission nor the project’s national funding agency DAAD are responsible for the content or liable for any losses or damage resulting of the use of these resources.

In the additional comments, respondents indicated that "implementation of new methods is often quite risky as it may negatively effect on some existing processes". However, a set of other skills was also suggested; for example, "ability to prototype", and. ability to "analyze a currently applied methodological process and change it if needed".

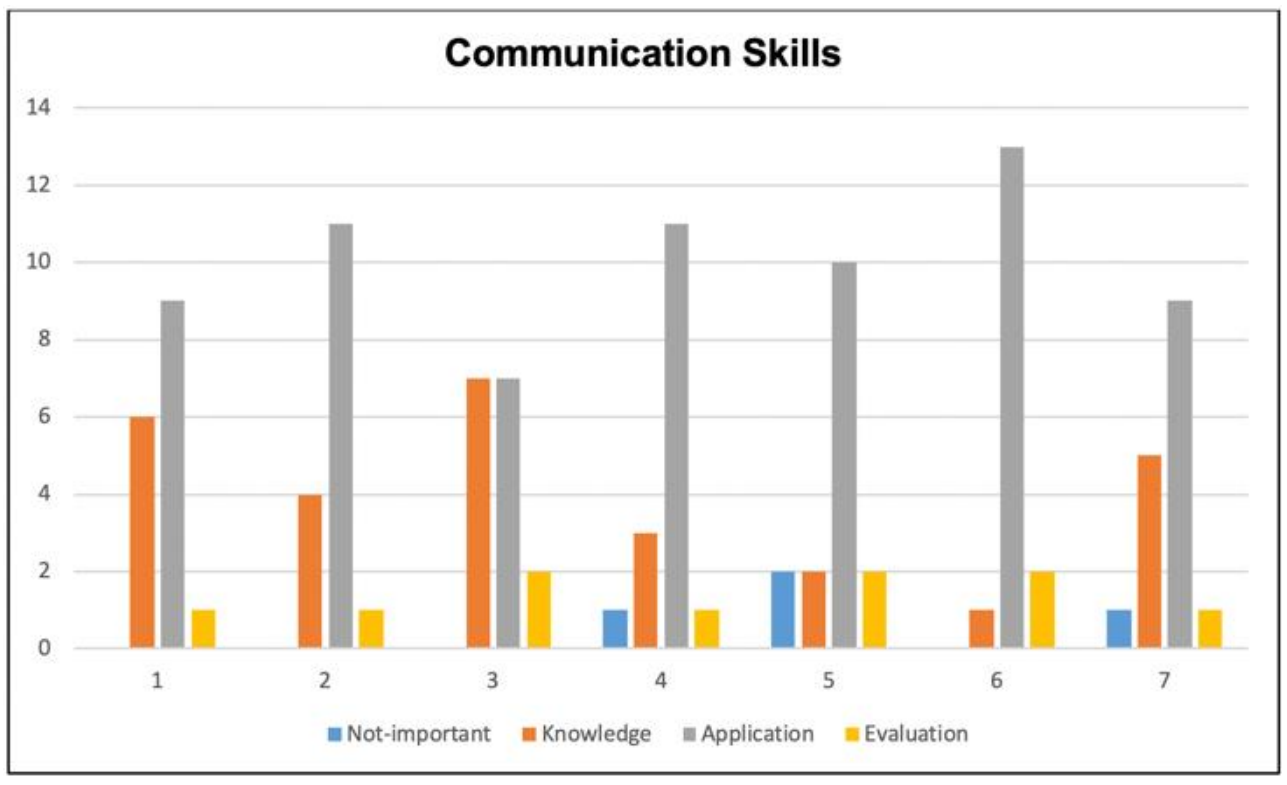

1. Demonstrate communication skills to external stakeholders (external users, customers, advisors, and etc.)
2. Demonstrate communication skills to the internal stakeholders (internal users, customers, advisors, and etc.)
3. Practice good written communication (documentation) skills
4. Demonstrate good verbal communication skills
5. Practice good skills to educate (e.g., internal and external) stakeholders
6. Operate good ability to communicate complex problems
7. Demonstrate good presentation skills

**Fig. 18**: Feedback on Communication skills

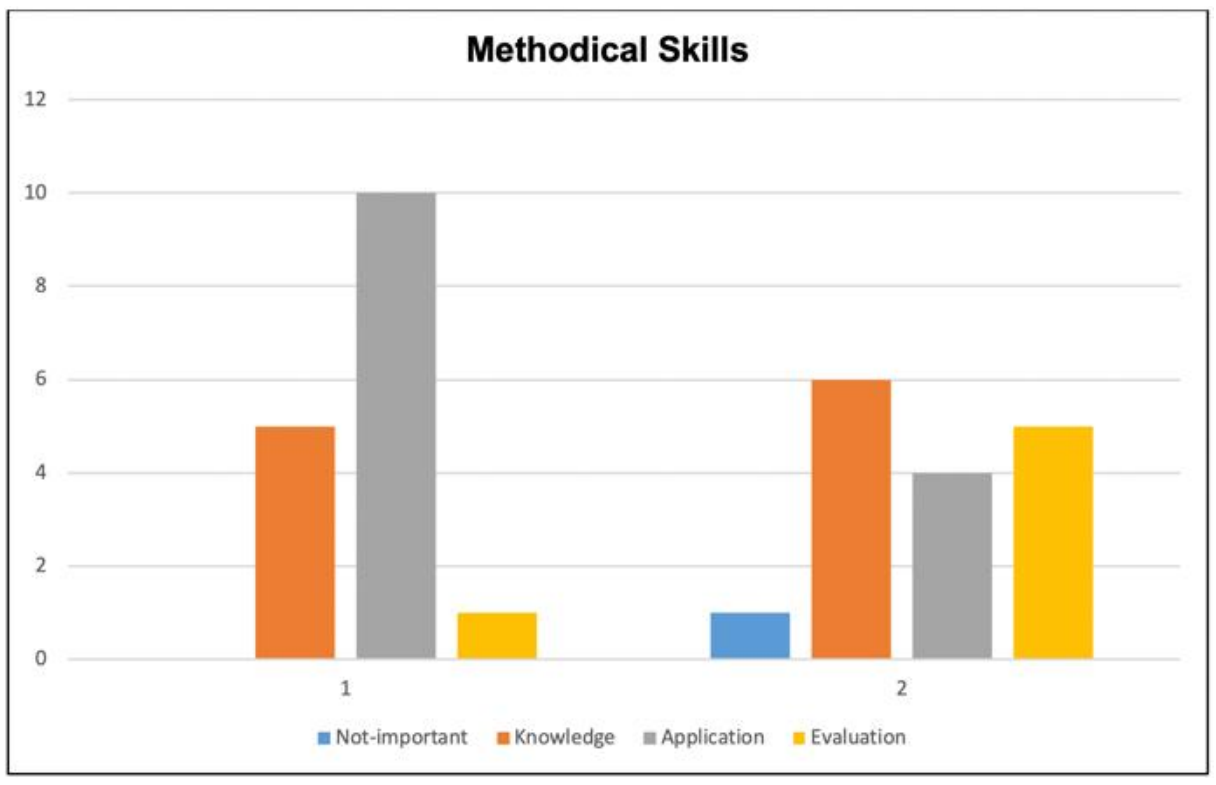

1. Apply innovative development methods
2. Research and design new methods of working

**Fig. 19**: Feedback on Methodical skills

**Learning skills**. Fig. 20 shows the feedback collected regarding *Learning skills*. Respondents assessed three learning items higher regarding the application criteria. Only the skill on (2) *interest in new technology* is found rather equal along all three criteria.

In the additional comments, respondents indicated importance of "sharing of knowledge/good experience" and "learning and applying learning methods". It was also indicated that these skills are "very general, can apply to any modern ICT topic".

**Professional decision making**. Fig. 21 shows the feedback collected regarding *Professional decision making*. Responses suggest that (3) *demonstration of ability to prioritise* is slightly higher regarding the evaluation criteria. Other professional decision making issues are found higher in the application criteria.

The respondents additionally indicated that learners should be able to "monitor solution and find the real options in changing environment", that should have "ability to recognise a problem". In addition these skills are "dependent on the work profile". Also, "good software design requires the ability to propose a range of possible solutions for a given problem, and analyse their advantages and disadvantages. This can be done in dialogue, but better in lucid writing".

Disclaimer
The creation of these resources has been (partially) funded by the ERASMUS+ grant program of the European Union under grant no. 2018-1-LT01-KA203-047044.
Neither the European Commission nor the project's national funding agency DAAD are responsible for the content or liable for any losses or damage resulting of the use of these resources.

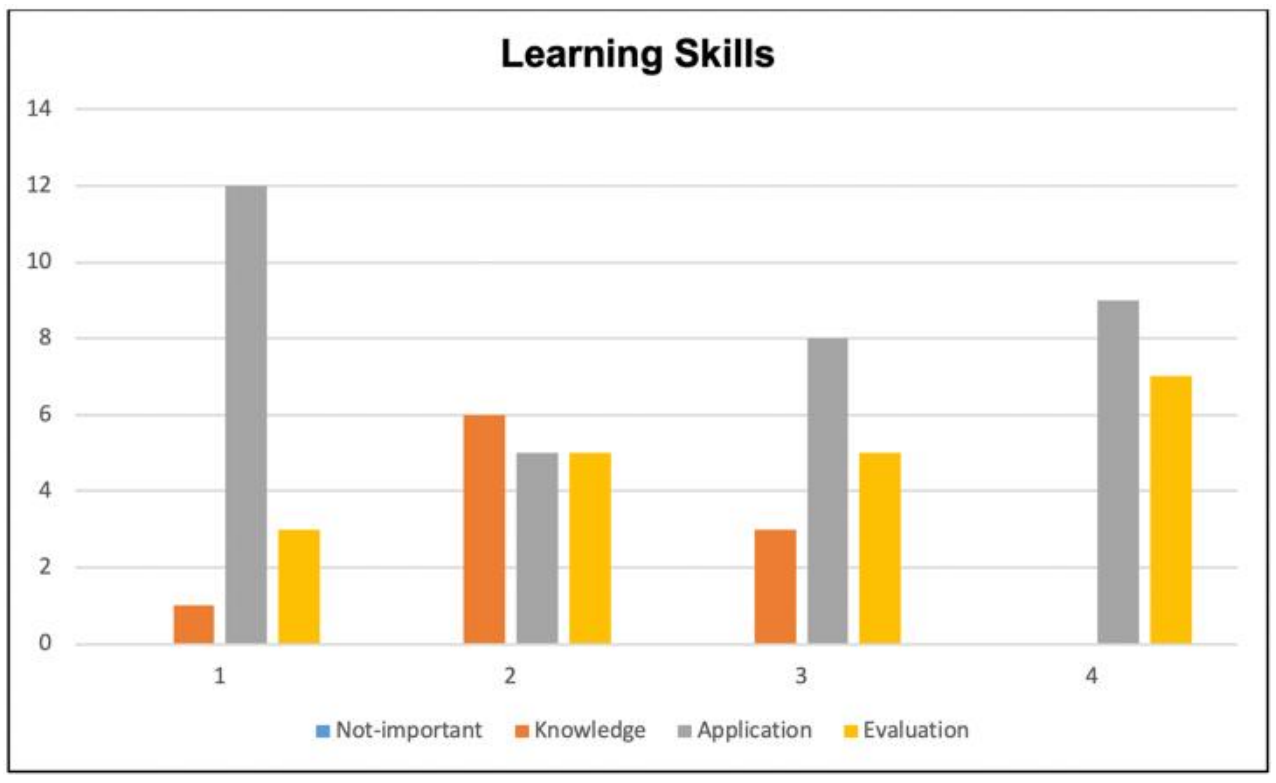

1. Ability to learn quickly
2. Interest in new technology
3. Interest in continuing learning
4. Be open-minded (take into account feedback)

**Fig. 20**: Feedback on Learning skills

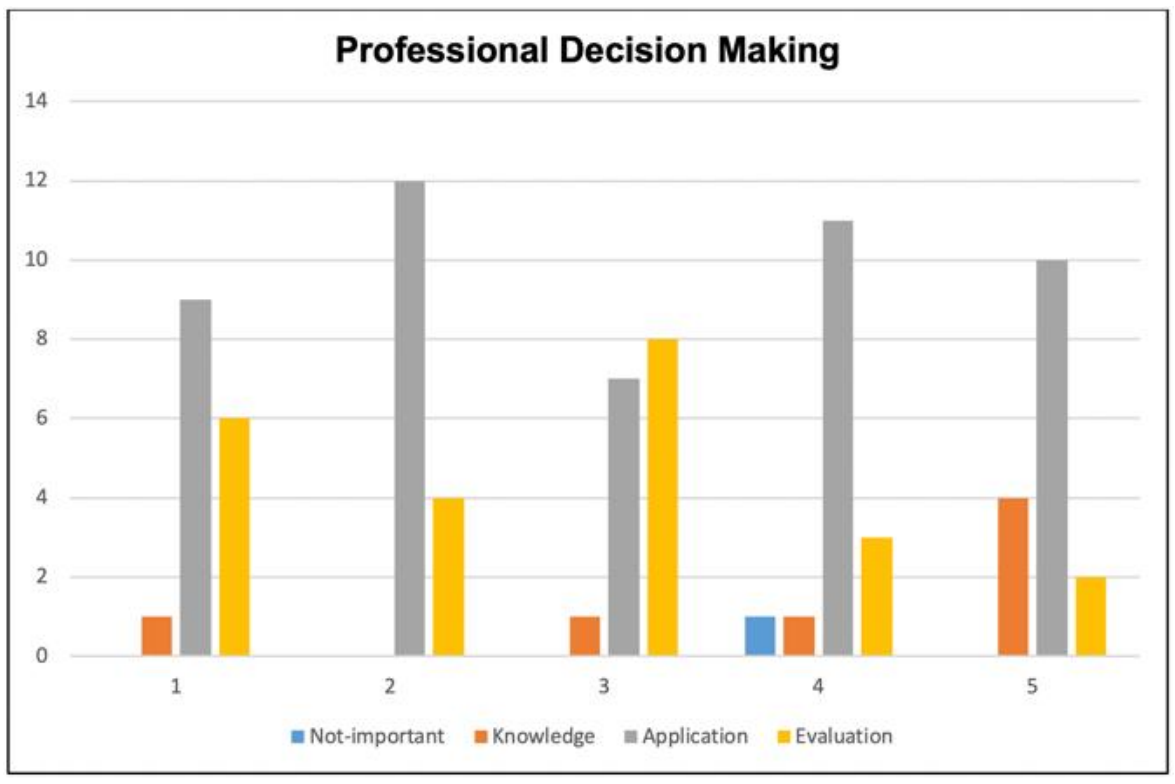

1. Apply analytical methods to solve problem
2. Apply evidence-based approaches for problem solving
3. Demonstrate ability to prioritise
4. Practice creative solution
5. Maintain and support solution

**Fig. 21**: Feedback on Professional decision making

## 1.6. Consolidation of Skills Concept

In this Section we discuss the skills concept consolidation activity (see Fig. 1). Firstly, we will consider the way how the consolidation activity is performed (see Section 1.6.1). Next, in Section 1.6.2, we will discuss the BlockNet competence concept itself.

### 1.6.1. Competence Consolidation

As already explained in the previous chapters, the systematic literature review (step 1), the screening of job advertisements (step 2) and the following survey (step 4) generated a comprehensive database, which is why the number of individual derived competence items (61) is correspondingly large. Each item represents a sufficiently defined competence requirement.

In the following step, the individual items were used to create a domain-specific competence model that describes the competence requirements for future experts in the interdisciplinary blockchain environment. As Leinweber (2013) points out, grouping the identified competence in predefined groups is essential to ensure further clarity and transparency of the model. For this reason, an aggregation of the predefined competence list is necessary in order to avoid possible redundancies (Hecklau et al. 2016). For this purpose, the project team discursively clustered the collected items. Since no competence model has yet been developed for the BCT area, the clustering was open to new results. The number of items was reduced by eliminating redundant competence items and carefully merging items with the same meaning. The results were discussed several times within the project consortium. This led to the competence concept that is described in the next chapter.

### 1.6.2. Competence Concept

For the BlockNet competence concept a domain-specific competence model for the BCT area was created, which consists of four competence fields. In addition to technical competence, we have methodological-, social- and personal competences. In particular, the technical competence was also divided into domain-specific competence clusters and can be seen in the following Table 17, which is an extract of the BlockNet competence catalogue.

In the first column of Table 17 we can see the competence field of the technical competence. The afore mentioned domain-specific competence clusters and their descriptions are presented in the subsequent columns. The last column contains the actual competence items. The competence items were identified using the KMK competence model for wide screening and were associated with each individual competence cluster. To allocate the items in the presented way, the items were analyzed and tested by means of the cluster description to ensure they would be applicable for the corresponding cluster.

Furthermore, methodological-, social- and self-competence items were aggregated in the BlockNet competence catalogue by means of seven clusters as can be seen in Table 18.

Disclaimer
The creation of these resources has been (partially) funded by the ERASMUS+ grant program of the European Union under grant no. 2018-1-LT01-KA203-047044.
Neither the European Commission nor the project's national funding agency DAAD are responsible for the content or liable for any losses or damage resulting of the use of these resources.

**Table 17:** Technical Skills Extract from BlockNet Competence Catalogue

| Competence field | Competence Cluster | Definition of Competence Cluster | Competence item |
|---|---|---|---|
| Technical Competences | Technical BCT Basics | Be able to explain the general BCT capabilities and functionalities of smart contracts. Be able to compare and explain different BCT solutions and plattforms. | Knowledge of the foundations, components, principles (e.g., cryptocurrencies, wallets, smart contracts, separate platforms) of blockchain systems |
| | | | Explain trust management principles |
| | | | Define ways to maintain blockchain-based systems |
| | | | Demonstrate blockchain technology capabilities and apply them to business-related challenges |
| | | | Compare blockchain platforms to enable understanding of different system design choices |
| | | | Discuss and compare different blockchain models, scheme and solutions with constructed/illustrated application (suggestions, proposals, methods for blockchain use in economics, business and finance) |
| | Business, Economics and Finance | Be able to explain processes in Business, Economics and Finance and understand the impact of BCT. | Knowledge of auditing, accounting and taxation processes as blockchain application fields |
| | | | Knowledge of financial operations, sales, payments, and transactions impacted by blockchain solutions |
| | | | Knowledge of regulatory standards, rules, laws, regulations, management standards relevant for blockchain implementations |
| | Supply Chain Management | Be able to explain processes in SCM and within corporate networks and understand the impact of BCT. | Knowledge of the general capabilities of blockchains in SCM |
| | | | Explain the interoperability of blockchain technology and possible collaboration between unknown or untrusted parties in SCM |
| | | | Demonstrate blockchain capabilities and apply them to counterfeit and fraud prevention problem statements |
| | | | Demonstrate blockchain capabilities and apply them to provenance and track&trace problem statements in SCM |
| | | | Analyze how information asymmetry in corporate networks can be addressed by the blockchain-based applications |
| | Computer Science and application development | Be able to explain the functionalities of the technical elements BCT consists of and understand development requirements in BCT environment. | Comprehension of development processes for blockchain solutions |
| | | | Knowledge of system architectures, frameworks, different layers |
| | | | Discuss software quality goals and their impact on blockchain system development |
| | | | Describe the software requirements elicitation and engineering process of blockchain systems |
| | | | Develop and manage databases using data management systems (also use of SQL, etc.) |
| | | | Knowledge of network protocols |
| | | | Apply (different) programming languages |
| | | | Develop a testing plan for concrete blockchain solutions |
| | Security Engineering and Privacy Management | Be able to explain the impacts BCT applications have in relation to security and privacy management. | Describe privacy management principles using the blockchain solutions |
| | | | Explain identity management principles using the blockchain solutions |
| | | | Explain how data, information and processes can be secured by the use of the blockchain technology |
| | | | Recognise security countermeasure implications |
| | | | Explain access control (authentication, authorization and identity) models |
| | | | Describe transaction protection and validation principles |
| | | | Underline major encryption and signature schemes |
| | | | State major fair mining principles |
| | | | Identify security errors in smart contracts |

As described above, all in all twelve competence clusters were developed and structured in the already presented four main fields of competence (technical, social, personal, methodological) according to Hecklau et al. (2017). The clusters and fields that constitute the final BlockNet competence concept can be presented in an integrated way as illustrated in Fig. 22.

Following the clustering and grouping tasks, the competence structure model was complemented by the use of Bloom's Taxonomy to describe the competence level of each item for the future online course. Hence, the competence levels as a whole, build a pattern of learning objectives and clarify the goals of the course. Fig. 23 shows an exemplary allocation of general blockchain competence items from technical competence cluster to the six Bloom's levels (knowledge, understanding, application, analysis, synthesis and evaluation). On top, we also have an allocation to the three levels (Reproduction, Application, Transfer) of the KMK model to further determine the competence level of the respective items.

Disclaimer
The creation of these resources has been (partially) funded by the ERASMUS+ grant program of the European Union under grant no. 2018-1-LT01-KA203-047044.
Neither the European Commission nor the project's national funding agency DAAD are responsible for the content or liable for any losses or damage resulting of the use of these resources.

**Table 18:** Other skills extract from BlockNet competence catalogue

| | | | |
|---|---|---|---|
| Methodical Competences | Efficiency orientation | Be able to work efficiently in a BCT project. | Ability to transfer knowledge to internal (e.g., colleagues, developers, testers, etc.) and external (e.g., customers, support teams, and etc.) stakeholders |
| | | | Demonstrate ability to prioritise |
| | | | Ability to organise interdisciplinary work |
| | Creativity | Be able to use creative approaches in BCT projects. | Sketch/imply creative solutions |
| | | | Apply innovative application development methods |
| | | | Practice creative solutions |
| | | | Knowledge of new interdisciplinary working methods |
| | Problem solving and Decision making | Be able to find the right decision and solve problems in BCT projects. | Apply analytical methods to solve problems |
| | | | Apply evidence-based approaches for problem solving |
| | | | Apply critical thinking |
| Social Competences | Leadership and Colloboration | Be able to lead a team and network to successfully run a BCT project. | Lead and manage the team |
| | | | Demonstrate strong (inter-) organisational networking competences |
| | | | Demonstrate ability to work in international and interdisciplinary teams |
| | | | Practice to support colleagues with expert knowledge |
| | | | Establish good social relationships with the customers |
| | Communication | Be able to communicate well in order to successfully run a BCT project. | Practice good written communication (documentation) competences |
| | | | Demonstrate good verbal communication competences |
| | | | Operate good ability to communicate complex and interdisciplinary problems |
| | | | Use social media means |
| | | | Demonstrate communication skills to internal and external stakeholders (colleagues, users, customers, advisors, and etc.) |
| | | | Demonstrate good presentation skills |
| Personal Competences | Willingness to learn | Be able to learn effectively in order to successfully run a BCT project. | Ability to learn quickly |
| | | | Demonstrate Interest in new technology |
| | | | Interest in continuing learning |
| | | | Take into account feedback |
| | | | Practice new ideas (open-minded) |
| | Abilitiy to work effectively | Be able to work effectively in order to successfully run a BCT project. | Demonstrate ability to work independentnly and self-organisation |
| | | | Be proactive and take initiative |
| | | | Be responsible, trusted, and committed |
| | | | Construct high quality results |

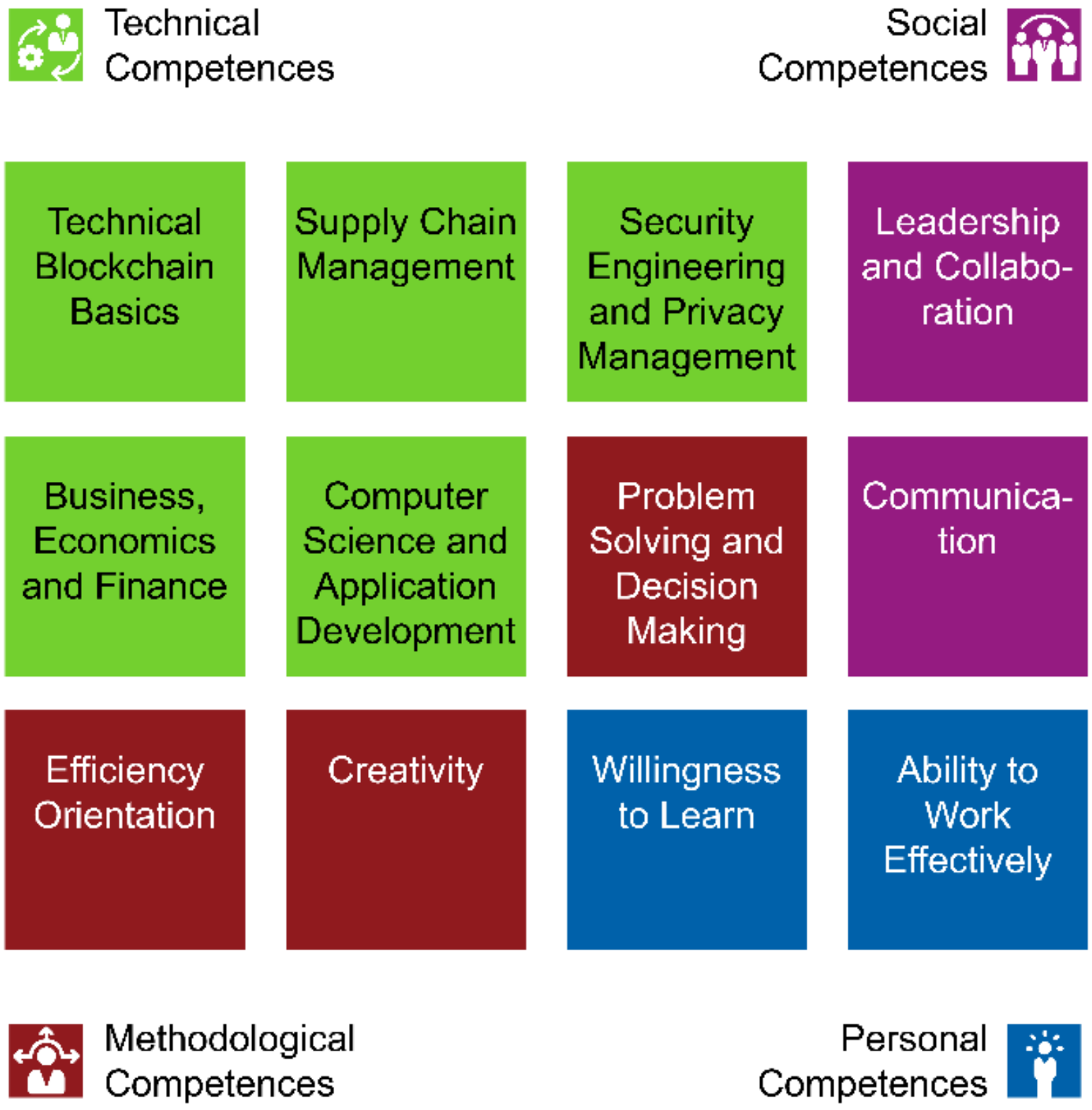

**Figure 22**: BlockNet Competence Model

Disclaimer
The creation of these resources has been (partially) funded by the ERASMUS+ grant program of the European Union under grant no. 2018-1-LT01-KA203-047044.
Neither the European Commission nor the project's national funding agency DAAD are responsible for the content or liable for any losses or damage resulting of the use of these resources.

| Competence item | Reproduction |  | Application | Transfer |  |  |
|---|---|---|---|---|---|---|
|  | Knowledge | Comprehension | Application | Analysis | Synthesis | Evaluation |
| Knowledge of the foundations, components, principles (e.g., cryptocurrencies, wallets, smart contracts, separate platforms) of blockchain systems | x |  |  |  |  |  |
| Explain trust management principles |  | x |  |  |  |  |
| Define ways to maintain blockchain-based systems |  | x |  |  |  |  |
| Demonstrate blockchain technology capabilities and apply them to business-related challenges |  |  | x |  |  |  |
| Compare blockchain platforms to enable understanding of different system design choices |  |  |  | x |  |  |
| Discuss and compare different blockchain models, scheme and solutions with constructed/illustrated application (suggestions, proposals, methods for blockchain use in economics, business and finance) |  |  |  |  | x |  |

**Fig. 23**: Exemplary excerpt from the BlockNet Bloom's taxonomy mapping

## 1.7. Concluding remarks

In the first part of this document, we report on the development of the interdisciplinary blockchain competence concept. The interdisciplinary blockchain competence concept is developed based on results of the following project research actions: 1) a literature survey, 2) job description screening, 3) validating the elicitation of competence concept with the BlockNet advisory board members and representatives from the leading universities of the partner countries. The BlockNet interdisciplinary blockchain competence concept includes interdisciplinary skills related to technical, methodological, social, and personal competences expected (or required) from graduates of higher education institutions for working in Blockchain-related projects/jobs.

In the following part of this document we report on the development of the Blockchain Best Practices activity, where the focus is placed on the qualitative empirical research based on case studies to be conducted with industry representatives.

Disclaimer
The creation of these resources has been (partially) funded by the ERASMUS+ grant program of the European Union under grant no. 2018-1-LT01-KA203-047044.
Neither the European Commission nor the project's national funding agency DAAD are responsible for the content or liable for any losses or damage resulting of the use of these resources.

Disclaimer
The creation of these resources has been (partially) funded by the ERASMUS+ grant program of the European Union under grant no. 2018-1-LT01-KA203-047044.
Neither the European Commission nor the project's national funding agency DAAD are responsible for the content or liable for any losses or damage resulting of the use of these resources.

Disclaimer
The creation of these resources has been (partially) funded by the ERASMUS+ grant program of the European Union under grant no. 2018-1-LT01-KA203-047044.
Neither the European Commission nor the project's national funding agency DAAD are responsible for the content or liable for any losses or damage resulting of the use of these resources.

# Appendix 1.1. Papers Selected for Systematic Literature Mapping

This appendix includes the papers selected for the systematic literature mapping. The lists are grouped according to the studied research fields: (*i*) economics, finance (FinTech) and business, (*ii*) supply chain management, (*iii*) security risk and security risk management and engineering, and (*iv*) software engineering.

## A1.1.1. Economics, finance (FinTech) and business

Disclaimer
The creation of these resources has been (partially) funded by the ERASMUS+ grant program of the European Union under grant no. 2018-1-LT01-KA203-047044.
Neither the European Commission nor the project's national funding agency DAAD are responsible for the content or liable for any losses or damage resulting of the use of these resources.

Disclaimer
The creation of these resources has been (partially) funded by the ERASMUS+ grant program of the European Union under grant no. 2018-1-LT01-KA203-047044.
Neither the European Commission nor the project's national funding agency DAAD are responsible for the content or liable for any losses or damage resulting of the use of these resources.

## A1.1.2. Supply chain management

Disclaimer
The creation of these resources has been (partially) funded by the ERASMUS+ grant program of the European Union under grant no. 2018-1-LT01-KA203-047044.
Neither the European Commission nor the project's national funding agency DAAD are responsible for the content or liable for any losses or damage resulting of the use of these resources.

Disclaimer
The creation of these resources has been (partially) funded by the ERASMUS+ grant program of the European Union under grant no. 2018-1-LT01-KA203-047044.
Neither the European Commission nor the project's national funding agency DAAD are responsible for the content or liable for any losses or damage resulting of the use of these resources.

Disclaimer
The creation of these resources has been (partially) funded by the ERASMUS+ grant program of the European Union under grant no. 2018-1-LT01-KA203-047044.
Neither the European Commission nor the project's national funding agency DAAD are responsible for the content or liable for any losses or damage resulting of the use of these resources.

## A1.1.3. Security risk and security risk management and engineering

Disclaimer
The creation of these resources has been (partially) funded by the ERASMUS+ grant program of the European Union under grant no. 2018-1-LT01-KA203-047044.
Neither the European Commission nor the project’s national funding agency DAAD are responsible for the content or liable for any losses or damage resulting of the use of these resources.

Disclaimer
The creation of these resources has been (partially) funded by the ERASMUS+ grant program of the European Union under grant no. 2018-1-LT01-KA203-047044.
Neither the European Commission nor the project's national funding agency DAAD are responsible for the content or liable for any losses or damage resulting of the use of these resources.

Disclaimer
The creation of these resources has been (partially) funded by the ERASMUS+ grant program of the European Union under grant no. 2018-1-LT01-KA203-047044.
Neither the European Commission nor the project's national funding agency DAAD are responsible for the content or liable for any losses or damage resulting of the use of these resources.

## A1.1.4. Software engineering

Disclaimer
The creation of these resources has been (partially) funded by the ERASMUS+ grant program of the European Union under grant no. 2018-1-LT01-KA203-047044.
Neither the European Commission nor the project's national funding agency DAAD are responsible for the content or liable for any losses or damage resulting of the use of these resources.

Disclaimer
The creation of these resources has been (partially) funded by the ERASMUS+ grant program of the European Union under grant no. 2018-1-LT01-KA203-047044.
Neither the European Commission nor the project's national funding agency DAAD are responsible for the content or liable for any losses or damage resulting of the use of these resources.

Disclaimer
The creation of these resources has been (partially) funded by the ERASMUS+ grant program of the European Union under grant no. 2018-1-LT01-KA203-047044.
Neither the European Commission nor the project's national funding agency DAAD are responsible for the content or liable for any losses or damage resulting of the use of these resources.

BLOCKNET

# Appendix 1.2. Mapping of the Papers and Competence

This appendix provides the mapping of the interdisciplinary blockchain competence and the literature sources in the Economics, finance (FinTech) and business, Supply chain management, Security risk and security risk management and engineering, and Software engineering research fields.

## A1.2.1. Supply chain management

| | **Explain general capabilities for the use in SCM** | **Explain the interoperability of BCT and possible collaboration between unknown or untrusted parties** | **Explain how data can be secured by the use of BCT** | **Explain how information asymmetry can be addressed by BCT applications** | **Demonstrate BCT capabilities and apply them to business-related challenges** | **Demonstrate BCT capabilities and apply them to counterfeit and fraud prevention problem statements** | **Demonstrate BCT capabilities and apply them to provenance and track&trace problem statements** |
|---|---|---|---|---|---|---|---|
| **SCM01** | | | X | X | | X | X |
| **SCM02** | | | | X | X | | |
| **SCM03** | X | | | | | | |
| **SCM04** | | | | X | | | |
| **SCM05** | | | X | | X | X | X |
| **SCM06** | | | | X | X | | |
| **SCM07** | | | X | X | X | | X |
| **SCM08** | | | X | X | X | | |
| **SCM09** | | | X | X | X | | X |
| **SCM10** | | | | | | X | X |
| **SCM11** | | | X | | | X | X |
| **SCM12** | | | | | X | | X |
| **SCM13** | | | | X | | | |
| **SCM14** | | | | X | | | X |
| **SCM15** | | | | X | X | | |
| **SCM16** | | | | X | X | | |
| **SCM17** | | X | | X | X | | |
| **SCM18** | | | | X | X | | |
| **SCM19** | | | | X | X | | X |
| **SCM20** | | | X | X | | | |
| **SCM21** | | | | | | X | |
| **SCM22** | | | | | | | X |
| **SCM23** | | | | X | X | | X |
| **SCM24** | | | | | | X | |
| **SCM25** | | | | X | | | X |
| **SCM26** | X | | | | | | |
| **SCM27** | | | X | X | X | X | X |
| **SCM28** | | | | X | | | X |

Disclaimer
The creation of these resources has been (partially) funded by the ERASMUS+ grant program of the European Union under grant no. 2018-1-LT01-KA203-047044.
Neither the European Commission nor the project's national funding agency DAAD are responsible for the content or liable for any losses or damage resulting of the use of these resources.

Erasmus+

| | | | | | | | |
|---|---|---|---|---|---|---|---|
| **SCM29** | **X** | | | | | | |
| **SCM30** | **X** | | | | | | |
| **SCM31** | | | | **X** | | **X** | |
| **SCM32** | | | | **X** | **X** | | X |
| **SCM33** | | **X** | | **X** | **X** | | |
| **SCM34** | | **X** | | **X** | **X** | | |
| **SCM35** | | | **X** | **X** | **X** | | X |
| **SCM36** | | **X** | **X** | | | **X** | |
| **SCM37** | | | | **X** | **X** | | X |
| **SCM38** | | | **X** | **X** | | | X |
| **SCM39** | | **X** | | **X** | | | |
| **SCM40** | | | | **X** | **X** | **X** | X |
| **SCM41** | **X** | | | | | | |
| **SCM42** | | | **X** | | | | X |
| **SCM43** | | | | **X** | | | X |
| **SCM44** | | **X** | **X** | **X** | **X** | **X** | X |
| **SCM45** | | | | | | | X |
| **SCM46** | | | **X** | **X** | **X** | | |
| **SCM47** | | | | **X** | | | X |
| **SCM48** | | | | **X** | **X** | | |
| **SCM49** | **X** | | | | | | |
| **SCM50** | | | | **X** | | | X |

## A1.2.2. Security risk management and engineering

| **Paper ID** | **Explain blockchain technology principles** | **Compare blockchain platforms to enable understanding of different system design choices** | **Discuss and compare different blockchain models and solutions to secure data and information** | **Describe privacy management principles using the blockchain solutions** | **Recognise security countermeasure implications** | **Explain access control (authentication, authorization and identity) models** | **Explain identity management principles using the blockchain solutions** | **Describe transaction protection and validation principles** | **Explain trust management principles** | **Underline major encryption and signature schemes** | **State major fair mining principles** | **Identify security errors in smart contracts** |
|---|---|---|---|---|---|---|---|---|---|---|---|---|
| **SEC01** | | | | | | | | X | | | | |
| **SEC02** | | | | | X | | | | | | | |
| **SEC03** | | X | | | | | | | | | | |
| **SEC04** | X | | | | | | | | | | | |
| **SEC05** | | | X | | | | | | | | | |
| **SEC06** | | | | X | | | | | | | | |
| **SEC07** | | | X | | | | | | | | | |
| **SEC08** | | | | | | | | X | | | | |
| **SEC09** | | | | | | | X | | | | | |
| **SEC10** | | | | | | | | | | | X | |

Disclaimer
The creation of these resources has been (partially) funded by the ERASMUS+ grant program of the European Union under grant no. 2018-1-LT01-KA203-047044.
Neither the European Commission nor the project's national funding agency DAAD are responsible for the content or liable for any losses or damage resulting of the use of these resources.

| | | | | | | | | | | | | |
|---|---|---|---|---|---|---|---|---|---|---|---|---|
| **SEC11** | | | | | | X | | | | | | |
| **SEC12** | | | | | | X | | | | | | |
| **SEC13** | | X | | | | | | | | | | |
| **SEC14** | | | | X | | | | | | | | |
| **SEC15** | | | | | X | | | | | | | |
| **SEC16** | | | X | | | | | | | | | |
| **SEC17** | X | | | | | | | | | | | |
| **SEC18** | | X | | | | | | | | | | |
| **SEC19** | | | | X | | | | | | | | |
| **SEC20** | | | X | | | | | | | | | |
| **SEC21** | | | | | | | | | | | | X |
| **SEC22** | | | | | | | | | | | | X |
| **SEC23** | | | X | | | | | | | | | |
| **SEC24** | | | X | | | | | | | | | |
| **SEC25** | | | X | | | | | | | | | |
| **SEC26** | | | | | | X | | | | | | |
| **SEC27** | | | | | X | | | | | | | |
| **SEC28** | | | | | | | | | | | X | |
| **SEC29** | | | X | | | | | | | | | |
| **SEC30** | | | X | | | | | | | | | |
| **SEC31** | | | | X | | | | | | | | |
| **SEC32** | | | | | X | | | | | | | |
| **SEC33** | | | X | | | | | | | | | |
| **SEC34** | | | | | | X | | | | | | |
| **SEC35** | | | X | | | | | | | | | |
| **SEC36** | | | X | | | | | | | | | |
| **SEC37** | | | X | | | | | | | | | |
| **SEC38** | | | X | | | | | | | | | |
| **SEC39** | | | X | | | | | | | | | |
| **SEC40** | | | | | | | X | | | | | |
| **SEC41** | | | | | | | | | | X | | |
| **SEC42** | | | | X | | | | | | | | |
| **SEC43** | | | | | X | | | | | | | |
| **SEC44** | | | X | | | | | | | | | |
| **SEC45** | | X | | | | | | | | | | |
| **SEC46** | | | X | | | | | | | | | |
| **SEC47** | | | | | | | | X | | | | |
| **SEC48** | | | | | | | | | | X | | |
| **SEC49** | | | | | | | | | X | | | |
| **SEC50** | | | | | | X | | | | | | |
| **SEC51** | | | X | | | | | | | | | |
| **SEC52** | | | X | | | | | | | | | |
| **SEC53** | | | X | | | | | | | | | |
| **SEC54** | | | X | | | | | | | | | |
| **SEC55** | | | | | | X | | | | | | |

Disclaimer
The creation of these resources has been (partially) funded by the ERASMUS+ grant program of the European Union under grant no. 2018-1-LT01-KA203-047044.
Neither the European Commission nor the project's national funding agency DAAD are responsible for the content or liable for any losses or damage resulting of the use of these resources.

| | | | | | | | | | | | | |
|---|---|---|---|---|---|---|---|---|---|---|---|---|
| **SEC56** | | | X | | | | | | | | | |
| **SEC57** | | | | | | | | | | X | | |
| **SEC58** | | | | X | | | | | | | | |
| **SEC59** | | | | X | | | | | | | | |
| **SEC60** | | | | | | | | | X | | | |
| **SEC61** | | | X | | | | | | | | | |
| **SEC62** | | | | | X | | | | | | | |
| **SEC63** | | | X | | | | | | | | | |
| **SEC64** | | | X | | | | | | | | | |
| **SEC65** | | | X | | | | | | | | | |
| **SEC66** | | | X | | | | | | | | | |
| **SEC67** | | | X | | | | | | | | | |

Disclaimer
The creation of these resources has been (partially) funded by the ERASMUS+ grant program of the European Union under grant no. 2018-1-LT01-KA203-047044.
Neither the European Commission nor the project's national funding agency DAAD are responsible for the content or liable for any losses or damage resulting of the use of these resources.

## Appendix 1.3. Feedback Collection Form

BlockNet: BlockChain Network Online Education for Interdisciplinary European Competence Transfer

# Interdisciplinary Blockchain Skills Concepts

This document assembles Skill Concepts defined after the literature review identifying the state of the art of blockchain in Supply Chain Management and Logistics, Finance, Business & Economics, Software Engineering, and Security Management and Engineering as well as adjoining fields. In addition, the job descriptions and advertisements were analysed in order to specify today's competence requirements from enterprises.

The objective of this questionnaire is to evaluate the developed skill concepts. Please give us recommendations on what employee's skills are important in the interdisciplinary environment where blockchain technology is used or developed. For each of the given skill concept, assess the level of skill in terms of

- **Knowledge** (1- lowest level of skills) – ability to remember previously learned information and demonstrate an understanding of the facts.
- **Application** (2 – medium level of the skills) - ability to apply knowledge to actual situations.
- **Evaluation** (3 – the highest level of skills) - ability to evaluate (i.e., make and defend judgments) based on internal evidence or external criteria.
- **Not important** (0) – mark this if the skill concept is not important.

The questionnaire consists of 8 sections. At first, we wish to learn about you as a respondent. Then, the rest of the questionnaire sections consider professional (33 questions), self-competence (8 questions), social (7 questions), communication (8 questions), methodical (3 questions), learning skills (5 questions) and professional decision making issues (6 skills).

It would take no more than 30 minutes to complete in this questionnaire. It is highly recommended that you would complete the questionnaire without interruption as intermediate results could not be saved.

We will inform you about the study results, once they are aggregated an analysed.

Disclaimer
The creation of these resources has been (partially) funded by the ERASMUS+ grant program of the European Union under grant no. 2018-1-LT01-KA203-047044.
Neither the European Commission nor the project's national funding agency DAAD are responsible for the content or liable for any losses or damage resulting of the use of these resources.

# Respondent's information

**Are you working at**

o University
o Industry company
o Other ____________________

**What is your job position** (e.g., professor, assoc. professor, manager, project leader, etc):
______________________________________

**What is your role regarding the blockchain technology teaching, usage or development** (e.g., blockchain technology teacher, blockchain miner, blockchain architect, blockchain hacker, blockchain developer, etc. )?

______________________________________

**Where is your job institution (university, company) located?**

o Denmark
o Estonia
o Germany
o Latvia
o Lithuania
o Other ____________________

**What is your experience in working with / teaching the Blockchain technology?**

o Novice (0-6 months)
o Beginner (7-12 months)
o Specialist (1-4 years)
o Expert (more than 4 years)

Disclaimer
The creation of these resources has been (partially) funded by the ERASMUS+ grant program of the European Union under grant no. 2018-1-LT01-KA203-047044.
Neither the European Commission nor the project's national funding agency DAAD are responsible for the content or liable for any losses or damage resulting of the use of these resources.

For each of the given skill concept, assess the level of skill in terms of

- **Knowledge** (1- lowest level of skills) – ability to remember previously learned information and demonstrate an understanding of the facts.
- **Application** (2 – medium level of the skills) - ability to apply knowledge to actual situations.
- **Evaluation** (3 – the highest level of skills) - ability to evaluate (i.e., make and defend judgments) based on internal evidence or external criteria.
- **Not important** (0) – mark this if the skill concept is not important.

For example:

| **Professional skills** | **Knowledge** | **Application** | **Evaluation** | | **Not important** |
|---|---|---|---|---|---|
| | 1 | 2 | 3 | | 0 |
| **Lead and manage** the team | o | o | o | | o |

If you mark Knowledge (e.g., ⊗ ), it would mean that the learner should be able to remember and understand how to *lead and manage the team*.
If you mark Application (e.g., ⊗ ), it would mean that the learner should be able to *lead and manage the team*.
If you mark Evaluation (e.g., ⊗ ), it would mean that the learner should be able to evaluate how well the *leading and managing the team* was executed.

If you mark Not important (e.g., ⊗ ), it would mean that skill of leading and managing the team is not important for the learner.

Please provide your additional comments (1-2 sentences are OK) at the end of each section of skills concepts.

| **Professional skills**<br>the willingness and ability, on the basis of technical knowledge and skills, to solve tasks and problems in a goal-oriented, appropriate, method-oriented and independent manner and to assess the result. | Knowledge | Application | Evaluation | | Not important |
|---|---|---|---|---|---|
| | 1 | 2 | 3 | | 0 |
| *Leadership* | | | | | |
| **Lead and manage** the team | o | o | o | | o |
| **Maintain blockchain-based systems** | o | o | o | | o |
| Perform **auditing, accounting** and **taxation** processes | o | o | o | | o |
| **Educate/train** internal (e.g., colleagues, developers, testers, etc.) and external (e.g., customers, support teams, and etc.) stakeholders. | o | o | o | | o |
| Demonstrate knowledge of **financial operations, sales, payments,** and **transactions** | o | o | o | | o |
| *General* | | | | | |
| Explain the **foundations, algorithms**, **components** and **principles** (e.g., cryptocurrencies, wallets, smart contracts, separate platforms) of blockchain systems | o | o | o | | o |
| Demonstrate blockchain technology capabilities and apply them to **business-related** challenges | o | o | o | | o |
| **Compare blockchain platforms** to enable understanding of different system design choices | o | o | o | | o |
| *Supply chain management, Finance, Business and Economics* | | | | | |
| Explain **general capabilities of** blockchains in supply chain management | o | o | o | | o |
| Discuss and compare different **blockchain models, scheme and solutions** with constructed/illustrated application (suggestions, proposals, methods for blockchain use in economics, business and finance) | o | o | o | | o |
| Explain the **interoperability** of blockchain technology and possible collaboration between unknown or untrusted parties | o | o | o | | o |
| Explain how **information asymmetry** can be addressed by the blockchain-based applications | o | o | o | | o |
| Demonstrate blockchain capabilities and apply them to **counterfeit and fraud prevention** problem statements | o | o | o | | o |
| Demonstrate blockchain capabilities and apply them to **provenance and track&trace** problem statements | o | o | o | | o |
| *Blockchain-based Application Development* | | | | | |
| Describe **development processes** for blockchain solutions | o | o | o | | o |
| Demonstrate knowledge of **system architectures, frameworks, different layers** | o | o | o | | o |
| Discuss software quality goals and their impact on blockchain system development | o | o | o | | o |
| Knowledge of **regulatory standards, rules, laws, regulations, management standards** | o | o | o | | o |

Disclaimer
The creation of these resources has been (partially) funded by the ERASMUS+ grant program of the European Union under grant no. 2018-1-LT01-KA203-047044.
Neither the European Commission nor the project's national funding agency DAAD are responsible for the content or liable for any losses or damage resulting of the use of these resources.

| | | | | |
|---|---|---|---|---|
| Describe the software **requirements** elicitation and engineering process of blockchain systems | o | o | o | o |
| Develop and manage **databases** using data management systems (also use of SQL, etc.) | o | o | o | o |
| Knowledge of **network protocols** | o | o | o | o |
| Apply (different) **programming languages** | o | o | o | o |
| Develop a **testing** plan for concrete blockchain solutions | o | o | o | o |
| *Privacy Management* | | | | |
| Describe **privacy management** principles using the blockchain solutions | o | o | o | o |
| Explain **identity management** principles using the blockchain solutions | o | o | o | o |
| *Security Engineering and Security Management* | | | | |
| Explain how data, information and processes can be **secured** by the use of the blockchain technology | o | o | o | o |
| Recognise **security countermeasure** implications | o | o | o | o |
| Explain **access control** (authentication, authorization and identity) models | o | o | o | o |
| Describe **transaction protection and validation** principles | o | o | o | o |
| Underline major **encryption and signature** schemes | o | o | o | o |
| Explain **trust management** principles | o | o | o | o |
| State major **fair mining** principles | o | o | o | o |
| Identify **security errors** in smart contracts | o | o | o | o |

| Please provide your additional comments (1-2 sentences are OK) how the current list of the **professional skills** should be further refined? |
|---|
| |

| **Self-competence skills**<br>the willingness and ability as an individual personality to clarify, think through and assess the development opportunities, demands and restrictions in family, career and public life, to develop one's own talents and to draw up and develop life plans. A sense of responsibility and duty, reliability, the ability to criticise and independence are characteristics of self-competence. It also includes the development of well-considered values and the self-determined attachment to values. | **Knowledge** | **Application** | **Evaluation** | | **Not important** |
|---|---|---|---|---|---|
| | 1 | 2 | 3 | | 0 |
| Demonstrate **work independence and self-organisation** | o | o | o | | o |
| Practice **new ideas** (**open-minded**) | o | o | o | | o |
| Sketch/imply **creative solutions** | o | o | o | | o |
| Apply **critical thinking** | o | o | o | | o |
| Be **proactive** and take **initiative** | o | o | o | | o |
| Be **responsible, trusted, and committed** | o | o | o | | o |
| Construct **high quality results** | o | o | o | | o |

| Please provide your additional comments (1-2 sentences are OK) how the current list of the **self-competence skills** should be further refined? |
|---|
| |

| **Social skills**<br>the willingness and ability to live and shape social relationships, to grasp and understand gifts and tensions and to deal with and communicate with others in a rational and responsible manner. Characteristics also include solidarity and the development of social responsibility. | **Knowledge** | **Application** | **Evaluation** | | **Not important** |
|---|---|---|---|---|---|
| | 1 | 2 | 3 | | 0 |
| Demonstrate strong (inter-) **organisational** skills | o | o | o | | o |
| Demonstrate ability to work in **international teams** | o | o | o | | o |
| Practice to **support colleagues** with expert knowledge | o | o | o | | o |
| Use **social media** means | o | o | o | | o |
| Establish good social **relationships with the customers** | o | o | o | | o |
| **Manage team and organise work** | o | o | o | | o |

Disclaimer
The creation of these resources has been (partially) funded by the ERASMUS+ grant program of the European Union under grant no. 2018-1-LT01-KA203-047044.
Neither the European Commission nor the project's national funding agency DAAD are responsible for the content or liable for any losses or damage resulting of the use of these resources.

| Please provide your additional comments (1-2 sentences are OK) how the current list of the **social skills** should be further refined? |
|---|
| |

| **Communication skills**<br>the willingness and ability to understand and shape communicative situations. This includes perceiving, understanding and presenting one's own intentions and needs as well as those of the partners. | **Knowledge** | **Application** | **Evaluation** | | **Not important** |
|---|---|---|---|---|---|
| | 1 | 2 | 3 | | 0 |
| Demonstrate **communication skills to external stakeholders** (external users, customers, advisors, and etc.) | o | o | o | | o |
| Demonstrate **communication skills to the internal stakeholders** (internal users, customers, advisors, and etc.) | o | o | o | | o |
| Practice good **written communication** (documentation) skills | o | o | o | | o |
| Demonstrate good **verbal communication** skills | o | o | o | | o |
| Practice good **skills to educate** (e.g., internal and external) stakeholders | o | o | o | | o |
| Operate good ability to **communicate complex problems** | o | o | o | | o |
| Demonstrate **good presentation** skills | o | o | o | | o |

| Please provide your additional comments (1-2 sentences are OK) how the current list of the **communication skills** should be further refined? |
|---|
| |

Disclaimer
The creation of these resources has been (partially) funded by the ERASMUS+ grant program of the European Union under grant no. 2018-1-LT01-KA203-047044.
Neither the European Commission nor the project's national funding agency DAAD are responsible for the content or liable for any losses or damage resulting of the use of these resources.

| **Methodical skills** the willingness and ability to take a goal-oriented, planned approach to the processing of tasks and problems (for example, when planning work steps) | **Knowledge** | **Application** | **Evaluation** | | **Not important** |
|---|---|---|---|---|---|
| | 1 | 2 | 3 | | 0 |
| Apply **innovative development methods** | o | o | o | | o |
| **Research and design new methods** of working | o | o | o | | o |

| Please provide your additional comments (1-2 sentences are OK) how the current list of the **methodical skills** should be further refined? |
|---|
| |

| **Learning skills** the willingness and ability to understand and evaluate information about facts and contexts independently and together with others and to classify it into mental structures. In addition, learning competence includes the ability and willingness to develop learning techniques and learning strategies at work and beyond the workplace and to use them for lifelong learning. | **Knowledge** | **Application** | **Evaluation** | | **Not important** |
|---|---|---|---|---|---|
| | 1 | 2 | 3 | | 0 |
| Ability to **learn quickly** | o | o | o | | o |
| Interest in **new technology** | o | o | o | | o |
| Interest in **continuing learning** | o | o | o | | o |
| Be **open-minded** (take into account feedback) | o | o | o | | o |

Disclaimer
The creation of these resources has been (partially) funded by the ERASMUS+ grant program of the European Union under grant no. 2018-1-LT01-KA203-047044.
Neither the European Commission nor the project's national funding agency DAAD are responsible for the content or liable for any losses or damage resulting of the use of these resources.

| Please provide your additional comments (1-2 sentences are OK) how the current list of the **learning skills** should be further refined? |
|---|
| |

| **Professional Decision Making** is the process of choosing an option out of a variety of possibilities and is a competence that is composed of sub-areas of the six prior described competence. | **Knowledge** | **Application** | **Evaluation** | | **Not important** |
|---|---|---|---|---|---|
| | 1 | 2 | 3 | | 0 |
| Apply **analytical methods** to solve problem | o | o | o | | o |
| Apply **evidence-based approaches** for problem solving | o | o | o | | o |
| Demonstrate ability to **prioritise** | o | o | o | | o |
| Practice **creative solution** | o | o | o | | o |
| **Maintain and support solution** | o | o | o | | o |

| Please provide your additional comments (1-2 sentences are OK) how the current list of the **professional decision making issues** should be further refined? |
|---|
| |

Disclaimer
The creation of these resources has been (partially) funded by the ERASMUS+ grant program of the European Union under grant no. 2018-1-LT01-KA203-047044.
Neither the European Commission nor the project's national funding agency DAAD are responsible for the content or liable for any losses or damage resulting of the use of these resources.

# PART 2: Blockchain Best Practice Use Cases in SCM and Logistics, Production and Smart Maintenance, Finance

Disclaimer
The creation of these resources has been (partially) funded by the ERASMUS+ grant program of the European Union under grant no. 2018-1-LT01-KA203-047044.
Neither the European Commission nor the project's national funding agency DAAD are responsible for the content or liable for any losses or damage resulting of the use of these resources.

## 2.1. Research Design

In this section, we will present the research design for the second part of the working package – "Blockchain Best Practice Use Cases in SCM and Logistics, Production and Smart Maintenance, Finance". First, we will explain our research goal and the used methodology. Then we will present our research questions and show how we ensure a practical-oriented extension of the first step literature outcomes. Finally, we will present the aggregated results.

### 2.1.1 Research Goal

In this part of the working package, we conduct a case study to analyze lockchain use cases in different fields. The study uses expert interviews as a form of data collection and aims at a validation and extension of the literature analysis results (see Part I). Especially in terms of Blockchain related competences, the case study should be used to prioritize existent competences and to identify new ones. As the list of theoretical competences was evaluated by universities and advisory board members with academical background, the aim of this part is to give a practice-oriented view on it. Furthermore, the previous findings mostly covered professional compentences lacking soft skills. Hence, identifiying additional social, personal and methodical compentences forms another research goal. Last but not least, we have another goal in adding competence descriptions to the already-derived items.

Another point that requires further investigation through our case study, is the examination of interdisciplinarity and interdisciplinary cooperation in Blockchain projects. In this area we seek answers to two problem statements defined at the beginning of the project. Firstly, which team members are involved in Blockchain Projects and which methods are used for the interdisciplinary work. Secondly, we consider the didactical methods and models, which exist to support interdisciplinary work. (Gürpinar et al. 2019)

### 2.1.2 Definition of Research Questions

Based on the results of the preceding literature review, we present research questions relevant for the further analysis. All research questions are based on the following overaching question:

"**Which blockchain related competences are seen relevant by practitioners for coping with future blockchain projects?"**

Further research questions:

- Which competences are necessary for employees involved in blockchain operations?
- Which ones are deemed the most important and why?
- How may an interdisciplinary team composition influence blockchain operations?

In order to work with the developed research questions, we identified challenges we would have to take into account in order to achieve valid and usable research outcomes.

Disclaimer
The creation of these resources has been (partially) funded by the ERASMUS+ grant program of the European Union under grant no. 2018-1-LT01-KA203-047044.
Neither the European Commission nor the project's national funding agency DAAD are responsible for the content or liable for any losses or damage resulting of the use of these resources.

First, we need to consider the interviewees' and projects' level of blockchain maturity. This ensures that received responses can be taken into account with a critical appraisal and emphasis. It is expected that advanced projects, will be able to show a variety of necessary competences. We can pay an attention to the answers of those projects, especially for the second and third research question. Thereby, the main focus is on identifying not only professional skills, but also soft skills which are of decisive importance for the success of blockchain project.

In the field of interdisciplinarity it is a challenge to combine the term with the rest of the interview, in a way that interviewees understand its meaning and give adequate answers. Finally, both by creating the questionnaire and by conducting the interview, it must be ensured that we identify and cover also negative and critical aspects. In order to ensure that we generate the targeted results in a scientifically valid manner, we follow a methodological approach, which is described in the next section.

### 2.1.3 Research Methodology

To describe the research methodology, first, the term case study must be defined. Following Yin, a case study is "an empirical inquiry that investigates a contemporary phenomenon within its real-life context, especially when the boundaries between phenomenon and context are not clearly evident" (2003, p. 13). Case studies are a central methodology for transferring practical experience into a generally valid model (Gemmel 2014).

The procedure of a case study can be realized with the help of a single case ("single case study") or multiple cases ("multiple case study"). If more than one case is to be considered in an investigation, a multiple case study is mandatory. The investigation of several cases is used to identify differences. But it also helps discover similarities between various cases. (Stake 1995) In addition, a broad coverage of the theoretical foundation as well as a wide-ranging investigation of the research questions can also take place (Gustafsson 2017). However, since the execution of a multiple case study is time consuming and the effort should not be underestimated. It is advisable to carry out such a study in a collaborative multidisciplinary manner with divergent academic departments rather than by a single researcher (Yin 2014). Due to the fact that multiple case studies ensure that several different perspectives are taken into account, the results of multiple case studies are considered more robust than single-case studies (Herriott & Firestone 1983).

When working on the research question regarding necessary skills for future Blockchain specialists, it is important that we produce generally valid findings that are beneficial for students of different disciplines. The goal is a broad analysis in order to discover complete sets of competences. In the case of the BlockNet project, we therefore ensure that we, on the one hand, not only analyse cases from different industries, but also interview people from diverse backgrounds, for example IT specialists as well as managers. In this way, we ensure that an explorative investigation of the so far little investigated research object of Blockchain education is carried out.

The implementation of an "exploratory" study is useful if the knowledge base and the information from the literature are not developed into a theoretical statement of adequate quality (Yin 2014 p.39). A researcher begins with a general question, which leads to further questions that are investigated in subsequent studies. The difference between exploratory studies and alternatively conclusive studies lies in the fact that exploratory results show a

Disclaimer
The creation of these resources has been (partially) funded by the ERASMUS+ grant program of the European Union under grant no. 2018-1-LT01-KA203-047044.
Neither the European Commission nor the project's national funding agency DAAD are responsible for the content or liable for any losses or damage resulting of the use of these resources.

number of possible causes and alternative solutions, whereas conclusive studies show the final or the only solution for a certain research problem.

A multiple case study research can be divided into three main parts, as seen in Fig. 24. The first part “Define & Design” focuses on the development of a theoretical premise. This premise can be confirmed by selected cases and a systematic approach, which is here the data collection protocol. The most important aspect of this part is the execution of each case. In the second part “Prepare, Collect & Analyse” the analysis of each case is presented in the form of an individual case study report. After the collection of the individual cases, it is important to further analyse the collected data basis in the part “Analyse & Conclude”. It is possible to draw a cross-case conclusion here, which results in a modification of the previously established theory. This then forms the basis for developing possible policy implications. The result of the methodology is a generalized case study report that is valid across all case studies.

The procedure presented is representative for a multiple case study as described by Yin (2014).

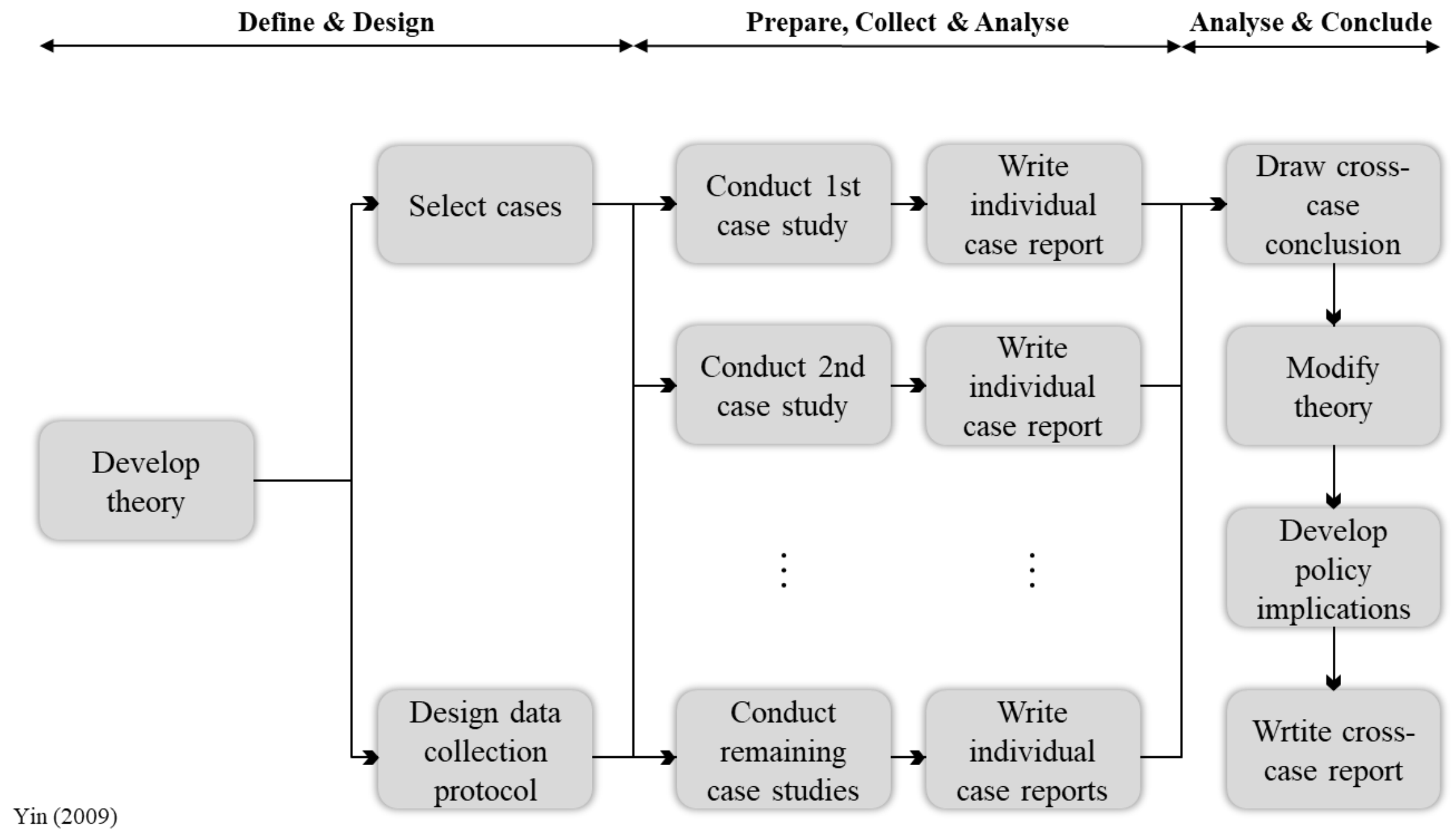

**Figure 24**: Case Study Approach according to **Yin** (2009).

An important instrument for data collection within a case study is the expert interview. Qualitative, guideline-supported interviews are thereby a differentiated and methodologically well elaborated method of generating qualitative data (Helfferich 2014). Expert interviews are carried out when information is to be collected that can be provided by a selected group of people. In many cases much of the knowledge of high interest, is only available in the minds of particular people – the so called experts (Runeson et al. 2012). Experts are persons who have exclusive knowledge and are usually members of a particular organisation or institution (Pickel & Pickel 2009). It is important to stress that the experts give personal insights and their view on the analyzed problems. They can be asked about facts but also about their opinion (Ridder 2016).

Disclaimer
The creation of these resources has been (partially) funded by the ERASMUS+ grant program of the European Union under grant no. 2018-1-LT01-KA203-047044.
Neither the European Commission nor the project’s national funding agency DAAD are responsible for the content or liable for any losses or damage resulting of the use of these resources.

The Framework we use for the interviews in the case study is a semi-structured interview. These are characterized by the fact that an interview guide is prepared and elaborated, but the questions are not necessarily asked in the same order as they are listed. Ideally, the interviewer should lead the conversation in a way that creates a natural conversation flow during which all relevant questions are asked without compulsively remaining in a predetermined order. In addition, improvised questions and questions adapted to the situation can also be used. The aim is to find a balance between asking questions, listening to the interviewee and monitoring the questions already asked. On the one hand, it is important to ask a certain number of questions and thus cover all relevant blocks. On the other hand, the interviewee should be given enough time to think about his answer rather than being pushed in a certain direction by a subsequent question. The flexible approach of semi-structured interviews ensures that relevant context information is not lost, especially in such an innovative field as Blockchain education. These are especially helpful when the interviewees speak from their experience and may therefore not be clear which information is of decisive interest or not. (Runeson et al. 2012)

In order to ensure a scientifically well-founded analysis, we evaluate the interviews using the qualitative content analysis. The qualitative content analysis follows a predefined systematic and is guided by rules and theory by creating categories. According to Mayring (2008) there are three basic methods that can be used for the qualitative analysis of interviews: The summary, the explication and the structuring. In the presented case, the form of summary content analysis was chosen, in which the material is reduced to a manageable short text by preserving only essential content (Mayring 2015). At the beginning of the summary, a uniform procedure is defined for the subsequent steps so that the comparability of the results can be guaranteed. According to Mayring (2003), there are four steps for the abstraction of a text, e.g. a transcribed interview: Paraphrase, Generalization, First Reduction, and Second Reduction.

The first step (paraphrase) is used mainly to summarise individual coding units by using shortened formulations and a grammatically consistent short form to adjust the language level. In this process, all text components with little or no content are excluded. Subsequently, an abstraction level is defined for the generalization, which specifies how general or how specific a term or statement must be. In this inductive thinking process, details are excluded so that the content can be generalized in a meaningful way. During the first reduction, the paraphrases that are identical are deleted so that only paraphrases that are of central importance are used. In the last step, the second reduction, the remaining paraphrases with similar statements are summarized and, if necessary, reformulated. A decisive advantage of the qualitative content analysis according to Mayring is that the method is transparent, as one can proceed step by step according to rules. (Mayring 2003)

In the following chapter we will discuss in more detail how the presented scientific principles are used during the BlockNet project. Thereby a multiple case study will be presented, which we cyrry out to extend the knowledge regarding necessary skills for future Blockchain specialists.

## 2.2 Prepare, Collect & Analyze

The following chapter describes how we apply the scientific procedures presented in Section 2.1. First, we discuss the interview approach. Building up on this, we describe how the

Disclaimer
The creation of these resources has been (partially) funded by the ERASMUS+ grant program of the European Union under grant no. 2018-1-LT01-KA203-047044.
Neither the European Commission nor the project's national funding agency DAAD are responsible for the content or liable for any losses or damage resulting of the use of these resources.

questionnaire was designed. Then we introduce the procedure for identifiying and reaching out to interview partners as well as the approach for conducting the interviews.

### 2.2.1 Interview Approach

We use the methodology of semi-structured interviews. In this context, we analyze the data using state of the art qualitative data analysis (see Part 2.3). The result is a skills overview and a landscape of current education and training in the field of Blockchain. We funnel findings into the design of the course curricula and disseminate them to the advisory board. Regarding the task division the Technical University Dortmund developes the case study design and the interview structure. This is shared with the project partners within the consortium who conduct interviews independently. Regional and interregional specifics are taken into account. Once the interviewees agree, a predefined preparation plan is followed to ensure that the interviews can be conducted in the best possible manner. The individual steps will be presented later in chapter 2.2.3.

The aim of the interviews is to further explore the question which Blockchain related skills are seen relevant by the practice to cope with future requirements. For this purpose, we structure the questionnaire in such a way that two research areas can be covered. The first one deals with necessary skills of future Blockchain experts. The aims are to expand the existing skill list, especially with regard to "other skills", and to give practical priorities. Possible categories for coding the answers are provided by the competence clusters of the KMK model that we already used during the first part of the work package. In the research area interdisciplinarity, the second research area, our aim is to build an understanding of how teams are structured in Blockchain projects and which methods are used to ensure interdisciplinary collaboration. During the aggregation of the answers we use an inductive procedure for the analysis of the team composition, so that the 4 disciplines of the consortium members already defined can be supplemented by new ones.

We followe a previously specified plan to ensure the best possible results. For this purpose, an interview guide for case studies is prepared by the Technical University of Dortmund. It is divided into three parts. Part A describes Instructions for interviewers. On the one hand, this part contains detailed information on how to best get in touch with possible interview partners. It also includes information on what materials should be sent out before the interviews, and what information should be obtained to ensure the best possible preparation. It is followed by section B, in which the five sets of questions are presented and the individual questions can be found. At the beginning this section contains a series of pre-formulated sentences which can be used by the interviewer to initiate the interview. For example the interviewer introduces himself, he presents general information about the project BlockNet and shows the benefits of participation for the company. In addition, the interview approach is presented and important terms are defined at the beginning. The last part contains a spider diagram to visualize the level of different skills of the interview partners. We use this as a method of self-evaluation of the interviewee’s relevant skills.

In order to be legally secure, we formulate a confidentiality agreement and use it for each interview. We sent this agreement to the interview partner in time so that it is available in signed form at the latest on the day of the interview. This point is particularly important, since the interviews had to be captured as a sound recordings in order to enable later transcription.

Disclaimer
The creation of these resources has been (partially) funded by the ERASMUS+ grant program of the European Union under grant no. 2018-1-LT01-KA203-047044.
Neither the European Commission nor the project’s national funding agency DAAD are responsible for the content or liable for any losses or damage resulting of the use of these resources.

We divide the second part of the overall work package into three temporally separated phases in which the individual consortium partners has different tasks. We based the three phases on the steps involved in the execution of a multiple case study defined by Yin (2009). In the first step, ‘Design & Define’, all consortium partners select suitable companies and cases. In addition the Technical University Dortmund identifies starting points from the given skill concept and develops the research methodology and the interview guidelines. This phase begins with the conduction of the pilot case by the Technical University of Dortmund. The aim of the pilot case is to refine the data collection plans with respect to the content of the data and the procedures to be followed (Yin 2018, p.106). Subsequently, all partners conduct the remaining interviews and prepare individual case reports. The last phase includes the items Analysis and Conclude. In this phase, the Technical University of Dortmund draws a cross-case conclusion to extend the existing skill list. Based on the cross-case report, we prepare the whitepaper. The final results are presented to a larger audience during the first multiplier event hosted by the Technical University of Dortmund. Several keynote speeches are given in this session, presenting the topic of Blockchain education in different domains. Furthermore, a workshop is offered in which both novices and Blockchain experts work together on a problem within the area of Distributed Ledger Technologies.

### 2.2.2 Interview Questions

As already described, we develop an interview guide which includes all relevant questions for the case study analysis. We group the questions into four main categories so that the previously defined research objectives can be achieved. The respective blocks of questions are divided into questions which must be asked in any case and back-up questions which can be asked if the necessary information is not obtained in the context of the other questions. Figure 25 presents the four sections of the interview guide and shows why they were integrated.

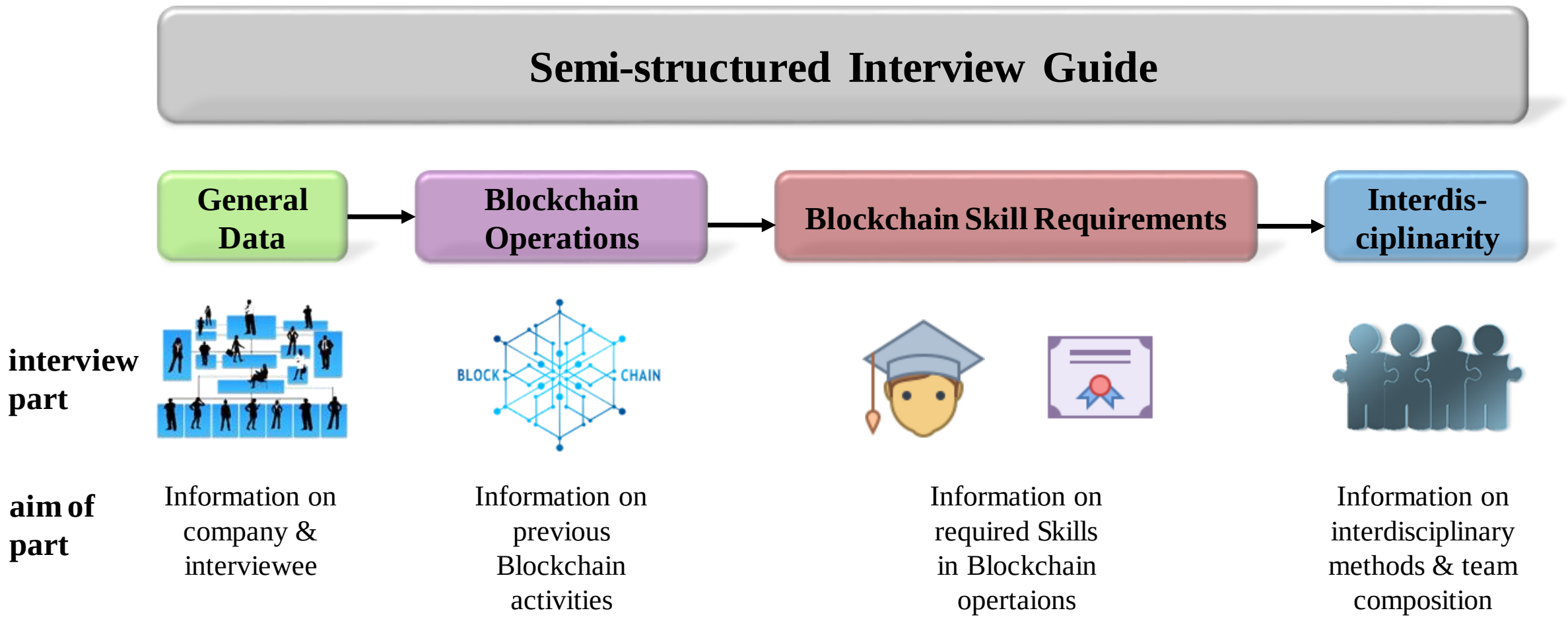

**Figure 25**: Parts of the interview guide and the reasons why they were seleted.

The first category is named *General Data*. The interviewer is free to fill in the questions of the first block by himself/herself with the help of a prior research. Nevertheless, the data should be verified by the interview partner. This questions are being asked because we need some general data about the company and the participants of the interview. We need this information

Disclaimer
The creation of these resources has been (partially) funded by the ERASMUS+ grant program of the European Union under grant no. 2018-1-LT01-KA203-047044.
Neither the European Commission nor the project’s national funding agency DAAD are responsible for the content or liable for any losses or damage resulting of the use of these resources.

to assess whether certain skills are related to specific industries. The questions of this questionnaire block are as follows:

1.1 *Institution name, sector/industry (only if not already clear)*
1.2 *Interviewee role, department, years of tenure in position / institution*

BACK-UP Questions

1.3 *Company facts & figures (no. of employees, turnover, …) (only if not already clear)*
1.4 *Business model of the company, their geographical scope of operating*

The second block of questionnaires specifies the previous Blockchain activities of the enterprise in which the interview partner is employed. General questions on Blockchain operations need to be verified only once per case and data partially might be collected before or after the interview. During the interview it is important to gain insights from the perspective of companies that already have experience with Blockchain applications. The questions of this questionnaire block are as follows:

2.1 *First, we would like to ask: Do you have BCT implementations, implementation preparations, or plans/discussions to implement BCT in the future? (decision for internal / external)*
2.2 *Why are you interested in BCT? What is your intention and aim in utilizing BCT?*
2.3 *Can you give us some general information about the Use Case (you have in mind)? You can focus on the use case you feel most comfortable with. Is it a PoC or a Pilot?*
2.4 *Which business processes are (would be) affected by the implementation?*
2.5 *What kind of Blockchain solution / framework is used (or would be appropriate)?*

BACK-UP Questions

2.6 *Do you have plans for the future? For the utilization of the pilot? For the use of the technology?*

In the third section, we ask focused questions about the skills required for employees in the area of Blockchain. This block of questions is structured in such a way that we ask for both professional skills and soft skills. In addition, there are questions that aim to investigate specific skills of individual specialists and to survey missing skills of university graduates. To ensure that these essential information are collected, we also ask playful questions. The questions of this questionnaire block are as follows:

3.1 *Which roles/responsibilities exist among the team members of your Blockchain project?*
3.2 *Which professional skills are necessary for (team member XY)? Why are they important?*
3.3 *Which skills other than professional skills are necessary for (team member XY)? Why are they important?*
3.4 *Let us pretend: You instruct the HR department to recruit a Blockchain expert: What specific points included in the requirements would you prioritize?*
3.5 *Which skills do you see particularly important for the fields (Business / SCM / Computer Science / IT Security)?*

Disclaimer
The creation of these resources has been (partially) funded by the ERASMUS+ grant program of the European Union under grant no. 2018-1-LT01-KA203-047044.
Neither the European Commission nor the project's national funding agency DAAD are responsible for the content or liable for any losses or damage resulting of the use of these resources.

3.6 *Which needed skills do you feel university graduates lack when they start work in a BCT role?*

BACK-UP Questions

3.7 *We have found out in previous analyses that (XY) is the most important professional skill for (team member XY). Would you agree? What other skills do you think should be added?*
3.8 *Are there skills that need to be newly acquired, or can existing employees be trained?*
3.9 *Which skills were important when putting together the Blockchain team? Why?*
3.10 *Do training programs (for BCT) already exist? Does an overall knowledge repository for BCT processes/ roles exist? (Can you provide us with job descriptions/ profiles?)*
3.11 *How do the skill profiles differ between different countries/ cultures your company is located in? How is BCT training embedded in the company? (in-house or external providers, training academy,…)*
3.12 *Can you provide us with your training plan/ program? What are your KPIs on employee training?*

The last section deals with interdisciplinary work in Blockchain environment that might also become more evident in the future. We try to evaluate the skills that are needed to cope with these challenges in the future. The questions of this questionnaire block are as follows:

4.1 *Which (professional / academic) fields do team members of the Blockchain project come from?*
4.2 *How is the cooperation between the team members organized?*
4.3 *What can/should be improved in team collaboration to increase project success?*
4.4 *Which methods are already used to enable good cooperation between team members?*

BACK-UP Questions

4.5 *Do team members need to acquire cross-disciplinary knowledge and skills?*
4.6 *What helps the specific person to address interdisciplinary problems?*

The interview ends with the question whether there is anything the interviewee would like to add or emphasize and the request that we may contact him again if clarification needs should arise.

### 2.2.3 Acquisition of the Interview Partners

We follow a predetermined plan in the search for suitable interview partners. At the beginning, all consortium members use a standardized email template including the BlockNet Project Sheet. This ensures that initial information on the research project are available, but also that the benefits of participating are clear to the interviewees. After a confirmation, an appointment is made for a phone call, during which the participation is reconfirmed and open questions can be clarified. These included, for example, agreeing on a suitable date for the interview, a

Disclaimer
The creation of these resources has been (partially) funded by the ERASMUS+ grant program of the European Union under grant no. 2018-1-LT01-KA203-047044.
Neither the European Commission nor the project's national funding agency DAAD are responsible for the content or liable for any losses or damage resulting of the use of these resources.

suitable location and, in some cases, the preferred way of conducting the interview, either a personal interview or an interview via the Internet (e.g. via skype).

Each individual consortium member looks for and contacts suitable interview partners from their field of expertise. This ensures that we generate the broadest possible and most meaningful picture of necessary competences in the field of Blockchain. In total, the four consortium members conduct fourteen interviews. How these are conducted will be discussed in the next section.

### 2.2.4 Conducting the Interviews

Of the fourteen interviews carried out, the Technical University of Dortmund, which is also the leader in this work package, conducts seven. The University of Tartu is conducting three more interviews. With the two interviews each from Vilnius University and the University of Copenhagen, we reach the fourteen.

We set a time of 60 to 90 minutes for the interviews. One person conducts the interview at a time, although another person can also be called in, for example, if additional help is needed with the transcription. The interview partners can be project leaders and managers but also project representatives from involved fields or other exeperts. In addition, employees from Human Resources can also be consulted for questions relating to skills or education. Nevertheless, we make sure that all interviewees have the necessary experience and can talk well-founded about their work with the Blockchain technology.

To ensure the best possible results, we choose the previously created guide for semi-structured interviews as our foundation. The interviews were conducted according to Yin’s (2008) guiding principle, which defines five characteristics as crucial. First we should ensure that we should only ask clear and relevant questions whose answers are interpreted fairly. This describes someone who can be considered a "good listener". However, it is equally important to be a “good” listener, who is free of existing ideologies or preconceptions. The third point stresses that it is important to stay adaptive so that new situations can be seen as opportunities, not as threats. You should have a fixed overview of the topics to be examined, even if you are in explorative mode. Last, we ensured to conduct research ethically, from a professional standpoint but also by being sensitive to contrary evidence. We conduct the interviews on the basis of the guidelines described above. The results of this are discussed in the following chapter.

## 2.3 Analyze and Conclude

The following section presents the findings from the second part of the work package. We introduce the results of the analysis of the individual interviews, before a cross-case analysis is carried out.

### 2.3.1 Interview Analysis following Mayring

The first step in the analysis of the generated data is to transcribe the interviews. As emphasised earlier, with the consent of the interviewees, the interviews are recorded using a sound recording device. The interviews are therefore available to us as audio files. The aim of the

Disclaimer
The creation of these resources has been (partially) funded by the ERASMUS+ grant program of the European Union under grant no. 2018-1-LT01-KA203-047044.
Neither the European Commission nor the project’s national funding agency DAAD are responsible for the content or liable for any losses or damage resulting of the use of these resources.

transcription is to transform the audio data into a form that allows a temporally relieved and methodically systematic and comprehensive evaluation (Kruse et al. 2015). We transcripted the interviews according to the clean verbatim method. Given this method, speech errors, filler words, conversational affirmations are not taken into account. False starts are only considered if they add information. Words like "Yeah", "Yep", "Yap" were transcribed as "Yes".

The transcribed interviews form the basis for the data analysis. We based the evaluation of the interviews on the previously described qualitative content analysis following Mayring. By choosing this framework be ensure that, in contrast to free interpretation, each step of the analysis can be traced back to a well-founded rule. The results of the Mayring analysis are also replicable. In order to meet these requirements, we use the general content analytical process model proposed by Mayring (2015). The qualitative content analysis follwing Mayring can be divided into the steps preparation, content analysis and interpretation as can seen in Figure 26. These can later be divided into 9 further steps.

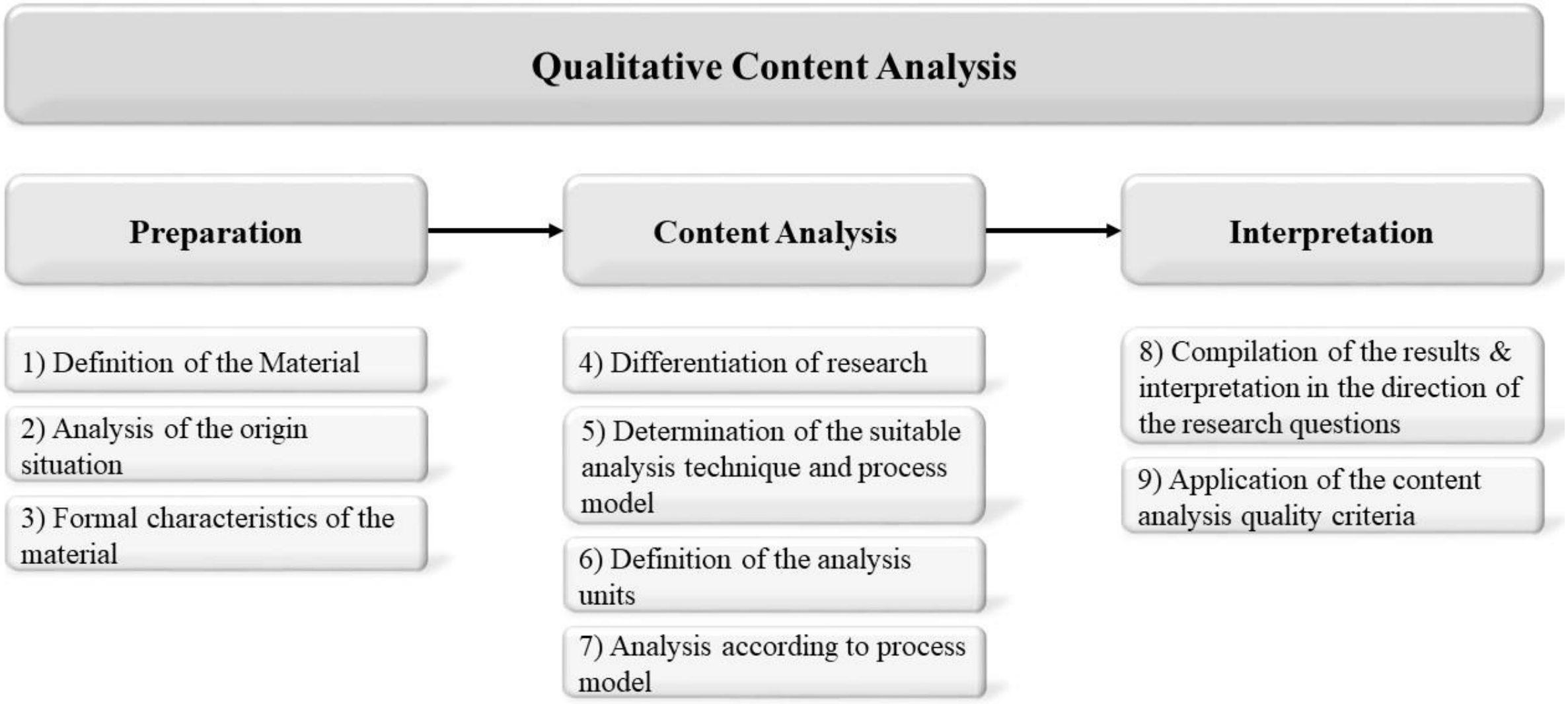

**Figure 26**: Three steps and nine sub-items of the qualitative content analysis. (Mayring 2015)

We apply the methodology as follows:

**Step 1 Definition of the Material:** The material we use for the qualitative content analysis are the 14 transcribed interviews. This are renamed into cases A to N. This is necessary to ensure anonymity.

**Step 2 Analysis of the origin situation:** We always carry out the interviews under the same conditions. The given time frame is set the same for each interview. Due to the use of the interview guide, the interview results are comparable with each other. In addition, the interviewers, who come from the participating consortium partners, have been trained equally in order to enable them to conduct interviews despite their different backgrounds. However, it has to be taken into account that by using the semi-structured guideline it is possible to have slightly different approaches to the interview and due to the possibility of using back-up questions, a different number of questions can be asked.

Disclaimer
The creation of these resources has been (partially) funded by the ERASMUS+ grant program of the European Union under grant no. 2018-1-LT01-KA203-047044.
Neither the European Commission nor the project's national funding agency DAAD are responsible for the content or liable for any losses or damage resulting of the use of these resources.

**Step 3 Formal characteristics of the material:** With regard to the formal characteristics, it can be said that we transcribe the interviews using the clean verbatim Method.

**Step 4 Differentiation of research:** The research questions of the qualitative content analysis are the same as of the whole part of this project. As already presented in chapter 2.1.2 the three main research questions are:

- What skills are necessary for employees involved in Blockchain operations?
- Which ones are deemed the most important? Why are they important?
- How may an interdisciplinary team composition influence Blockchain operations?

**Step 5 Determination of the suitable analysis technique and process model:** According to Mayring, the analytical technique is selected in the 5th step depending on the research question and the material.

For research question 1 we select a deductive approach. A deductive approach determines the evaluation instrument through theoretical considerations, from preliminary studies, from the current state of research, from newly developed theories or theoretical concepts. The categories are also developed towards the material in an operationalization process. For the subcategories of question 1 we choose an inductive approach. The inductive approach has the advantage that the object is recorded in the text without the influence of assumptions or distortions of the researcher (Mayring 2015). For research question 2 and research question 3 we choose a completely inductive approach.

**Step 6 Definition of the analysis units:** Before we form the categories inductively or deductively, it is important that we determine analysis units. The aim of these analysis units is to guarantee the precision and repeatability of an analysis. The units used for coding determine the smallest possible text component. Since a statement of the interviewees partly extends over several sentences, we define exactly one sentence as the minimum unit. In the other direction, the context unit describes the largest text component that can fall into a category. In this case, this is the entire interview of a person. The evaluation unit includes which material we evaluate. In the case of this analysis the evaluation is the entire material, that is, all fourteen interviews.

**Step 7 Analysis according to process model:** In this step the qualitative content analysis takes place. The results of the analysis are presented in the chapter 2.3.2.

**Step 8 Compilation of the results and interpretation in the direction of the research questions:** For the compilation of the results and the interpretation in the direction of the Research Questions we use Excel tables. We use the model provided by Mayring and explained in the chapter 2.1.3 which consists the steps transcription, paraphrase, generalisation, and category. The step Transcription contains the transcription of the interview. For the Paraphrase we translate the words of the column "Transcription" into our own understanding. It must be ensured that the own words do not stray far from those originally stated. The Generalisation takes the paraphrasing one step further and lists the general ideas and concepts stated in the case study. It must be ensured that the generalisation falls in line with the initially stated ideas.

Disclaimer
The creation of these resources has been (partially) funded by the ERASMUS+ grant program of the European Union under grant no. 2018-1-LT01-KA203-047044.
Neither the European Commission nor the project's national funding agency DAAD are responsible for the content or liable for any losses or damage resulting of the use of these resources.

We keep this section at one or two sentences. In the part Category we write down the categorised interview results.

**Step 9 Application of the content analysis quality criteria:** In order to meet requirementents of a scientific research method we applied the quality criteria provided by Yin (2011) and Creswell (2009). We included the criteria Construct validity, internal validity, external validity and reliability.

The results of the qualitative content analysis of all fourteen interviews are presented in the following. In doing so, we show above all which insights we were able to gain and to what extent we are able to find answers to the research questions posed.

### 2.3.2 Overview of the Interview Results in Numbers

By conducting a total of fourteen interviews, we have an enormous amount of data available. Due to the fact that the interviews each have a length of at least 60 minutes to sometimes more than 90 minutes, relevant points for each object of investigation are identified by each interview.

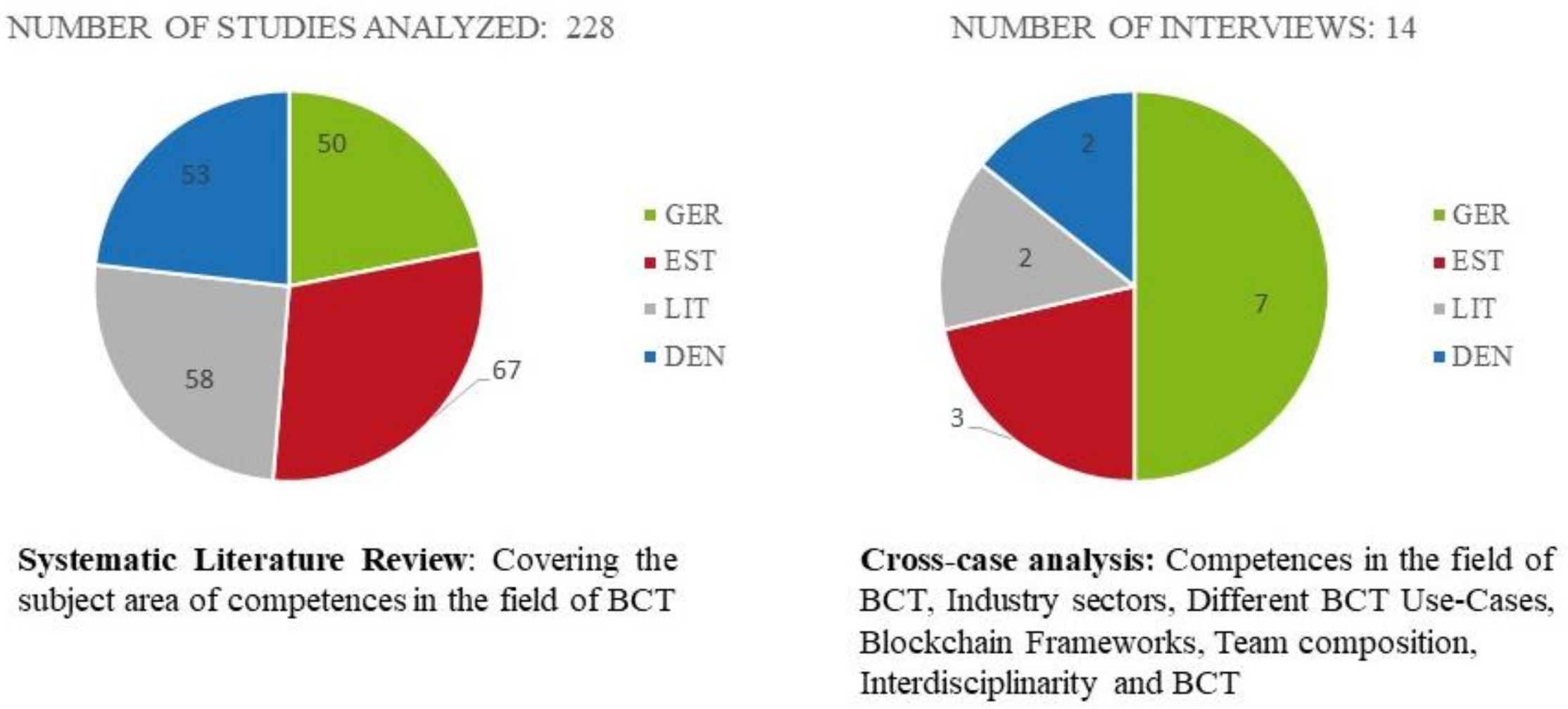

**Figure 27**: Number of papers analysed and interviews conducted by the consortium partners

Figure 27 compares the outcome of the first step, the systematic literature review, with that of the second step. This makes it clear that we were able to analyse a considerable number of sources in the first step. In contrast, the number of sources in the second step, which includes the interviews, is significantly lower. However, we use the results of the interviews not only to answer the basic question of specific competences in the Blockchain environment, but also to gain information about the industry sectors, interdisciplinary team compositions or the Blockchain frameworks used. This information is particularly important in the following section where the number of items analysed is compared.

Disclaimer
The creation of these resources has been (partially) funded by the ERASMUS+ grant program of the European Union under grant no. 2018-1-LT01-KA203-047044.
Neither the European Commission nor the project's national funding agency DAAD are responsible for the content or liable for any losses or damage resulting of the use of these resources.

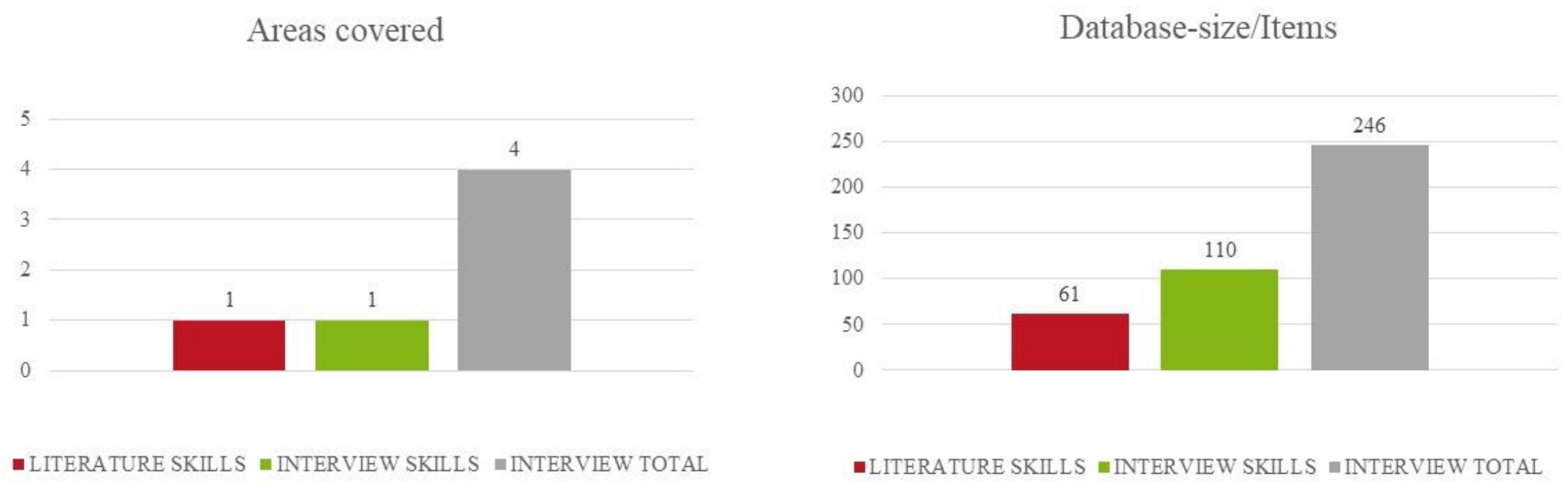

**Figure 28**: Display of the areas Coverd and the number of items surveyed.

Figure 28 also compares the first results of the systematic literature review with the results of the interviews. We compare three key figures with each other, so that a logical comparison between the results can be made. On the left side we present for instance the areas covered. In the first literature-based analysis step, we examine, as already emphasized, only the area of Blockchain-specific competences. This is also the content of the second interview-based analysis step, which is represented by the middle bar. The right bar represents the complete content, which we analyzed using the interview guide and covers the four previously presented objects of investigation. Due to the fact that several areas were investigated, the absolute number of items developed by conducting the interviews (246) is four times higher than the number developed by the literature search (61). Nevertheless, the number of items developed in the area of competencies, at 110, is also much higher than the originally defined items of the compiled skill list, which suggests that we were not only able to confirm the existing items of the list, but also to add items not previously considered.

This is also made clear by looking at Figure 29. For the presentation and classification of the interview results we use the same categorisation as we have chosen in the systematic literature review. As a result, the items generated by the qualitative content analysis, which represent the most granular source of information, were grouped into sub-categories, which in turn were finally aggregated into categories. In the area of Blockchain specific competences, the items of the Four Fields Competence model, which was explained in chapter 1.1.4, form the categories. Since we have deductively analyzed into these, the number of categories, as well as the number of sub-categories, remained the same.

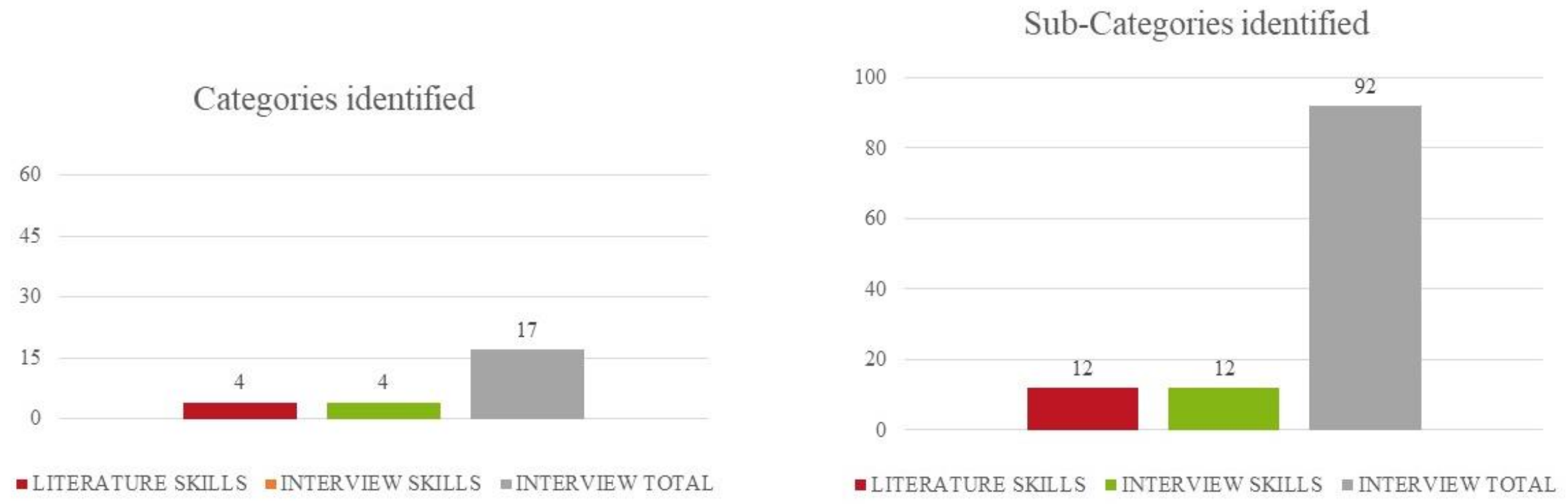

Disclaimer
The creation of these resources has been (partially) funded by the ERASMUS+ grant program of the European Union under grant no. 2018-1-LT01-KA203-047044.
Neither the European Commission nor the project's national funding agency DAAD are responsible for the content or liable for any losses or damage resulting of the use of these resources.

**Figure 29**: Display of the categories and sub-categories surveyed.

From the presentation of the raw data it is clear that the amount of data collected is too large to be integrated in future curricula. As shown in Figure 28, the number of items concerning Blockchain-specific competences alone has grown to 110. Therefore, we had to apply a methodology to clean up the raw data, e.g. by removing duplications or performing meaningful aggregation steps. This procedure is presented in the following after the. Before we do so, however, we will briefly discuss the results regarding the three blocks of questions, which deal with general information about the interview partner's company, their previous Blockchain activities and the topic of interdisciplinarity.

Disclaimer
The creation of these resources has been (partially) funded by the ERASMUS+ grant program of the European Union under grant no. 2018-1-LT01-KA203-047044.
Neither the European Commission nor the project's national funding agency DAAD are responsible for the content or liable for any losses or damage resulting of the use of these resources.

### 2.3.3 Results regarding Interview Information, Blockchain Operations and Interdisciplinarity

In the following we present the first results of the analysis. In doing so, we deal with each individual part separately. Special attention is paid to the third part, which deals with competences in the area of Blockchain. This chapter provides a brief overview of the data volumes collected.

**General Data**

By means of the first block of questions we collect some general information about the company and the participants of the interview. We need this information to assess whether certain skills are related to specific industries. We attach great importance to acquiring interview partners and thus also data from as many different areas as possible. As Figure 32 also shows, we have succeeded in this.

**Institution Industry**

| | |
|---|---|
| 1 | Technical Wholesale |
| 1 | Social Security Services |
| 1 | Timber |
| 1 | Agriculture |
| 1 | Fashion |
| 1 | Energy |
| 1 | Blockchain Consulting |
| 1 | IT Security |
| 3 | Crypto Currencies |
| 1 | Mobility |
| 1 | Digital Identities |
| 1 | Blockchain for Digital Twins |

**Interviewee Role**

| | |
|---|---|
| 7 | Manager |
| 4 | IT Specialist |
| 4 | External Consultant |

**Function / Department**

| | |
|---|---|
| 2 | Management |
| 1 | Procurement |
| 3 | IT Implementation |
| 1 | CSR |
| 1 | Research |
| 2 | Consulting |

**Figure 32**: Number of answers depending on the question of the part General Data.

The companies participating in the interviews come from twelve different industries. Only from the sector of crypto-currencies there are several interview partners. In addition, there are companies from IT-related industries such as IT security or digital identities as well as from non-adjacent sectors such as fashion or agriculture. In most cases, the information we collect comes from managers who bear the main responsibility for the Blockchain venture and can therefore report well on their experiences. In addition, we conducted interviews with IT specialists, as well as external consultants who are specialized in the field of Blockchain. Overall, as shown in Figure 32, the interview partners come from different departments or work in various functions.

Disclaimer
The creation of these resources has been (partially) funded by the ERASMUS+ grant program of the European Union under grant no. 2018-1-LT01-KA203-047044.
Neither the European Commission nor the project's national funding agency DAAD are responsible for the content or liable for any losses or damage resulting of the use of these resources.

## Blockchain Operations

The second block serves to determine information about the Blockchain projects in which the interview partners work or have worked. On the one hand, we investigate which use cases have been developed by the companies or in which context they want to use the Blockchain technology. On the other hand, we also investigated the reasons why the companies invest in the Blockchain. We also determine the project stage and the solution that was used. The results are shown in Figure 33, which shows how often a specific answer is mentioned in the respective areas.

**General Information on Use Case**

| | |
|---|---|
| 1 | Data Exchange |
| 1 | Social Security |
| 2 | Payments via BCT |
| 1 | Tracking & Tracing |
| 1 | Integrity of Health records |
| 1 | Mobility |
| 3 | Crypto Currencies |
| 1 | Smart Grids |
| 3 | Certification on BCT |

**Interest in BCT**

| | |
|---|---|
| 9 | Monetary Business Reasons |
| 4 | Innovation |
| 1 | Trust Issues |
| 2 | Social Impact |
| 1 | Automatization |
| 1 | Interoperability |
| 1 | Transparency |

**Project Stage**

| | |
|---|---|
| 8 | Implementation |
| 5 | Preparations |
| 1 | Plans |

**Blockchain solution used**

| | |
|---|---|
| 3 | Hyperledger |
| 2 | Corda |
| 2 | Multichain |
| 2 | Different Cryptocurrencies |
| 3 | Own Solutions |

**Figure 33**: Number of answers depending on the question of the part Blockchain Operations.

Comparable to the survey of the industries in which the interview partners are active, a multitude of different use cases can be found in the collected data. This also enables us to gain a comprehensive overview of the necessary competencies of future Blockchain experts. Interestingly, the interviewed companies see several reasons for investing in the Blockchain. Although Monetary Business Reaons clearly dominates the survey with nine mentions, we could find six more reasons including the hope to become a more innovative company or to use the Blockchain to address social problems. Most of the projects that are reported are also in the implementation stage. The most popular solution is Hyperledger, which suggests that it could be of great interest to include this in future university curricula.

## Blockchain and Interdisciplinarity

In the following we will discuss the correlation between the involvement with Blockchain technology and interdisciplinary team compositions and working methods. The analysis of the interviews clearly shows that teams working in the Blockchain environment are usually very heterogeneous, since they consist of members with different backgrounds. In addition, the BlockNet project will design a course that will be used by an interdisciplinary group of students. Thus, there is also a high interest in which methods are already used in practice to improve interdisciplinary cooperation

Disclaimer
The creation of these resources has been (partially) funded by the ERASMUS+ grant program of the European Union under grant no. 2018-1-LT01-KA203-047044.
Neither the European Commission nor the project's national funding agency DAAD are responsible for the content or liable for any losses or damage resulting of the use of these resources.

| Fields of Team Members | |
|---|---|
| 14 | Computer Science |
| 8 | Business Administration |
| 5 | Engineering & SCM |
| 3 | Natural Sciences |
| 3 | Marketing |
| 2 | Sales |
| 2 | Law |
| 2 | Finance |
| 1 | Insurance |

| How work is organized | |
|---|---|
| 5 | Constant communication |
| 5 | Meetings |
| 4 | Communication via Mail/Phone/Messenger |
| 3 | Face-to-face communication |
| 2 | Desktop sharing |
| 2 | Video/Phone conferences |
| 1 | Workshops |
| 1 | Seminars |
| 1 | Document Sharing |
| 1 | Shared code revision control |
| 1 | Concrete work plan and steps |

| What can be inproved | |
|---|---|
| 1 | Better Communication |
| 1 | Common Understanding |
| 1 | Problems between disciplines / different team members |

| Methods to enable good cooperation | |
|---|---|
| 4 | Regular Meetings |
| 3 | Constant Communication |
| 3 | Informal Communication and Events |
| 2 | Scrum |
| 2 | Clear Roles and Processes |
| 1 | Mind Mapping |
| 1 | Ad-hoc Meetings |
| 1 | Workshops |

**Figure 34**: Results regarding Blockchain and interdisciplinarity.

First, it becomes clear from Figure 34 that members in current Blockchain Teams are actually composed in an interdisciplinary manner. Although all fourteen interview partners confirm that they have people from the field of computer science in their teams, people from eight other fields were also mentioned. Interestingly, persons from the fields of law and natural sciences were listed separately. In the field of work organisation of interdisciplinary teams the point of good and constant communication is of decisive importance. This is achieved classically face-to-face or with the help of technical tools such as mail, messengers or video conferences. Occasionally, methods such as sharing the display content or collaborative work on the code were also mentioned.

Although good communication has been defined as a decisive success factor, the problems that currently still exist in collaboration are mainly directed towards poor communication. For example, it is mentioned that a common understanding is not given in certain cases. Nevertheless, due to the low number of responses in this area, we can conclude that at least in the teams studied there are only few problems with collaborative work. In the following, we present the results regarding specific competences in the Blockchain environment and the process of preparing the raw data.

### 2.3.4 Proccessing of the Interview Results

To reasonably reduce the amount of data collected, we choose a procedure to clean up the raw data. This procedure, which consists of removing duplications and the sensible linking of existing and new items, is illustrated below using a subcategory as an example. The example we choose is the first subcategory of technical competencies called Technical BCT Basics.

Disclaimer
The creation of these resources has been (partially) funded by the ERASMUS+ grant program of the European Union under grant no. 2018-1-LT01-KA203-047044.
Neither the European Commission nor the project's national funding agency DAAD are responsible for the content or liable for any losses or damage resulting of the use of these resources.

| | | |
|---|---|---|
| Technical BCT Basics | Knowledge of general functionality of BCT | A, B, D, E, F, G, H, I, J, M |
| | Knowledge of the foundation, components, principles of Blockchain systems | A, B, G, M |
| | Knowledge of Blockchain Architecture | H |
| | Understanding of components of the BC architectures like oracles or reverse verifiers | J |
| | Comprehension of smart contracts | A, B, C, J |
| | Comprehension of BCT application development | A, B, D |
| | Knowledge of BC Use Cases and potential applications | J |
| | Evaluation of possible business fields of application for BCT | B, D, F, M |
| | Evaluation of different Blockchain platforms and designs | C, G |
| | Evaluation of the BCT solution regarding applicability, meaningfulness and profitability | B, G |
| | Synthesis of a Blockchain platform | A, D |

| | | | Exists in Skill List |
|---|---|---|---|
| Technical BCT Basics | Knowledge of general functionality of BCT | A, B, D, E, F, G, H, I, J, M | (T-1-1) |
| | Knowledge of the foundation, components, principles of Blockchain systems | A, B, G, M | T-1-1 |
| | Knowledge of Blockchain Architecture | H | (T-1-1) |
| | Understanding of components of the BC architectures like oracles or reverse verifiers | J | (T-1-1) |
| | Comprehension of smart contracts | A, B, C, J | T-1-1 |
| | Comprehension of BCT application development | A, B, D | x |
| | Knowledge of BC Use Cases and potential applications | J | x |
| | Evaluation of possible business fields of application for BCT | B, D, F, M | T-1-6 |
| | Evaluation of different Blockchain platforms and designs | C, G | T-1-5 |
| | Evaluation of the BCT solution regarding applicability, meaningfulness and profitability | B, G | x |
| | Synthesis of a Blockchain platform | A, D | T-1-1 |

Item already exists in previous list

Item is existent but can be combined

Item missing completely

**Figure 30**: First step for preparing the raw data of the interview results.

Figure 30 shows the first step in processing the raw data. As shown at the beginning of the figure, the raw data consists of the items obtained from the qualitative content analysis and the label from which interview they were derived. The first step is to compare each item with the elements of the existing skill list. The aim is to find out whether these already exist in the original list, or whether they are comparable in content with existing items or could be added to it. In addition, we record with which existing element we compare the item to be examined. The note "T-1-1" in the figure shows as an example that we link the item with the first element, of the first subcategory in the category Technical Skills (T). "T-1-5" indicates that the fifth element of the first subcategory is chosen as the reference.

Disclaimer
The creation of these resources has been (partially) funded by the ERASMUS+ grant program of the European Union under grant no. 2018-1-LT01-KA203-047044.
Neither the European Commission nor the project's national funding agency DAAD are responsible for the content or liable for any losses or damage resulting of the use of these resources.

If the skill already exists, we highlight the item in yellow. This means that we do not need to consider the item any further. If there is a partial match, we mark the item orange. This could mean that an existing item can be modified and enriched in such a way that the original wording becomes clearer or more meaningful. A red marking indicates the presence of an item which has not been considered by the systematic literature review. This suggests that we can include this item as a new element in the final skill list. An example is the item "Comprehension of BCT application development". Although the development of Blockchain solutions already plays a decisive role in the literature-based skill list, it is only considered in the context of the specific requirements of future IT specialists. However, it became clear from the interviews that a basic understanding of the Blockchain development process can be beneficial for all employees involved and therefore must be part of the Blockchain Basics.

| | | | Exists in Skill List |
|---|---|---|---|
| Technical BCT Basics | Knowledge of general functionality of BCT | A, B, D, E, F, G, H, I, J, M | (T-1-1) |
| | Knowledge of the foundation, components, principles of Blockchain systems | A, B, G, M | T-1-1 |
| | Knowledge of Blockchain Architecture | H | (T-1-1) |
| | Understanding of components of the BC architectures like oracles or reverse verifiers | J | (T-1-1) |
| | Comprehension of smart contracts | A, B, C, J | T-1-1 |
| | Comprehension of BCT application development | A, B, D | x |
| | Knowledge of BC Use Cases and potential applications | J | x |
| | Evaluation of possible business fields of application for BCT | B, D, F, M | T-1-6 |
| | Evaluation of different Blockchain platforms and designs | C, G | T-1-5 |
| | Evaluation of the BCT solution regarding applicability, meaningfulness and profitability | B, G | x |
| | Synthesis of a Blockchain platform | A, D | T-1-1 |

Item already exists in previous list — Item is existent but can be combined — Item missing completely

| Skill Cluster | Definition of Skill Cluster | Skill item |
|---|---|---|
| **Technical BCT Basics** | Be able to explain the general BCT capabilities and functionalities of smart contracts. Be able to compare and explain different BCT solutions and plattforms. | Knowledge of the foundations, **general functionality, architecture,**<br>components, principles (e.g., cryptocurrencies, wallets, smart contracts, separate<br>Explain trust management principles<br>Define ways to maintain blockchain-based systems<br>Demonstrate blockchain technology capabilities and apply them to business-<br>Compare blockchain platforms to enable understanding of different system<br>Discuss and compare different blockchain models, scheme and solutions with<br>constructed/illustrated application (suggestions, proposals, methods for<br>**Comprehension of BCT application development**<br>**Knowledge of BC Use Cases and different application fields** |

**Figure 31**: Second step for preparing the raw data of the interview results.

Disclaimer
The creation of these resources has been (partially) funded by the ERASMUS+ grant program of the European Union under grant no. 2018-1-LT01-KA203-047044.
Neither the European Commission nor the project's national funding agency DAAD are responsible for the content or liable for any losses or damage resulting of the use of these resources.

We then use the findings of the first step of the preparation to extend the existing skill list as shown in Figure 31. The items marked in red are added as new items at the end of the skill list. In the given example these would be the items "Comprehension of BCT application development" and "Knowledge of BC Use Cases and different application fields". The item "Evaluation of the BCT solution regarding applicability, meaningfulness and profitability" no longer appears in the given sub-category, since it has been added to the sub-category "Business, Economics and Finance". The orange marked items have been linked to existing elements according to their correspondence. This is highlighted by the red inserts in the existing points.

This procedure has been carried out for all 110 items. This allows us to successfully shorten the existing data set and link it to existing results. The final skill list contains 87 items in the twelve sub-categories that were previously created. A more detailed presentation of the results is given below.

### 2.3.5 Final Skill List

The biggest section of the interviews was used to answer the first two research questions directly. These deal with the issues: What skills are necessary for employees involved in Blockchain operations, which ones are deemed the most important and why are they important? In contrast to the investigation of the research objectives “General Data”, “Blockchain Operations” and “Interdisciplinarity” we can asses the database, which we created at the beginning of the research project. This is the skill list that we created with the systematic literature review as presented in the first part of this document.

Based on these results, we created the BlockNet Competence Model, which is shown in Figure 22. This describes the necessary competences for future Blockchain experts with sixteen competence clusters. Five clusters were developed in the area of technical competencies, three in the area of methodological competencies, two in the area of social competencies and two in the area of personal competencies. By means of the questionnaire, we deductively collect further competencies within these question blocks, but can also validate the existing ones one more time. Since the raw data collected were of enormous size, we use the methodology presented in chapter 2.3.4 to prepare the data.

In the following, the results in the four competence types Technical Competences, Methodological Competences, Social Competences and Personal Competences are presented.

Disclaimer
The creation of these resources has been (partially) funded by the ERASMUS+ grant program of the European Union under grant no. 2018-1-LT01-KA203-047044.
Neither the European Commission nor the project’s national funding agency DAAD are responsible for the content or liable for any losses or damage resulting of the use of these resources.

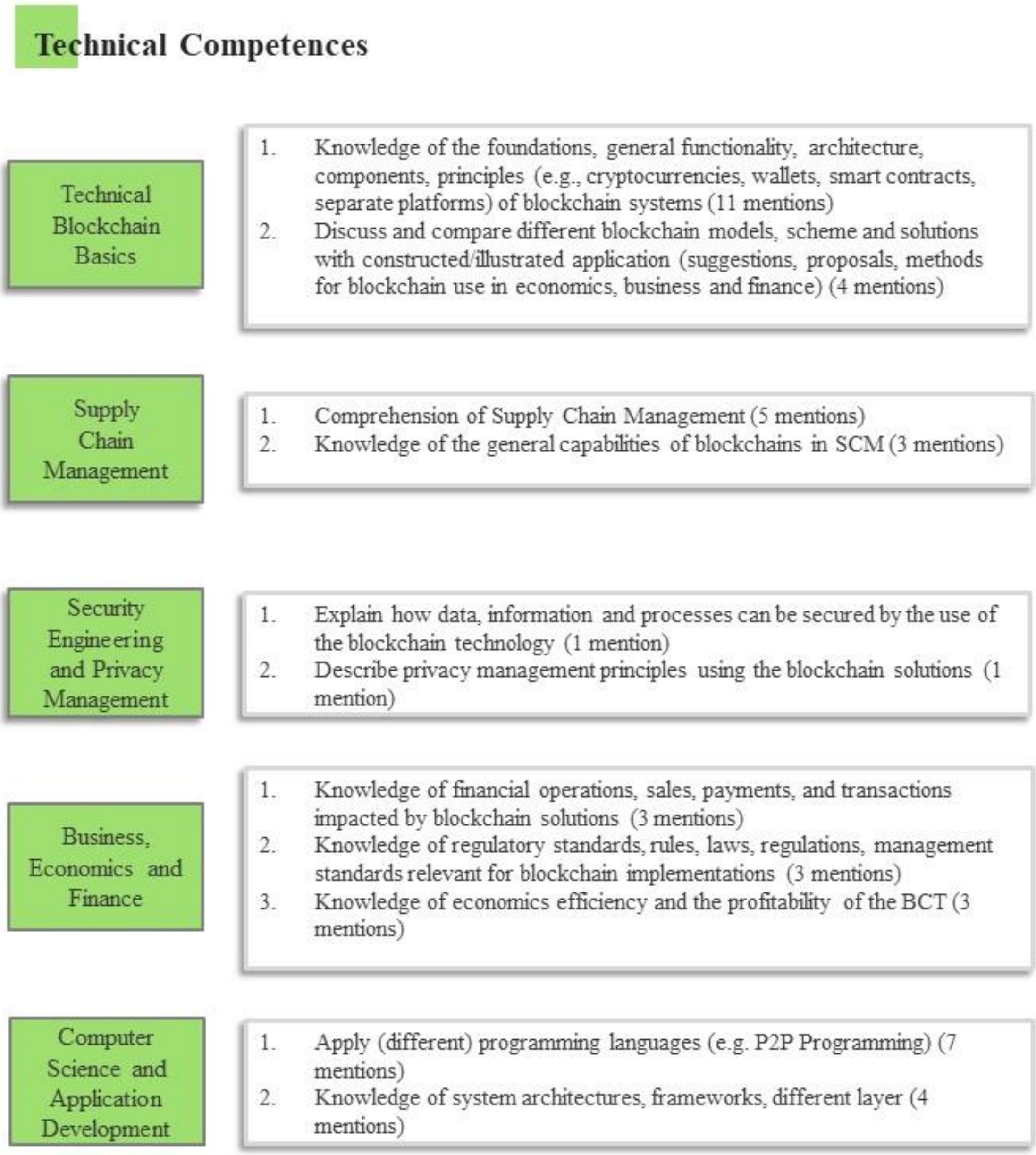

**Figure 35**: Most mentioned points regarding technical competences in the Blockchain environment.

At the beginning we present the most frequently mentioned answers from the area of technical competences. When looking at the results, it is noticeable that the point of general knowledge about the foundations, the functionality, architecture, components and the principles of the Blockchain technology is confirmed by almost all interview partners. Therefore, this point is established as an essential skill for all future Blockchain employees and therefore an important part of our future curriculum. We can further emphasize the demand for programming skills, which was confirmed by seven participants.

Disclaimer
The creation of these resources has been (partially) funded by the ERASMUS+ grant program of the European Union under grant no. 2018-1-LT01-KA203-047044.
Neither the European Commission nor the project's national funding agency DAAD are responsible for the content or liable for any losses or damage resulting of the use of these resources.

Furthermore, it is also positive to note that we can add points to each subcategory defined in the systematic literature review. These do not necessarily have to be Blockchain specific. In the area of supply chain management, for example, the most frequently mentioned item is the general understanding of SCM. This suggests that in the future, team members in this area will need to have a profound knowledge of their domain in order to be able to work successfully in Blockchain projects. The area of business, economics and finance is the area where we see the greatest diversity. Experts from these areas must not only have specific knowledge in the area of Blockchain, but also from different areas of economics such as law or profitability. In summary, it can be said that the many different competences that can be assigned to specific fields once again confirm the interdisciplinarity in Blockchain projects.

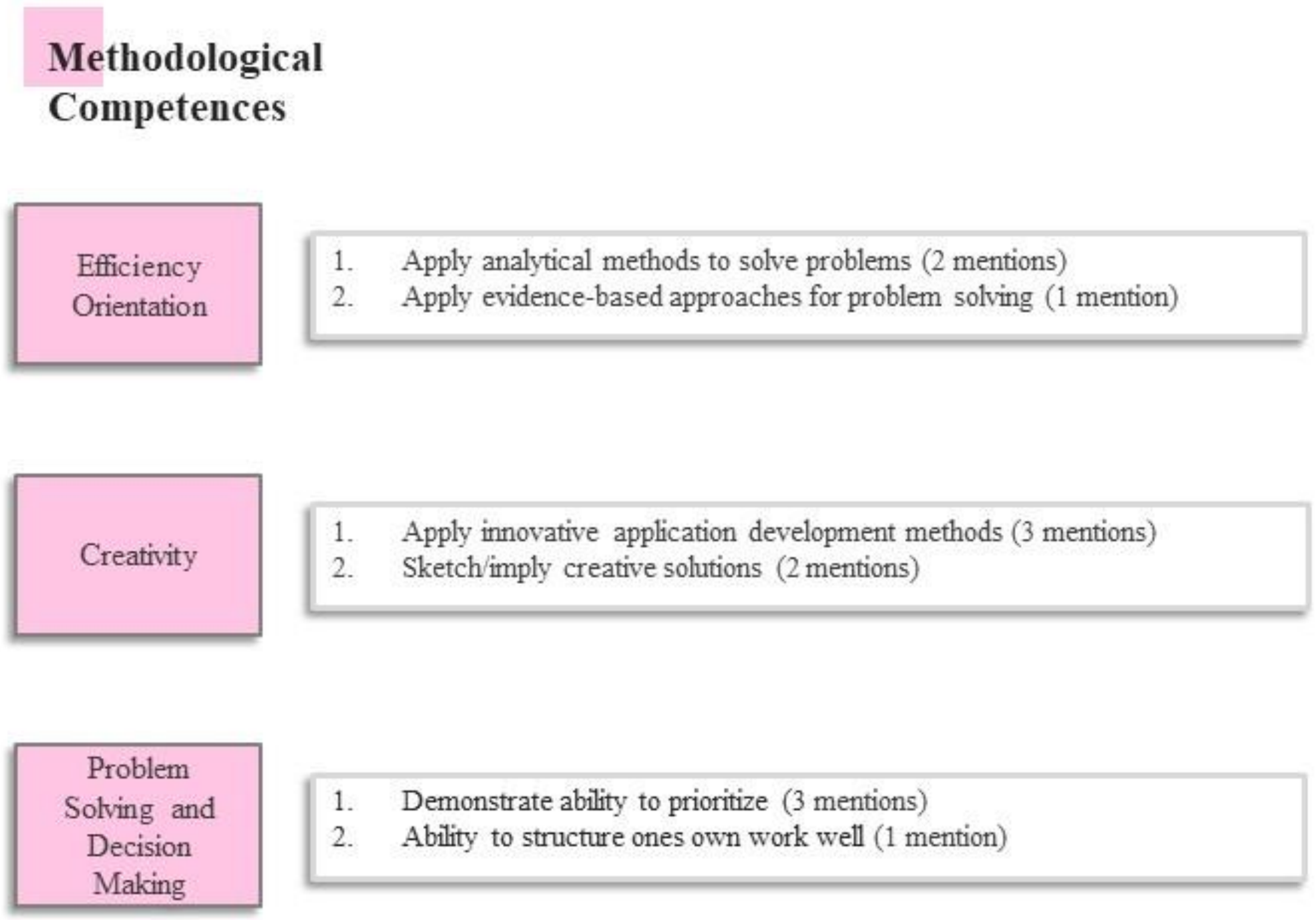

**Figure 36**: Most mentioned points regarding methodological competences in the Blockchain environment.

In the area of methodological skills, the analytical ability to solve problems can be emphasized. In addition, special attention is paid to the fact that employees in the Blockchain area can master innovative application development methods and develop creative solutions. Finally, the ability to prioritize and to structure ones own work well is highlited as a relevant.

Disclaimer
The creation of these resources has been (partially) funded by the ERASMUS+ grant program of the European Union under grant no. 2018-1-LT01-KA203-047044.
Neither the European Commission nor the project's national funding agency DAAD are responsible for the content or liable for any losses or damage resulting of the use of these resources.

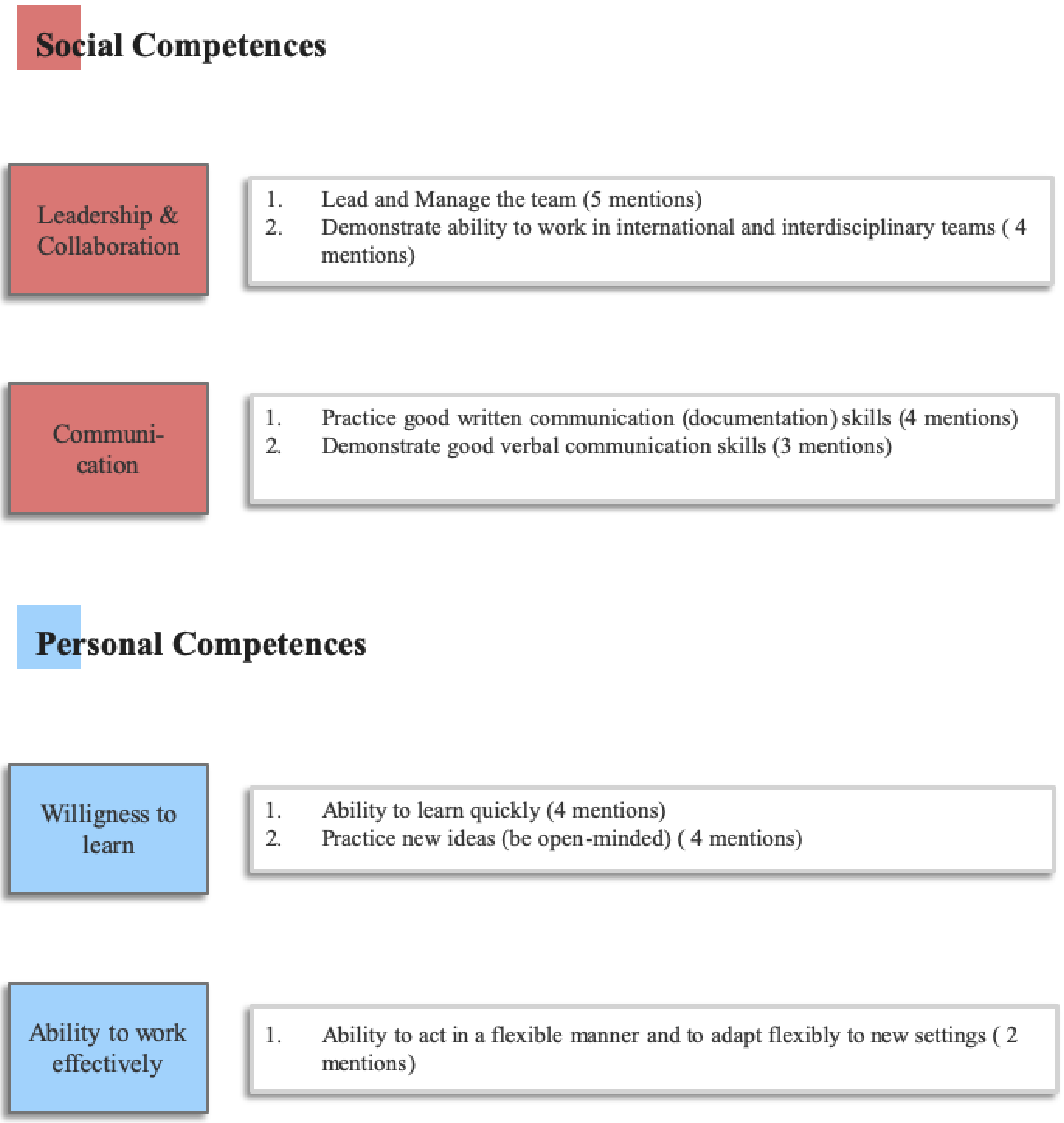

**Figure 37**: Most mentioned points regarding social and personal competences in the Blockchain environment.

As a final point we present the most important social and personal competences. The ability to lead and manage a team is considered particularly important in the case of social social competences, as this point is mentioned by five interview partners. Almost as important is the ability to lead international and interdisciplinary teams. Once again, the importance of good communication and consequently the ability to communicate well, both in written and verbal form, is stressed. As the Blockchain as a technology is still novel and therefore there Blockchain-related experience is a scarce resource, the ability to learn quickly and to be open-minded are mentioned as the most important personal competences.

In summary, we can state that the interviews not only confirm the existing items of the skill list, but that important items can also be added to it. In the interviews, some skills were also

Disclaimer
The creation of these resources has been (partially) funded by the ERASMUS+ grant program of the European Union under grant no. 2018-1-LT01-KA203-047044.
Neither the European Commission nor the project's national funding agency DAAD are responsible for the content or liable for any losses or damage resulting of the use of these resources.

mentioned which we did not include in the final skill list, as they do not represent competences in the classical sense. The most important point in this respect is experience. Many interview partners emphasize that they see it as a great advantage if people in Blockchain projects already have experience from other Blockchain initiatives. In addition, in the interviews there are also points mentioned which are important by all means but are too granular to be included as a single point. An example would be mathematical basics - a good understanding of linear algebra for computer scientists. The final skill list, which will be decisive for the creation of the BlockNet course, is shown below. The elements marked in red and bold form extensions, which we add to the list based on the interview analysis.

Disclaimer
The creation of these resources has been (partially) funded by the ERASMUS+ grant program of the European Union under grant no. 2018-1-LT01-KA203-047044.
Neither the European Commission nor the project's national funding agency DAAD are responsible for the content or liable for any losses or damage resulting of the use of these resources.

| Technical Competences | | |
|---|---|---|
| **Skill Cluster** | **Definition of Skill Cluster** | **Skill item** |
| **Technical BCT Basics** | Be able to explain the general BCT capabilities and functionalities of smart contracts. Be able to compare and explain different BCT solutions and plattforms. | Knowledge of the foundations, **general functionality, architecture,** components, principles (e.g., cryptocurrencies, wallets, smart contracts, separate platforms) of blockchain systems |
| | | Explain trust management principles |
| | | Define ways to maintain blockchain-based systems |
| | | Demonstrate blockchain technology capabilities and apply them to business-related challenges |
| | | Compare blockchain platforms to enable understanding of different system design choices |
| | | Discuss and compare different blockchain models, scheme and solutions with constructed/illustrated application (suggestions, proposals, methods for blockchain use in economics, business and finance) |
| | | **Comprehension of BCT application development** |
| | | **Knowledge of BC Use Cases and different application fields** |
| **Business, Economics and Finance** | Be able to explain processes in Business, Economics and Finance and understand the impact of BCT. | Knowledge of auditing, accounting and taxation processes as blockchain application fields |
| | | Knowledge of financial operations, sales, payments, and transactions impacted by blockchain solutions |
| | | Knowledge of regulatory standards, rules, laws, regulations, management standards relevant for blockchain implementations |
| | | **Comprehension of economic efficiency and ways to assess the profitability of BCT** |
| | | **Comprehension of market and customer needs in order to apply blockchain solutions** |
| | | **Ability to apply process designing methods (e.g. CMMN & BPMN)** |
| | | **Comprehension of Risk Management in BCT operations** |
| **Supply Chain Management** | Be able to explain processes in SCM and within corporate networks and understand the impact of BCT. | Knowledge of the general capabilities of blockchains in SCM |
| | | Explain the interoperability of blockchain technology and possible collaboration between unknown or untrusted parties in SCM |
| | | Demonstrate blockchain capabilities and apply them to counterfeit and fraud prevention problem statements |
| | | Demonstrate blockchain capabilities and apply them to provenance and track&trace problem statements in SCM |
| | | Analyze how information asymmetry in corporate networks can be addressed by the blockchain-based applications |
| | | **Comprehension of Supply Chain Management** |
| | | **Ability to design a concept for the use of BCT in SCM** |

Disclaimer
The creation of these resources has been (partially) funded by the ERASMUS+ grant program of the European Union under grant no. 2018-1-LT01-KA203-047044.
Neither the European Commission nor the project's national funding agency DAAD are responsible for the content or liable for any losses or damage resulting of the use of these resources.

| | | |
|---|---|---|
| **Computer Science and application development** | Be able to explain the functionalities of the technical elements BCT consists of and understand development requirements in BCT environment. | Comprehension of development processes for blockchain solutions |
| | | Knowledge of system architectures, frameworks, different layers |
| | | Discuss software quality goals and their impact on blockchain system development |
| | | Describe the software requirements elicitation and engineering process of blockchain systems |
| | | Develop and manage databases using data management systems (also use of SQL, etc.) |
| | | Knowledge of network protocols |
| | | Apply different programming languages **(e.g. P2P Programming)** |
| | | Develop a testing plan for concrete blockchain solutions |
| | | **Knowledge of Cryptocurrencies & Cryptocurrency Coding** |
| | | **Ability to apply methods of systems engineering** |
| | | **Comprehension of interface management** |
| | | **Ability to apply formal abstraction** |
| | | **Knowledge of the connection bewtween Data Science & BCT** |
| | | **Abbility to programm smart contracts** |
| **Security Engineering and Privacy Management** | Be able to explain the impacts BCT applications have in relation to security and privacy management. | Describe privacy management principles using the blockchain solutions |
| | | Explain identity management principles using the blockchain solutions |
| | | Explain how data, information and processes can be secured by the use of the blockchain technology |
| | | Recognise security countermeasure implications |
| | | Explain access control (authentication, authorization and identity) models |
| | | Describe transaction protection and validation principles |
| | | Underline major encryption and signature schemes |
| | | State major fair mining principles |
| | | Identify security errors in smart contracts |

Disclaimer
The creation of these resources has been (partially) funded by the ERASMUS+ grant program of the European Union under grant no. 2018-1-LT01-KA203-047044.
Neither the European Commission nor the project's national funding agency DAAD are responsible for the content or liable for any losses or damage resulting of the use of these resources.

BlockNet

| Methodological Competences | | |
|---|---|---|
| **Skill Cluster** | **Definition of Skill Cluster** | **Skill item** |
| **Efficiency orientation** | Be able to work efficiently in a BCT project. | Ability to transfer knowledge to internal (e.g., colleagues, developers, testers, etc.) and external (e.g., customers, support teams, and etc.) stakeholders |
| | | Demonstrate ability to prioritise and to have a good time-management |
| | | Ability to organise interdisciplinary work |
| | | **Ability to structure ones own work well** |
| **Creativity** | Be able to use creative approaches in BCT projects. | Sketch/imply creative solutions |
| | | Apply innovative application development methods |
| | | Practice creative solutions |
| | | Knowledge of new interdisciplinary working methods |
| **Problem solving and Decision making** | Be able to find the right decision and solve problems in BCT projects. | Apply analytical methods to solve problems |
| | | Apply evidence-based approaches for problem solving |
| | | Apply critical thinking |

| Social Competences | | |
|---|---|---|
| **Skill Cluster** | **Definition of Skill Cluster** | **Skill item** |
| **Leadership and Colloboration** | Be able to lead a team and network to successfully run a BCT project. | Lead and manage the team |
| | | Demonstrate strong (inter-) organisational networking skills |
| | | Demonstrate ability to work in international and interdisciplinary teams |
| | | Practice to support colleagues with expert knowledge |
| | | Establish good social relationships with the customers |
| | | **Ability to allocate team members according to their specific competences in different BC tasks** |
| | | **Ability to work effective and collaborative in a team** |
| | | **Willigness to behave socially and ethically correct** |
| | | **Ability to mediate between different team members and align the knowledge level of different team members** |
| **Commun-ication** | Be able to communicate well in order to successfully run a BCT project. | Demonstrate good written communication (documentation) skills |
| | | Demonstrate good verbal communication skills |
| | | Demonstrate ability to communicate complex and interdisciplinary problems |
| | | Use social media means |
| | | Demonstrate communication skills to internal and external stakeholders (colleagues, users, customers, advisors, and etc.) |
| | | Demonstrate good presentation skills |
| | | **Ability to communicate within intercultural teams** |
| | | **Ability to demonstrate negotiation skills considering different cultural background** |

Disclaimer
The creation of these resources has been (partially) funded by the ERASMUS+ grant program of the European Union under grant no. 2018-1-LT01-KA203-047044.
Neither the European Commission nor the project's national funding agency DAAD are responsible for the content or liable for any losses or damage resulting of the use of these resources.

Erasmus+

| Personal Competences | | |
|---|---|---|
| **Skill Cluster** | **Definition of Skill Cluster** | **Skill item** |
| **Willingness to learn** | Be able to learn effectively in order to successfully run a BCT project. | Ability to learn quickly |
| | | Demonstrate Interest in new technology |
| | | Demonstrate Interest in continuing learning |
| | | Ability to accept and take into account feedback |
| | | Apply new ideas and be open-minded |
| | | **Ability to educate oneself** |
| | | **Willigness to work without existing partners/examples in the area of running BCT Projects** |
| **Abilitiy to work effectively** | Be able to work effectively in order to successfully run a BCT project. | Demonstrate ability to work independently and self-organized |
| | | Be proactive and take initiative |
| | | Be responsible, trusted, and committed |
| | | Ability to determine qualitative results |
| | | **Ability to act in a flexible manner and to adapt easily to new settings** |
| | | **Demonstrate honesty and accurate working methods** |
| | | **Demonstrate resilience and the ability to continue working on tasks despite difficulties** |

## 2.4 Concluding Remarks

In this second part of the document, we report on the execution of a case study which analyzes Blockchain use cases in different business fields. By doing so we extend the existing Skill List of the first part of this document with practice orientated outcomes. The study consists mainly of expert interviews with the goal to validate and prioritise existing skills items, as well as to complement it with specific competences in the field of Blockchain obtained from the use cases analysis.

The identified competences will be essential in the development of a didactical concept for the interdisciplinary Blockchain SNOC. This will be developed in the next interlectual output to provide a scientifically sound basis for the BlockNet online course.

Disclaimer
The creation of these resources has been (partially) funded by the ERASMUS+ grant program of the European Union under grant no. 2018-1-LT01-KA203-047044.
Neither the European Commission nor the project's national funding agency DAAD are responsible for the content or liable for any losses or damage resulting of the use of these resources.

Some icons provided by https://icons8.de/

Disclaimer
The creation of these resources has been (partially) funded by the ERASMUS+ grant program of the European Union under grant no. 2018-1-LT01-KA203-047044.
Neither the European Commission nor the project's national funding agency DAAD are responsible for the content or liable for any losses or damage resulting of the use of these resources.

## Appendix 1.4

# NON-DISCLOSURE AGREEMENT

BETWEEN: Interviewer
Max Mustermann, born. XX.XX.XXXX in Musterstadt
Technical University Dortmund
Lehrstuhl für Unternehmenslogistik
Leonhard-Euler-Straße 5, 44227 Dortmund, Germany

AND: Interviewee
Erika Mustermann
[Name of Institution]
[Department]
[Address]

AGREEMENT:

We assure [Name of interviewee] that the information obtained through the interview on [Date] will only be used within the research project "BlockNet - BlockChain Network Online Education for interdisciplinary European Competence Transfer".

In order to be able to process the information gained appropriately, it is planned to document the interview with the help of a sound recording. If this is expressly not desired, please let us know before the interview begins. The integration of the obtained information will be carried out anonymously without any personal or company name.

At the end of the project (probably in February 2021), the results of the research project will be published and made available to the participating persons/companies.

………………………………………………………… ……………………………………………………………

Date, Signature interviewee Date, Signature interviewer

Disclaimer
The creation of these resources has been (partially) funded by the ERASMUS+ grant program of the European Union under grant no. 2018-1-LT01-KA203-047044.
Neither the European Commission nor the project's national funding agency DAAD are responsible for the content or liable for any losses or damage resulting of the use of these resources.

# Appendix 1.5 spider diagram

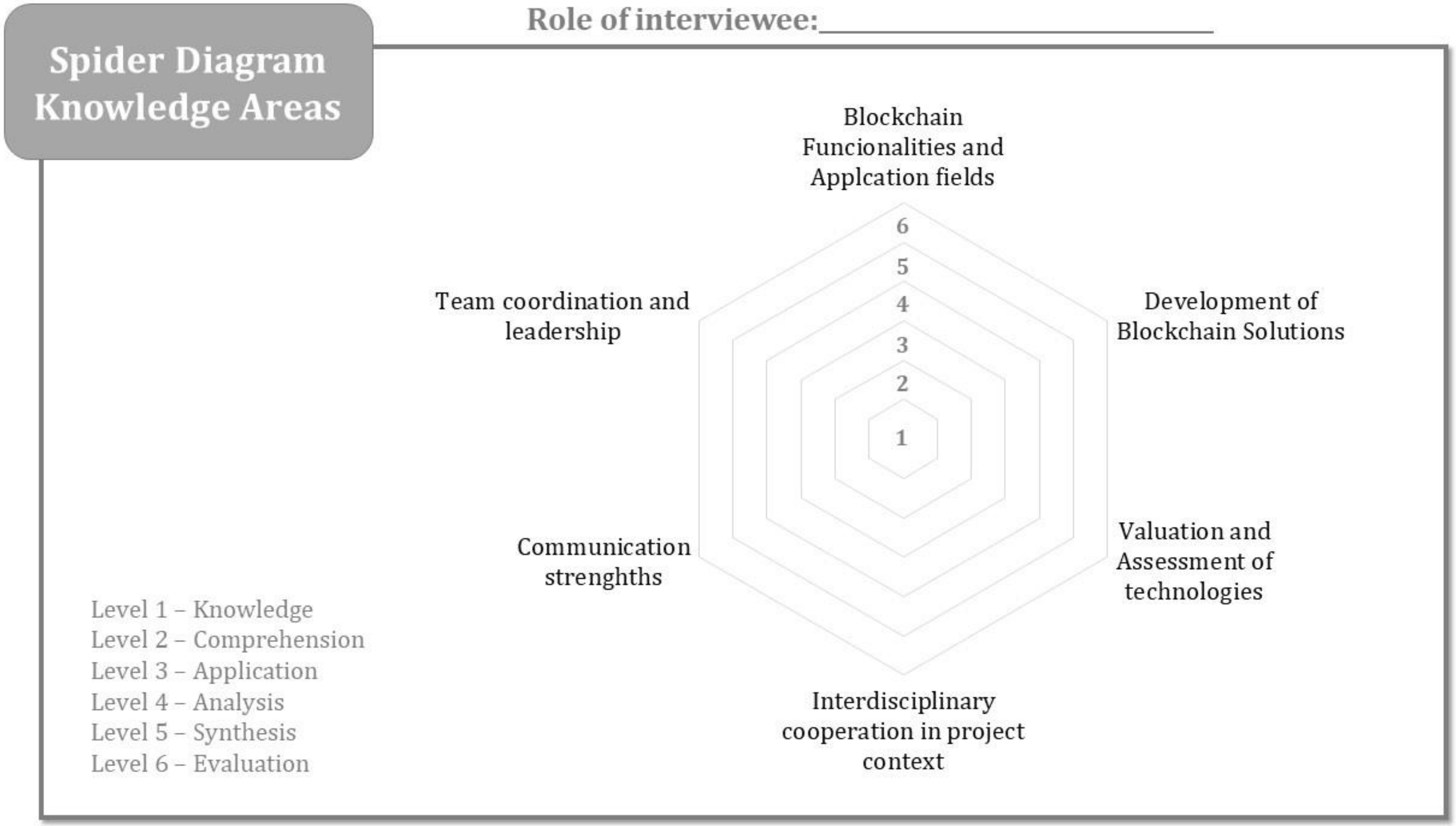

Disclaimer
The creation of these resources has been (partially) funded by the ERASMUS+ grant program of the European Union under grant no. 2018-1-LT01-KA203-047044.
Neither the European Commission nor the project’s national funding agency DAAD are responsible for the content or liable for any losses or damage resulting of the use of these resources.